\documentclass[usenatbib]{mn2e}
\usepackage{apjfonts,amsfonts,amsmath,amssymb,bm,ctable,verbatim}

\newcommand{\gizmourl}{\href{http://www.tapir.caltech.edu/~phopkins/Site/GIZMO.html}{\url{http://www.tapir.caltech.edu/~phopkins/Site/GIZMO.html}}}
\newcommand{\FIREurl}{\href{http://fire.northwestern.edu}{\url{http://fire.northwestern.edu}}}

\newcommand{\etal}{et al.}

\newcommand{\acknowledgments}[1]{\begin{small}\section*{Acknowledgments}\end{small}{\noindent #1}\vspace{5pt}}
\newcommand{\datastatement}[1]{\begin{small}\section*{Data Availability Statement}\end{small}{\noindent #1}\vspace{5pt}}

\title[New Softening, Now with Tides]{Novel Conservative Methods for Adaptive Force Softening in Collisionless and Multi-Species $N$-Body Simulations}

\author[Hopkins \etal]{
\parbox[t]{\textwidth}{
Philip F.~Hopkins$^1$, Ethan O.~Nadler$^{2,3}$, Michael Y.\ Grudi{\'c}$^2$, Xuejian Shen$^1$, \\ 
Isabel Sands$^1$, \&\ Fangzhou Jiang$^{1,2}$
}\vspace*{4pt} \\
$^1$ TAPIR, Mailcode 350-17, California Institute of Technology, Pasadena, CA 91125, USA. E-mail:phopkins@caltech.edu \\
$^2$ Carnegie Observatories, 813 Santa Barbara Street, Pasadena, CA 91101, USA \\
$^3$ Department of Physics $\&$ Astronomy, University of Southern California, Los Angeles, CA, 90007, USA}

\date{}
\begin{document}
\maketitle

\begin{abstract}
Modeling self-gravity of collisionless fluids (e.g.\ ensembles of dark matter, stars, black holes, dust, planetary bodies) in simulations is challenging and requires some force softening. It is often desirable to allow softenings to evolve adaptively, in any high-dynamic range simulation, but this poses unique challenges of consistency, conservation, and accuracy, especially in multi-physics simulations where species with different ``softening laws'' may interact. We therefore derive a generalized form of the energy-and-momentum conserving gravitational equations of motion, applicable to arbitrary rules used to determine the force softening, together with consistent associated timestep criteria, interaction terms between species with different softening laws, and arbitrary maximum/minimum softenings. We also derive new methods to maintain better accuracy and conservation when symmetrizing forces between particles. We review and extend previously-discussed adaptive softening schemes based on the local neighbor particle density, and present several new schemes for scaling the softening with properties of the gravitational field, i.e.\ the potential or acceleration or tidal tensor. We show that the `tidal softening' scheme not only represents a physically-motivated, translation and Galilean invariant and equivalence-principle respecting (and therefore conservative) method, but imposes negligible timestep or other computational penalties, ensures that pairwise two-body scattering is small compared to smooth background forces, and can resolve outstanding challenges in properly capturing tidal disruption of substructures (minimizing artificial destruction) while also avoiding excessive $N$-body heating. We make all of this public in the {\small GIZMO} code.
\end{abstract}


\section{Introduction}
\label{sec:intro}

Gravitational dynamics are ubiquitous across astrophysics and are commonly studied via numerical simulations. In a huge variety of astrophysical contexts, including e.g.\ cosmology, galaxy formation, stellar dynamics, star formation, dark matter physics, dust and particulate dynamics, planetesimal and planet formation and dynamics, it is often necessary to evolve the gravitational dynamics and self-gravity of collisionless or weakly-collisional fluids. These are systems where the effective collisional mean free paths are large compared to the resolution scale, and where the number of true micro-physical bodies (e.g.\ $10^{60}-10^{90}$ dark matter particles, in a cosmological simulation) is larger than the number of resolution elements, so each represents many micro-particles. In this limit one should, of course, evolve the full six-dimensional phase-space distribution function $f({\bf x},\,{\bf v},\,t)$ in time \citep{cheng:1976.vlasov.split,sonnedrucker:1999.semi.lagrangian.vlasov,binneytremaine,weinberg:bar.res.1,mitchell:2013.boltzmann.moments.methods.for.stars.dm}. But in most astrophysical  applications noted above, this would be prohibitively computationally expensive for all but highly idealized or specialized problems \citep[see e.g.][]{filbet:2001.conservative.vlasov,abel:phase.space.sheets.for.dm,hahn:2013.dm.phase.space.sheet.sims,yoshikawa:2013.6d.vlasov.direct,colombi:2014.waterbags,sousbie:2016.vlasov.poisson.tesselation,mocz:2017.integer.lattice.vlasov,tanaka:2017.6d.phase.space.solvers.finite.volume}.

Instead, by far the most common approach is to evolve the gravitational dynamics of these collisionless fluids via standard, computationally efficient and extremely well-studied $N$-body methods \citep{hernquistkatz:treesph,hernquist:1990.barnes.nbody.review,makino.aarseth.1992:hermite.integrator,aarseth:2003.nbody.review}, making it a Monte Carlo method. But in this case, preventing numerical divergences and catastrophically large errors or short timesteps, as well as ensuring physical behavior of the underlying system necessitates introducing some force softening \citep{dyer.ip:1993.softening.law,earn.sellwood:95.nbody.bar.stab,merritt:1996.optimal.softening,melott:1997.discreteness.error.nbody,kravtsov:1997.ART,steinmetz:1997.two.body.nbody.heating,romeo.1998:optimal.softening,athanassoula:2000.optimal.force.softening.collisionless.sims}.
While it has been recognized for decades that such softening terms for collisionless fluids in $N$-body methods cannot be derived strictly self-consistently, there are still commonly desired goals of such a method (discussed in the references above): namely that it should retain all the symmetries and conservation laws of the underlying system (e.g.\ be translation and Galilean and gauge-invariant, conserve momentum and energy and respect the equivalence principle, represent a positive-definite mass-energy density); it should minimize ``two body'' scattering (the energy/momentum scattering between two particles on close approach should be small compared to the collective background forces); it should be computationally efficient; and it should (ideally) be at least somewhat related to the physical distribution function (e.g.\ the ``extent'' or domain of softening, and resulting forces, should not be unphysical or wildly different from a reasonable model for the spatial extent and true forces of the matter represented by the $N$-body particles). 

In practice it is challenging to satisfy all of these conditions simultaneously. In simulations with large dynamic range, softenings fixed in time, while still the most common approach for collisionless fluids and easy to discretize in a manner satisfying the first condition, cannot adapt in different local environments as needed to satisfy the latter three conditions (except in some sort of global-average or median sense; see e.g.\ \citealt{romeo.1998:optimal.softening,athanassoula:2000.optimal.force.softening.collisionless.sims,dehnen:2001.optimal.softening}). A variety of ``adaptive force softening'' schemes have therefore been proposed in the literature, in which softening parameters (e.g.\ a softening length $\varepsilon$) for a given particle can evolve \citep[for some examples, see][]{1986desd.book.....S,hernquistkatz:treesph,merritt:1996.optimal.softening,dehnen:2001.optimal.softening}. For some time, these had serious problems with energy and phase space conservation, but \citet{price:2007.lagrangian.adaptive.softening} showed how a conservative form of the equations of motion could be derived, which was then implemented and expanded upon in \citet{springel:arepo,iannuzzi:2011.collisionless.adaptive.softening.gadget,barnes:2012.softening.is.smoothing,hopkins:gizmo} for a small number of {\em specific} schemes for setting $\varepsilon$. But those specific schemes, while widely used for the gravitational dynamics of {\em collisional} fluids where a self-consistent softening can be rigorously defined, have limitations that have largely restricted their use for collisionless fluid $N$-body dynamics.

In this paper, we therefore revisit this question and derive a general family of equations of motion for arbitrary schemes to define force softening in simulations with collisionless components (\S~\ref{sec:deriv}). We provide a number of concrete examples, and introduce several novel schemes, including one in particular based on the gravitational tidal tensor, which appear to avoid all of the major difficulties that have been cited for applications of previous schemes to collisionless fluids (\S~\ref{sec:ex}). We show how maximum and minimum softenings (often necessary) can be self-consistently included in these formulations (\S~\ref{sec:minmax}), and derive timestep criteria for any such scheme needed to ensure stability and accuracy (\S~\ref{sec:timesteps}). We derive a general formulation for how to handle ``mixed'' interactions between two different species which obey {\em different} softening rules, crucial for modern multi-physics simulations (\S~\ref{sec:multi}). We discuss different schemes to symmetrize the softened potential and derive a new scheme which ensures correct physical behaviors in situations with highly-unequal softenings between neighbor particles (\S~\ref{sec:symm}). We discuss various choices for the functional form of softening (\S~\ref{sec:kernels}), computational expense (\S~\ref{sec:cpu}), summarize advantages of different choices (\S~\ref{sec:pros}), test these different choices and examine their effects in commonly-studied applications to cosmological dark matter simulations (\S~\ref{sec:cosmo}), and conclude in \S~\ref{sec:conclusions}. In the Appendices we discuss subtleties of cross-application of the results here to cases where $f({\bf x},\,{\bf v},\,t)$ {\em is} evolved (e.g.\ fluid dynamics; \S~\ref{sec:gas}) and of the ``self'' potential terms (\S~\ref{sec:self}), give some estimates for reasonable normalization of the tidal schemes (\S~\ref{sec:tidal.norm}), and provide some additional tests (\S~\ref{sec:extra.tests}). And we provide the complete details of the numerical implementation, by making public a modular implementation of these methods in the {\small GIZMO}\footnote{\gizmourl} code \citep{hopkins:gizmo}.

\section{Derivation}
\label{sec:deriv}

A collisionless (noting for now that we neglect gas), non-relativistic system of some arbitrarily large number ($N_{\rm true}\gg1$ per macroscopic ``resolution element'') of spatially compact conserved masses in $D=3$ dimensions\footnote{It is straightforward to generalize the results here to $D\ne 3$ dimensions, but we will focus on this as the case of practical interest.} is described by the distribution function $f = \sum_{s} f_{s}({\bf x},\,{\bf v},\,t,\,s)$ (defined here as the phase-space density of mass) where $s$ denotes different species, which obeys the Vlasov-Poisson equations: 
$\partial_{t} f_{s} + {\bf v} \cdot \partial_{\bf x} f_{s} - (\partial_{\bf x} \Phi) \cdot \partial_{\bf v} f_{s} = 0$, in terms of time $t$, coordinate position ${\bf x}$, velocity ${\bf v} = \dot{\bf x} = {\rm d}_{t} {\bf x}$, and gravitational potential $\Phi$ defined by $\nabla^{2}_{\bf x} \Phi = 4\pi\,G \sum_{s} \int f_{s}\,d^{3}{\bf v}$. This is equivalent to a Lagrangian
\begin{align}
\mathcal{L} &\equiv \sum_{s}\,\int {\rm d}^{3}{\bf x}\,{\rm d}^{3}{\bf v}\,f_{s} \,\left(\frac{1}{2}\,{\bf v} \cdot {\bf v} - \Phi \right)
\end{align}
with $\Phi = \Phi({\bf x},\,t) = \sum_{s} \int {\rm d}^{3}{\bf x}^{\prime}\,{\rm d}^{3}{\bf v}^{\prime}\,G\,f_{s}({\bf x}^{\prime},\,{\bf v}^{\prime},\,t) / |{\bf x} - {\bf x}^{\prime} |$. 

Now we discretize this into a system of $N$ discrete $N$-body elements or ``particles'' in our simulation each with {\em total} $N$-body particle mass $m_{b}$, giving: 
\begin{align}
\label{eqn:lagrangian} \mathcal{L} &\equiv 
\sum_{b=1}^{N} m_{b}\, \frac{\mathbf{v}_{b} \cdot \mathbf{v}_{b}}{2} + \sum_{b=1}^{N}\sum_{c>b}^{N} m_{b}\,m_{c}\,\tilde{\phi}_{bc}({\bf x}_{b} - {\bf x}_{c},\,\varepsilon_{b},\,\varepsilon_{c})
\end{align}
with ${\bf v}_{b} = \dot{{\bf x}}_{b}$. This is the starting point for essentially {\em all} $N$-body gravity methods, regardless of whether the equations are solved via tree, particle-mesh, multipole, or other numerical methods. What matters is that we have already made the ``Monte Carlo'' approximation:  Eq.~\ref{eqn:lagrangian} can be simply obtained via approximating the distribution function (DF) $f_{s}({\bf x},\,{\bf v}) \rightarrow \sum_{b} m_{\rm b}\,\delta({\bf x}-{\bf x}_{b},\, {\bf v}-{\bf v}_{b})$ \citep{hernquistkatz:treesph,dehnen:2001.optimal.softening,barnes:2012.softening.is.smoothing}.\footnote{Note that even before considering gravity we have replaced the integral over $f_{s}{\bf v} \cdot {\bf v}$ with $m_{b}\,{\bf v}_{b}\cdot {\bf v}_{b}$. This means we have assumed the ``particle'' is dynamically cold, since we do not explicitly evolve higher order moments of the velocity distribution function \citep{hahn:2013.dm.phase.space.sheet.sims}.} Although the only definition of the function $\tilde{\phi}$ linked strictly-self-consistently to a unique DF evolving exactly according to the collisionless Vlasov-Poisson equation is the Keplerian (point-mass) $\tilde{\phi}_{bc} = G/|{\bf x}_{b}-{\bf x}_{c}|$, for the usual reasons given in \S~\ref{sec:intro} we will introduce ``softened'' gravity, via the function $\tilde{\phi}_{bc}$ which depends on some ``softening parameter'' $\varepsilon$. 

We do not need to specify the form of $\tilde{\phi}$ yet, but in order to ensure it is both physically {\em and numerically} well-behaved and computationally efficient to calculate, $\tilde{\phi}_{bc}$ must be (a) finite, (b) sufficiently smooth (have continuous and smooth third derivatives or higher, equivalent to a smooth and differentiable $f_{s}$), (c) exactly Keplerian (and independent of $\varepsilon$) at sufficiently large values of $|{\bf x}_{b} - {\bf x}_{c} | / \varepsilon$ (the ``exactly'' part here is equivalent to the statement that the non-Keplerian portion of $\tilde{\phi}$ has compact support\footnote{Most important, this is required  in order for the methods herein to be computationally tractable. But kernels with infinite support, like the classic Plummer softening, also entail a significant loss of accuracy \citep{dehnen:2001.optimal.softening}, and systematic bias or errors in different phenomena such as large-scale structure on scales arbitrarily large relative to the nominal ``softening'' \citep{joyce.marcos:2007.n.body.fx.shell.crossing,garrison:2018.abacus.sims.cosmo.nbody}, as well as being unphysical or incompatible with boundary conditions in certain specific situations.}), (d) spherically symmetric (depending just on $|{\bf x}_{b} - {\bf x}_{c}|$), and (e) translationally symmetric, i.e.\ $\tilde{\phi}_{bc} = \tilde{\phi}_{cb}$ (otherwise the Lagrangian would lose translational symmetry and the resulting equations-of-motion would not conserve momentum). Also note that the sum $\sum_{b}\sum_{c>b}$ in Eq.~\ref{eqn:lagrangian} is such that every unique pair $b\ne c$ appears exactly once (as they should), and importantly and correctly does not include a ``self'' ($c=b$) contribution which is fundamentally undefined for such a system and leads to numerous physical and numerical inconsistencies (see discussion in \S~\ref{sec:self}). However for notational convenience in what follows, we can simply define $\tilde{\phi}_{bb} = 0$ and use the symmetry properties of $\tilde{\phi}$ to replace $\sum_{b=1}^{N}\sum_{c>b}^{N}\,m_{b}\,m_{c} \tilde{\phi}_{bc} = (1/2)\, \sum_{b} \sum_{c} m_{b}\,\tilde{\phi}_{bc}$ (the $1/2$ accounting for the fact that every pair now appears exactly twice), and drop the explicit notation of summation from $1$ to $N$ (this is implicit).

Finally, we must define some rule for $\varepsilon$, which we can write in general as: 
\begin{align}
\varepsilon_{b} &\equiv \mathcal{E}\left( \mathbf{G}_{b} \equiv \sum_{c \ne b} \boldsymbol{\mathcal{G}}_{bc}\left( {\bf x}_{b} - {\bf x}_{c},\, \mathbf{U}_{b},\, \mathbf{U}_{c} \right) \ , \ \mathbf{U}_{b} \right)
\end{align}
where $\mathcal{E}$ is some arbitrary scalar function of the tensor $\mathbf{G}_{b}$ (itself given by a sum over the arbitrary tensor function $\boldsymbol{\mathcal{G}}_{bc}$ for each pair $b$, $c$) and some ``state vector'' $\mathbf{U}_{b}$ (which can include the value of $\varepsilon_{b}$ itself, making this an implicit equation for $\varepsilon$, but also particle mass and labels like the particle ``type'' or species in multi-species simulations).

Now we can immediately obtain the usual Euler-Lagrange equations:
\begin{align}
\label{eqn:ele} \frac{{\rm d}\mathbf{p}_{a} }{{\rm d}t} &= m_{a}\,\frac{{\rm d}\mathbf{v}_{a} }{{\rm d}t} = \frac{{\rm d} \mathcal{L}}{{\rm d} {\bf x}_{a}}  \\ 
&= -\frac{1}{2} \sum_{b}\sum_{c} m_{b} m_{c} \left[ \nabla_{a} \tilde{\phi}_{bc}  + \frac{\partial \tilde{\phi}_{bc}}{\partial \varepsilon_{b}}\,\frac{{\rm d} \varepsilon_{b}}{{\rm d} {\bf x}_{a}}  + \frac{\partial \tilde{\phi}_{bc}}{\partial \varepsilon_{c}}\,\frac{{\rm d} \varepsilon_{c}}{{\rm d} {\bf x}_{a}}  \right] \\
\label{eqn:ele.2} &= - \sum_{b} \frac{m_{a}\,m_{b}}{2}  {\nabla_{a} \left[  \tilde{\phi}_{ab}  +  \tilde{\phi}_{ba} \right]} \\
\nonumber & \ \ \ \ \ \ \ \ \ \  - \frac{1}{2} \sum_{b}\sum_{c} \,m_{b}\,m_{c} \left[ \frac{\partial \tilde{\phi}_{bc}}{\partial \varepsilon_{b}} \frac{{\rm d}  \varepsilon_{b}}{{\rm d} {\bf x}_{a}} +  \frac{\partial \tilde{\phi}_{cb}}{\partial \varepsilon_{c}} \frac{{\rm d} \varepsilon_{c}}{{\rm d} {\bf x}_{a}} \right] 
\end{align}
where we use the partial derivatives ``$\partial$'' and $\nabla_{a} \equiv \partial_{{\bf x}_{a}}$ to denote derivatives with fixed values of all other quantities, whereas ``${\rm d}$'' denotes the total derivative with respect to ${\bf x}_{a}$, ${\bf U}$, etc.\footnote{Though still at fixed ${\bf x}_{b}$.} 
The terms in ${\rm d}\varepsilon/{\rm d}{\bf x}_{a}$ in Eq.~\ref{eqn:ele.2} are the ``grad-$\varepsilon$'' terms as introduced by \citet{price:2007.lagrangian.adaptive.softening}, akin to the usual ``grad-$h$'' terms in smoothed-particle hydrodynamics (SPH) methods (e.g.\ \citealt{springel:entropy}).
Therefore, 
\begin{align}
\label{eqn:h.deriv.chain} \frac{{\rm d} \varepsilon_{b}}{{\rm d} {\bf x}_{a}} &= \frac{\partial \mathcal{E}_{b}}{\partial \mathbf{U}_{b}} \cdot \frac{{\rm d} \mathbf{U}_{b}}{{\rm d} {\bf x}_{a}} +  \frac{\partial \mathcal{E}_{b}}{\partial \mathbf{G}_{b}} \cdot \sum_{c} \left( \frac{\partial \boldsymbol{\mathcal{G}}_{bc}}{\partial {\bf x}_{a}}
+ \frac{\partial \boldsymbol{\mathcal{G}}_{bc}}{\partial \mathbf{U}_{b}} \cdot \frac{{\rm d} \mathbf{U}_{b}}{{\rm d} {\bf x}_{a}}
+ \frac{\partial \boldsymbol{\mathcal{G}}_{bc}}{\partial \mathbf{U}_{c}} \cdot \frac{{\rm d} \mathbf{U}_{c}}{{\rm d} {\bf x}_{a}}
 \right) \ .
\end{align}
The usual (as in SPH) complication arising here is that ${\rm d}\varepsilon/{\rm d}{\bf x}$ appears not only on the left-hand side of Eq.~\ref{eqn:h.deriv.chain}, but also as part of ${\rm d}\mathbf{U}/{\rm d}{\bf x}$ on its right-hand side.
For all cases we consider below, the only component of ${\bf U}$ which has non-trivially vanishing derivatives in Eq.~\ref{eqn:h.deriv.chain} is $\varepsilon$. So we can set ${\rm d}\mathbf{U}\rightarrow {\rm d}\varepsilon$ in Eq.~\ref{eqn:h.deriv.chain} (since all other derivatives of other terms in $\mathbf{U}$ vanish), and rearrange, giving:
\begin{align}
\label{eqn:presum} \frac{{\rm d} \varepsilon_{b}}{{\rm d} {\bf x}_{a}} &= \frac{1}{\Omega^{\prime}_{b}} \frac{\partial \mathcal{E}_{b}}{\partial \mathbf{G}_{b}} \cdot \sum_{c} \left( \frac{\partial \boldsymbol{\mathcal{G}}_{bc}}{\partial {\bf x}_{a}} + \frac{\partial \boldsymbol{\mathcal{G}}_{bc}}{\partial \varepsilon_{c}}  \frac{{\rm d} \varepsilon_{c}}{{\rm d} {\bf x}_{a}}
 \right) \\ 
 \Omega^{\prime}_{b} &\equiv 1 - \frac{\partial \mathcal{E}_{b}}{\partial \varepsilon_{b}} - \frac{\partial \mathcal{E}_{b}}{\partial \mathbf{G}_{b}} \cdot \sum_{c} \frac{\partial \boldsymbol{\mathcal{G}}_{bc}}{\partial \varepsilon_{b}}
\end{align}
As discussed in detail in Appendix~\ref{sec:deriv.details}, if $\partial\boldsymbol{\mathcal{G}}_{bc}/\partial\varepsilon_{c} \rightarrow 0$ vanishes, then Eq.~\ref{eqn:presum} is an explicit equation for ${{\rm d} \varepsilon_{b}}/{{\rm d} {\bf x}_{a}}$ which is straightforward to evaluate. We show in \S~\ref{sec:ex} that this can easily be ensured for all methods we discuss in this paper, by appropriate choice of the function $\boldsymbol{\mathcal{G}}_{bc}$, and these will be the default implementations we explore. 

If and only if $\partial\boldsymbol{\mathcal{G}}_{bc}/\partial\varepsilon_{c} \ne 0$, which can occur for e.g.\ schemes where $\boldsymbol{\mathcal{G}}_{bc}$ is chosen to be some averaged function of both $\varepsilon_{b}$ and $\varepsilon_{c}$, then Eq.~\ref{eqn:presum} becomes a more complicated implicit expression. We show in Appendix~\ref{sec:deriv.details} that for this case, Eq.~\ref{eqn:presum} can be inserted recursively into itself to derive an explicit expression for ${\rm d}\varepsilon_{b}/{\rm d}{\bf x}_{a}$ in terms of an infinite series of progressively higher-order (smaller) terms in $\partial\boldsymbol{\mathcal{G}}_{bc}/\partial \varepsilon_{c}$ times a term $\mathcal{S}_{2,\,bc}$ where $|\mathcal{S}_{2,\,bc}| \ll 1$ is small. Given such a series sum, it is notationally convenient to re-write Eq.~\ref{eqn:presum} as:
\begin{align}
\label{eqn:w.regular.omega} \frac{{\rm d} \varepsilon_{b}}{{\rm d} {\bf x}_{a}} &= \frac{1}{\Omega_{b}} \frac{\partial \mathcal{E}_{b}}{\partial \mathbf{G}_{b}} \cdot \sum_{c}  \frac{\partial \boldsymbol{\mathcal{G}}_{bc}}{\partial {\bf x}_{a}} \\ 
\nonumber \Omega_{b} &\equiv \Omega^{\prime}_{b} -  \frac{\partial \mathcal{E}_{b}}{\partial \mathbf{G}_{b}} \cdot \sum_{c} \frac{\partial \boldsymbol{\mathcal{G}}_{bc}}{\partial \varepsilon_{c}}\,\mathcal{S}_{2,\,bc} \\
 \label{eqn:omega}   &= 1 - \frac{\partial \mathcal{E}_{b}}{\partial \varepsilon_{b}} - \frac{\partial \mathcal{E}_{b}}{\partial \mathbf{G}_{b}} \cdot \sum_{c}\left[  \frac{\partial \boldsymbol{\mathcal{G}}_{bc}}{\partial \varepsilon_{b}}  +  \frac{\partial \boldsymbol{\mathcal{G}}_{bc}}{\partial \varepsilon_{c}}\,\mathcal{S}_{2,\,bc} \right] \ .
\end{align}
where the $\mathcal{S}_{2}$ term represents the higher-order series terms which appear if and only if $\partial\boldsymbol{\mathcal{G}}_{bc}/\partial\varepsilon_{c} \ne 0$. 
Conveniently, written this way, Eq.~\ref{eqn:w.regular.omega} becomes the correct exact expression from Eq.~\ref{eqn:presum} (and $\Omega_{b} = \Omega^{\prime}_{b}$) if $\partial\boldsymbol{\mathcal{G}}_{bc}/\partial\varepsilon_{c} = 0$ (which we again emphasize is the case for all the default implementations we explore below). If $\partial\boldsymbol{\mathcal{G}}_{bc}/\partial\varepsilon_{c} \ne 0$ (so the $\mathcal{S}_{2}$ are non-vanishing), we discuss multiple methods in Appendix~\ref{sec:deriv.details} for solving Eq.~\ref{eqn:w.regular.omega}, but we show that because $|\mathcal{S}_{2,bc}|\ll 1$, simply approximating $\mathcal{S}_{2} \approx 0$ (making Eq.~\ref{eqn:w.regular.omega} an approximate, rather than exact, equality) generally introduces only small errors.

With this, we can now expand the latter half of Eq.~\ref{eqn:ele} to obtain:
\begin{align}
\frac{1}{2} \sum_{b} & \sum_{c} \,m_{b}\,m_{c} \left[ \frac{\partial \tilde{\phi}_{bc}}{\partial \varepsilon_{b}} \frac{{\rm d} \varepsilon_{b}}{{\rm d} {\bf x}_{a}} +  \frac{\partial \tilde{\phi}_{cb}}{\partial \varepsilon_{c}} \frac{{\rm d} \varepsilon_{c}}{{\rm d} {\bf x}_{a}} \right] \\ 
\nonumber & = \sum_{b}\sum_{c} \,m_{b}\,m_{c}  \frac{\partial \tilde{\phi}_{bc}}{\partial \varepsilon_{b}} \frac{{\rm d} \varepsilon_{b}}{{\rm d} {\bf x}_{a}}  \\
\nonumber & \rightarrow \sum_{b} m_{b}\,\frac{\zeta_{b}}{\Omega_{b}} \frac{\partial \mathcal{E}_{b}}{\partial \mathbf{G}_{b}} \cdot \sum_{c}  \frac{\partial \boldsymbol{\mathcal{G}}_{bc}}{\partial {\bf x}_{a}}   \\
 &=  \sum_{b} m_{b} \frac{\zeta_{b}}{\Omega_{b}}\frac{\partial \mathcal{E}_{b}}{\partial \mathbf{G}_{b}}  \cdot \sum_{c}  \frac{\partial \boldsymbol{\mathcal{G}}_{bc}}{\partial {\bf x}_{a}}\,\left(\delta_{ba} + \delta_{ca} \right)   \\
 &= \sum_{b}\left[m_{a} \frac{\zeta_{a}}{\Omega_{a}}\frac{\partial \mathcal{E}_{a}}{\partial \mathbf{G}_{a}} \cdot \frac{\partial\boldsymbol{\mathcal{G}}_{ab}}{\partial {\bf x}_{a}} +  m_{b} \frac{\zeta_{b}}{\Omega_{b}}\frac{\partial \mathcal{E}_{b}}{\partial \mathbf{G}_{b}} \cdot \frac{\partial\boldsymbol{\mathcal{G}}_{ba}}{\partial {\bf x}_{a}}  \right]
\end{align}
where we have defined: 
\begin{align}
\label{eqn:zeta} \zeta_{b} &\equiv \sum_{c} m_{c}  \frac{\partial \tilde{\phi}_{bc}}{\partial \varepsilon_{b}} \ .
\end{align}
akin to \citet{price:2007.lagrangian.adaptive.softening}.
Note that the steps above use the various symmetry properties of $\tilde{\phi}_{bc}$ above, as well as the fact that $\boldsymbol{\mathcal{G}}_{bc}$ depends on ${\bf x}_{b} - {\bf x}_{c}$ but not other particle values of ${\bf x}$ (this allows the insertion of the $\delta_{ba}$ terms above). 

Finally we can combine all of this to obtain
\begin{align}
\frac{{\rm d}\mathbf{p}_{a} }{{\rm d}t} &=\sum_{b}  \frac{{\rm d}\mathbf{p}_{ab} }{{\rm d}t} \\
\frac{{\rm d}\mathbf{p}_{ab} }{{\rm d}t} &= - \frac{m_{a}\,m_{b}}{2}  {\nabla_{a} \left[  \tilde{\phi}_{ab}  +  \tilde{\phi}_{ba} \right]} 
\\ 
\nonumber & + \left[\frac{m_{a} \zeta_{a}}{\Omega_{a}}\frac{\partial \mathcal{E}_{a}}{\partial \mathbf{G}_{a}} \cdot \frac{\partial\boldsymbol{\mathcal{G}}_{ab}}{\partial {\bf x}_{a}} +  \frac{m_{b} \zeta_{b}}{\Omega_{b}}\frac{\partial \mathcal{E}_{b}}{\partial \mathbf{G}_{b}} \cdot \frac{\partial\boldsymbol{\mathcal{G}}_{ba}}{\partial {\bf x}_{a}}  \right] 
\end{align}
For convenience, we can use $\partial\boldsymbol{\mathcal{G}}_{ba} / \partial {\bf x}_{b} = -\partial\boldsymbol{\mathcal{G}}_{ba} / \partial {\bf x}_{a}$ (because $\boldsymbol{\mathcal{G}}_{ab}$ depends on ${\bf x}_{b}-{\bf x}_{a}$ rather than ${\bf x}_{b}$ or ${\bf x}_{a}$ separately) to re-write this once more as:
\begin{align}
\label{eqn:eom} \frac{{\rm d}\mathbf{p}_{a} }{{\rm d}t} &=-\sum_{b} m_{a}\,m_{b} \left[ \nabla_{a} \tilde{\phi}_{ab} + \boldsymbol{\Upsilon}_{ab} - \boldsymbol{\Upsilon}_{ba}  \right] 
\end{align}
where 
\begin{align}
\boldsymbol{\Upsilon}_{ab} &\equiv -\frac{\zeta_{a}}{m_{b}\,\Omega_{a}}\frac{\partial \mathcal{E}_{a}}{\partial \mathbf{G}_{a}} \cdot \frac{\partial\boldsymbol{\mathcal{G}}_{ab}}{\partial {\bf x}_{a}} \ .
\end{align}
Recall $\zeta$ is defined in Eq.~\ref{eqn:zeta} and $\Omega$ in Eq.~\ref{eqn:omega} above. This makes it more obvious that the momentum/acceleration equation has two components, the ``usual'' gradient of the potential $\nabla_{a}\tilde{\phi}_{ab}$ at fixed $\varepsilon$, plus the $\boldsymbol{\Upsilon}$ terms. The latter are sometimes called ``correction'' or ``grad-$\varepsilon$/grad-$h$'' terms, and they can be thought of as accounting for the additional terms that should appear in the gradient of the potential owing to gradients in $\varepsilon$ (as pointed out in e.g.\ \citealt{price:2007.lagrangian.adaptive.softening} for the specific example discussed in \S~\ref{sec:ex.rho}). Energy conservation of the system in Eq.~\ref{eqn:lagrangian} is automatically ensured up to integration error because this is derived discretely from the particle Lagrangian (Eq.~\ref{eqn:ele}; see \citealt{springel:entropy}), and we validate this in Appendix~\ref{sec:extra.tests}. But we can also see intuitively that conservation holds from the fact that the full set of gradient terms ensures that the change in kinetic energy (${\rm d}_{t} m_{b}\,{\bf v}_{b}^{2}/2$) is exactly equal to the gravitational work done ($\int \nabla \Phi \cdot d{\bf x}$).

\begin{figure}
	\includegraphics[width=0.95\columnwidth]{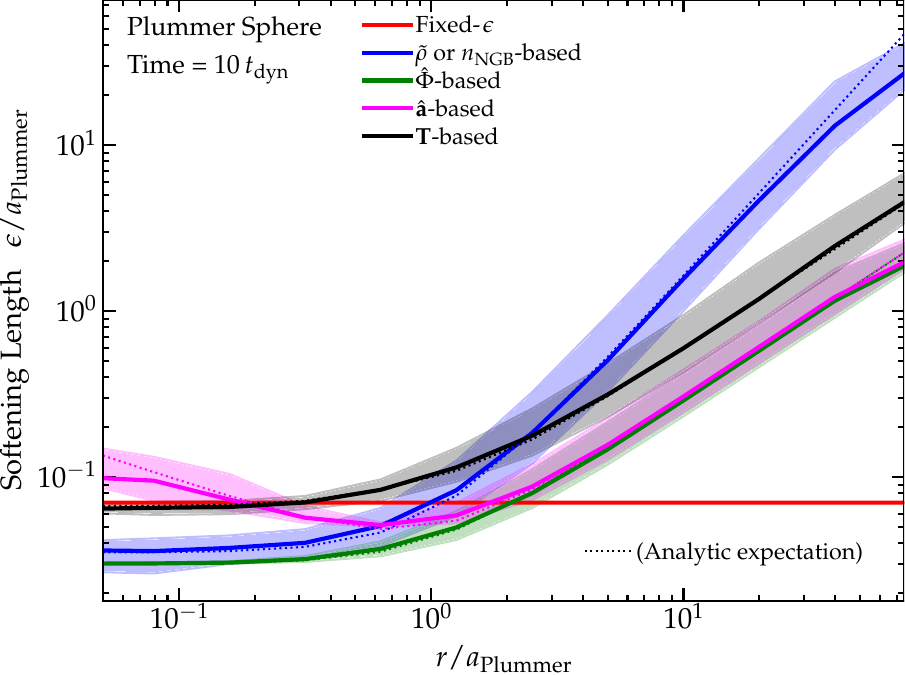}
	\includegraphics[width=0.95\columnwidth]{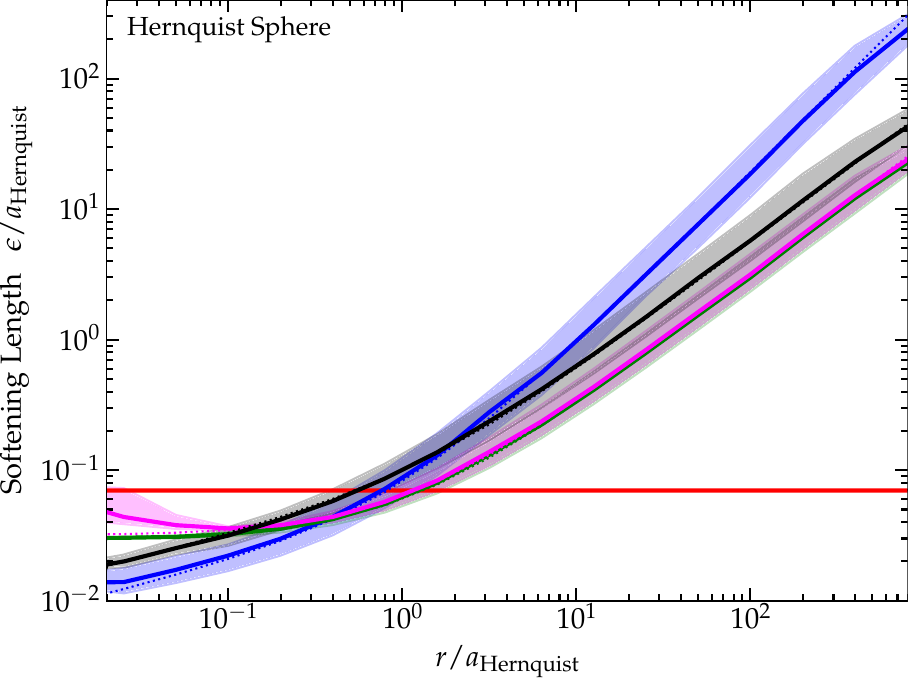}
	\vspace{-0.2cm}
	\caption{{\em Top:} Examples of different softening ``rules'' from \S~\ref{sec:ex}, in a collisionless Plummer sphere ($\rho=3\,M/(4\pi\,a_{\rm Plummer}^{3}\,[1+r^{2}/a_{\rm Plummer}^{2}]^{5/2})$) realized with $\sim 10^{5}$ equal-mass ($m$) $N$-body particles and evolved to $10$ dynamical times ($t_{\rm dyn} = (G\,M/a_{\rm Plummer}^{3})^{-1/2})$. We plot the median (thick line) and $90\%$ inclusion interval of the calculated force softening $\varepsilon$ (defined for the cubic spline softening) at each radius $r$. The normalization of each rule is somewhat arbitrary, but all of the adaptive rules produce a broadly similar behavior of larger $\varepsilon$ at $r\gg a_{\rm Plummer}$, with $\varepsilon$ (by construction) similar to the constant-softening (\S~\ref{sec:ex.fixed}) around the effective radius. For $\tilde{\rho}$ (\S~\ref{sec:ex.rho}) and $n_{\rm NGB}$ (\S~\ref{sec:ex.n}), identical here because particle masses are equal, we adopt $N_{\rm ngb}=32$; for the ${\bf T}$-based model (\S~\ref{sec:ex.T}), $\xi=3.5$. But for the $\hat{\Phi}$ (\S~\ref{sec:ex.phi}) and $\hat{\bf a}$ (\S~\ref{sec:ex.a}) models we must adopt somewhat more ``ad hoc'' normalizations to obtain reasonable $\varepsilon$, with $\xi \sim 3000$ and $\xi \sim 0.01$ respectively. Thin dotted lines show the analytically expected $\varepsilon$ for a perfect realization of the system.
	{\em Bottom:} Same, for a \citet{hernquist:profile} profile ($\rho = M\,a_{\rm Hernquist}/(2\pi\,r\,[r + a_{\rm Hernquist}]^{3})$). The qualitative differences between models are similar to the Plummer case.
	In all cases the system is stable in time and conserves energy and momentum to integration accuracy.
	\label{fig:plummer.h}}
\end{figure}

\section{Examples of Different Choices for the Softening}
\label{sec:ex}

Above, we left the definition of the softening parameter quite general. Of course, for a given simulation, we must make some specific choice to define $\varepsilon_{b} = \mathcal{E}_{b}(....)$. Here we review a number of choices, both those previously used in the literature and several additional choices as well, and give the discrete expressions for $\boldsymbol{\Upsilon}$ and $\Omega$ which arise.

\subsection{Example: ``Fixed'' Softenings}
\label{sec:ex.fixed}

Consider first the simplest case, $\varepsilon_{a} = \mathcal{E}_{a}(m_{a},\,s_{a},\,...)$ which depends only on intrinsic properties of $a$ (e.g.\ its mass, species, etc.) that {\em do not} vary under pure gravitational operations. This includes of course ``constant'' softenings $\mathcal{E}_{a} = C$, but also softening which depends on particle mass or other intrinsically conserved variables in an arbitrarily complicated manner. In this case we immediately have:
\begin{align}
 \boldsymbol{\Upsilon}_{ab}  &\rightarrow 0 
 \end{align}
 and the force is just given by the gradients of $\tilde{\phi}_{ab}$, as expected.

\subsection{Example: Softenings Based on a Kernel Density Estimator $\tilde{\rho}$}
\label{sec:ex.rho}

Now consider a definition where the softening length $\varepsilon_{a}$ is tied to some kernel density estimator (KDE) used to approximate the local density field $\tilde{\rho}$ of the collisionless species as e.g.\ $\tilde{\rho}_{a} \equiv \sum_{c} m_{c} W_{ac}({\bf x}_{a} - {\bf x}_{c},\,\varepsilon_{a})$ (where $W$ is some typical kernel function with compact support in terms of a kernel size parameter $\varepsilon_{a}$). One can then define a desired $\varepsilon_{a}$ in terms of some ``effective neighbor number'' or ``enclosed mass in the kernel'' as 
\begin{align}
\frac{4\pi}{3}\,\varepsilon_{a}^{3}\,\tilde{\rho}_{a} = M_{0}
\end{align}
or $=m_{a}\,N_{\rm ngb}$ (where $M_{0}$ or $N_{\rm ngb}$ are some arbitrary constants set as desired by the user).

With these definitions, ${\bf G}_{a} \rightarrow \tilde{\rho}_{a}$, $\boldsymbol{\mathcal{G}}_{ac} \equiv m_{c} W_{ac}({\bf x}_{a} - {\bf x}_{c},\,\varepsilon_{a})$, $\mathcal{E}_{a} \equiv (3\,m_{a}\,N_{\rm ngb}/4\pi\,{\bf G}_{a})^{1/3}$. 
This gives $\partial_{{\bf x}_{a}} {\boldsymbol{\mathcal{G}}_{bc}} \rightarrow m_{c} \, \partial_{{\bf x}_{a}} W_{bc}({\bf x}_{b}-{\bf x}_{c} \, , \, \varepsilon_{b})$ and $\partial_{{\bf G}_{b}} \mathcal{E}_{b} \rightarrow - \varepsilon_{b} / (3\,{\bf G}_{b}) = - \varepsilon_{b} / (3\,\tilde{\rho}_{b})$. So we have:
\begin{align}
 \boldsymbol{\Upsilon}_{ab}  &\rightarrow
  \frac{\zeta_{a}\,\varepsilon_{a}}{\Omega_{a}\,3\,\tilde{\rho}_{a}}\,\frac{\partial W_{ab}(\varepsilon_{a})}{\partial {\bf x}_{a}}
\end{align}
\begin{align}
\Omega_{a} &\rightarrow 1 + \frac{\varepsilon_{a}}{3\,\tilde{\rho}_{a}} \sum_{c} m_{c} \frac{\partial W_{ac}(\varepsilon_{a})}{\partial \varepsilon_{a}}
\end{align}

This is the adaptive softening function $\mathcal{E}$ proposed in \citet{1986desd.book.....S} and tested in \citet{price:2007.lagrangian.adaptive.softening}, and (reassuringly) our resulting equation of motion is identical to theirs in every term for this choice. 

However, we are careful to note that $\tilde{\rho}$ should not be confused with the actual physical density $\rho \equiv \sum_{s} \int f_{s} \, d^{3}{\bf v}$ of the collisionless system determined by the phase-space distribution function $f_{s}$. In \citet{price:2007.lagrangian.adaptive.softening}, the authors enforce the Poisson relation $\nabla^{2} \phi_{ab} = 4\pi\,G\,W_{ab}$ because of their specific application to collisional hydrodynamics (SPH; see \S~\ref{sec:gas}), but for collisionless fluids this is neither necessary nor more consistent than any independent choice of $W_{ab}$ and $\tilde{\phi}_{ab}$. The reason is that for collisionless $N$-body methods with softened gravity, no unique, rigorous reconstruction of the local density field $\rho({\bf x})$ which also obeys the Vlasov-Poisson equation and Newtonian gravity everywhere can be defined. Rather, only (non-unique) Monte Carlo estimators can be used to estimate some coarse-grained $\langle \rho \rangle$ over some larger volume $V$. So $\tilde{\rho}$ is better thought of as a ``post-processing'' KDE applied at each time to estimate a softening length which will roughly correspond to each radius of compact support enclosing a similar total $N$-body particle mass.\footnote{The numerical details of the practical calculation of quantities like $\tilde{\rho}$ are all provided in the public code, and follow \citet{springel:entropy} with the implementation in {\small GIZMO} described in \citet{hopkins:gizmo} and \citet{hopkins:fire2.methods}.}

\begin{figure}
	\includegraphics[width=0.95\columnwidth]{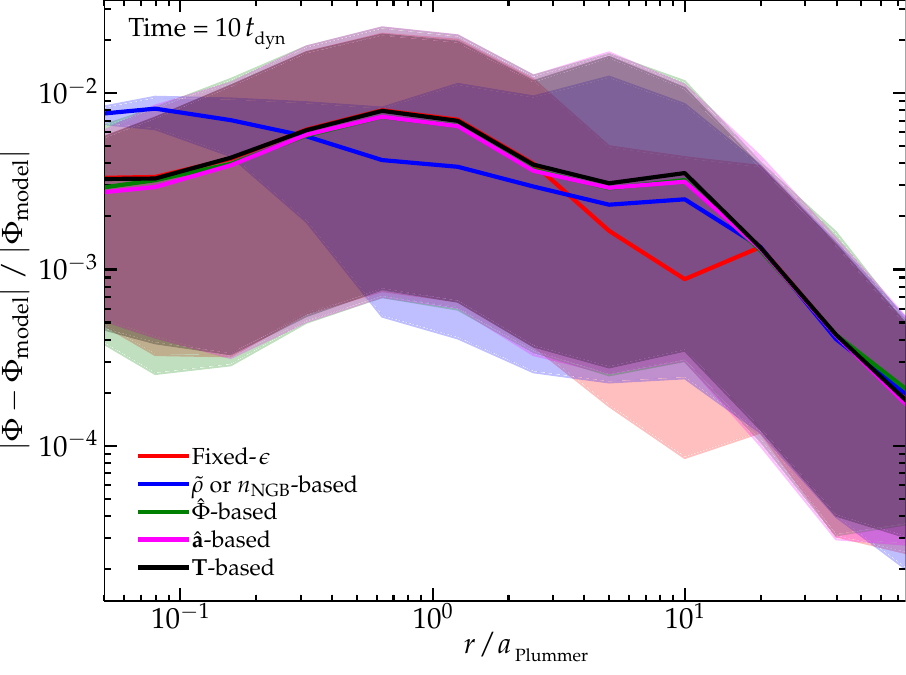}
	\includegraphics[width=0.95\columnwidth]{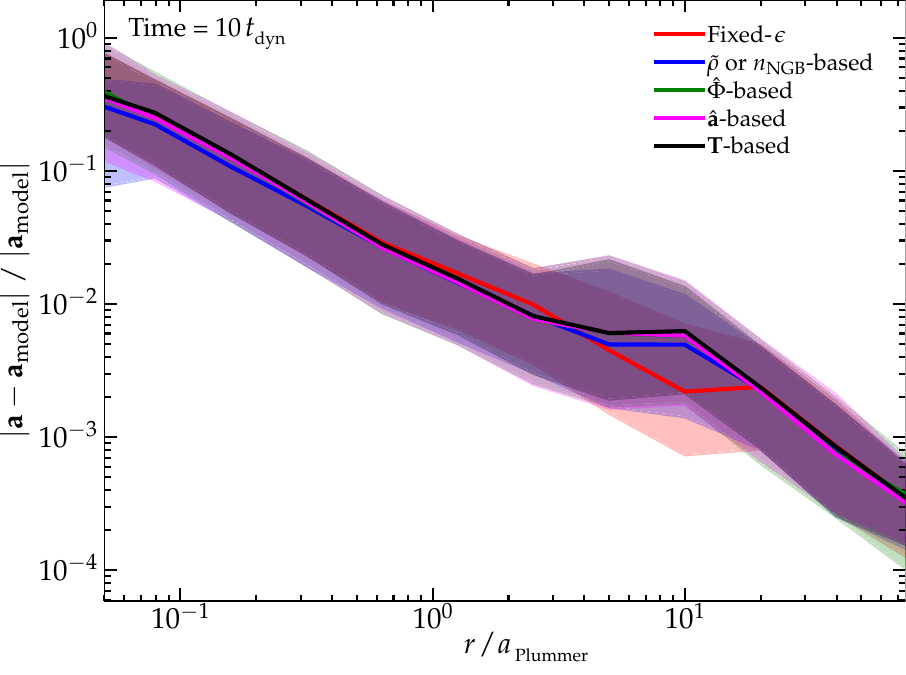}
	\vspace{-0.1cm}
	\caption{Errors in the reconstruction of the potential ({\em top}) and acceleration ({\em bottom}) of the Plummer sphere test from Fig.~\ref{fig:plummer.h}, where $\Phi$ and ${\bf a}$ are the potential and acceleration of each $N$-body particle and $\Phi_{\rm model}$, ${\bf a}_{\rm model}$ are the analytic Plummer values. We confirm the well-known result that force softening actually has a quite weak effect on these errors. For ${\bf a}$, the error is dominated by discreteness noise determined by the number of particles: we can accurately approximate the mean $|{\bf a}-{\bf a}_{\rm model}| / |{\bf a}_{\rm model}|$ as $\sim 2/\sqrt{N(<r)}$ in terms of the number of enclosed particles $N(<r)$. The result is very similar if we compare errors at $t=0$ (when the particle configuration is identical).
	\label{fig:plummer.err}}
\end{figure}

\subsection{Example: Softenings Based on the Number Density of Neighboring Particles $n_{\rm NGB}$}
\label{sec:ex.n}

Alternatively consider a definition where the softening length $\varepsilon_{a}$ is tied to an estimate of the some average ``inter-particle separation'' within the vicinity of the particle. This is often suggested as an ``optimal'' softening length choice for certain problems \citep[see][]{merritt:1996.optimal.softening,romeo.1998:optimal.softening,athanassoula:2000.optimal.force.softening.collisionless.sims,dehnen:2001.optimal.softening,rodionov:2005.optimal.force.softening}, especially those which use {\em fixed} softenings \citep{iannuzzi:2013.no.need.adaptive.softening.for.dm}. The number density of neighbor particles within some radius of compact support can be estimated as $n_{{\rm NGB},\,a} \equiv \sum_{c} W_{ac}({\bf x}_{a} - {\bf x}_{c},\,\varepsilon_{a})$, giving a mean inter-neighbor distance $\langle \Delta x \rangle_{\rm NGB} = n_{{\rm NGB},\,a}^{-1/3}$, so one can set $\varepsilon_{a} \propto \,\langle \Delta x \rangle_{\rm NGB}$ or 
\begin{align}
n_{\rm NGB}\,\varepsilon_{a}^{3} \propto n_{\rm NGB}\,\langle \Delta x \rangle_{\rm NGB}^{3} = {\rm constant} .
\end{align}
For e.g.\ $D=3$, one could adopt something like $(4\pi/3)\,\varepsilon_{a}^{3}\,n_{a} = N_{\rm NGB}$ in terms of some desired ``neighbor number'' $N_{\rm NGB}$.

This gives ${\bf G}_{a} \rightarrow n_{a}$, $\boldsymbol{\mathcal{G}}_{ac} \equiv W_{ac}({\bf x}_{a} - {\bf x}_{c},\,\varepsilon_{a})$, $\mathcal{E}_{a} \equiv (3\,N_{\rm ngb}/4\pi\,{\bf G}_{a})^{1/3}$. 
In turn, $\partial_{{\bf x}_{a}} {\boldsymbol{\mathcal{G}}_{bc}} \rightarrow \partial_{{\bf x}_{a}} W_{bc}({\bf x}_{b}-{\bf x}_{c} \, , \, \varepsilon_{b})$ and $\partial_{{\bf G}_{b}} \mathcal{E}_{b} \rightarrow - \varepsilon_{b} / (3\,{\bf G}_{b}) = - \varepsilon_{b} / (3\,{n}_{{\rm NGB},\,b})$. So we have:
\begin{align}
  \boldsymbol{\Upsilon}_{ab} &\rightarrow  
\frac{\zeta_{a}\,\varepsilon_{a}}{\Omega_{a}\,3\,m_{b}\,n_{{\rm NGB},\,a}}\,\frac{\partial W_{ab}(\varepsilon_{a})}{\partial {\bf x}_{a}}
\end{align}
\begin{align}
\Omega_{a} &\rightarrow 1 + \frac{\varepsilon_{a}}{3\,n_{{\rm NGB},\,a}} \sum_{c} \frac{\partial W_{ac}(\varepsilon_{a})}{\partial \varepsilon_{a}}
\end{align}
This is the adaptive softening scheme proposed and tested in \citet{hernquistkatz:treesph,dehnen:2001.optimal.softening,hopkins:gizmo,hopkins:fire2.methods}, and once again our final equations of motion are identical to the latter as expected. This is numerically very similar to the \citet{price:2007.lagrangian.adaptive.softening} $\tilde{\rho}$-based scheme, and in fact the two are identical if all particles have equal masses, but this particular expression makes it more obvious that the softening length is chosen to reflect the particle/grid/mesh configuration at a given time, without necessarily any reference to any {\em physical} density distribution for collisionless particles.

\subsection{Example: Softenings Scaling with the Local Potential $\Phi$}
\label{sec:ex.phi}

Now consider a definition of $\varepsilon_{a}$ scaling with the depth of the local potential given by some estimator $\hat{\Phi}_{a} \equiv -(1/2)\,\sum_{c} m_{c}\,\hat{\phi}_{ac}$ (which does not, strictly, have to be the same as $\tilde{\phi}_{ac}$), relative to some chosen background/constant value ($\Delta \Phi_{a} \equiv \Phi_{0} - \hat{\Phi}_{a}$), dimensionally given by 
\begin{align}
\Delta \Phi_{a} \equiv \Phi_{0} - \hat{\Phi}_{a} = \xi\,\frac{G\,m_{a}}{\varepsilon_{a}} 
\end{align}
(where $\xi$ is an arbitrary $\mathcal{O}(1)$ constant set as desired for the simulation). 
This is akin to saying that some crude estimate of the ``self-potential'' (which recall, is formally undefined) of the particle should scale as some multiple of the background potential. For a test particle in e.g.\ a background Keplerian potential $\Delta \Phi_{a} \sim -G\,M_{0}/r$, $\varepsilon_{a} \sim \xi\,(m_{a}/M_{0})\,r$ ensures that the maximum possible work done in an $N$-body encounter ``through'' the test particle $a$ can never be more than a fraction $\mathcal{O}(\xi)$ background potential depth. 

We then have $\mathcal{E}_{a} = \xi\, G\,m_{a} / (\Phi_{0} - \hat{\Phi}_{a})$, ${\bf G}_{a} \rightarrow \hat{\Phi}_{a}$, $\boldsymbol{\mathcal{G}}_{ac} = -(m_{c}/2)\,\hat{\phi}_{ac}$, so $\partial_{{\bf x}_{a}} {\boldsymbol{\mathcal{G}}_{bc}} \rightarrow -(m_{c}/2)\,\partial_{{\bf x}_{a}} \hat{\phi}_{bc}$ and $\partial_{{\bf G}_{b}} \mathcal{E}_{b} \rightarrow \varepsilon_{b} / \Delta \Phi_{b}$. Then:
\begin{align}
  \boldsymbol{\Upsilon}_{ab} &\rightarrow   
\frac{\zeta_{a}\,\varepsilon_{a}}{\Omega_{a}\,2\Delta \Phi_{a}}\frac{\partial \hat{\phi}_{ab}}{\partial {\bf x}_{a}}
\end{align}
\begin{align}
\label{eqn:phimodel.omega} \Omega_{a} &\rightarrow 1 + \frac{1}{2} \frac{\varepsilon_{a}}{\Delta \Phi_{a}}  \sum_{c} m_{c}  \left[\frac{\partial \hat{\phi}_{ac}}{\partial \varepsilon_{a}}  +  \frac{\partial \hat{\phi}_{ac}}{\partial \varepsilon_{c}}\mathcal{S}_{2,\,ac}  \right]
\end{align}

Note here that we have been careful to define $\hat{\Phi}$ in terms of $\hat{\phi}_{bc}$ (our ``estimator''), instead of $\tilde{\phi}_{bc}$ defined above (the precise numerical pair-wise potential function used for the force computation). This is because this is only a rule for choosing the already somewhat arbitrary function $\mathcal{E}_{a}$, and therefore $\hat{\Phi}$ only needs to be an {\em estimator} of the local potential -- there is no physical or numerical reason why this has to exactly match the discrete potential used in calculating the gravitational forces. Of course, by the same token, there is nothing wrong with choosing $\hat{\phi}_{bc} \rightarrow \tilde{\phi}_{bc}$, but there may be situations where it is numerically advantageous to use a different value. For example, we could choose $\hat{\phi}_{bc} \rightarrow \hat{\phi}({\bf x}_{b} - {\bf x}_{c},\,\varepsilon_{b})$ to be an {\em un-symmetrized} version of $\phi$ (i.e.\ $\hat{\phi}_{bc} \ne \hat{\phi}_{cb}$), which means that the ``$\mathcal{S}_{2}$ terms'' discussed in Appendix~\ref{sec:deriv.details} (the terms in $\partial\boldsymbol{\mathcal{G}}_{bc}/\partial\varepsilon_{c}$ in Eq.~\ref{eqn:presum}, or $\partial_{\varepsilon_{c}}\hat{\phi}_{ac}$ in $\Omega_{a}$ in Eq.~\ref{eqn:phimodel.omega}) vanish and the other term $\partial_{\varepsilon_{a}} \hat{\phi}_{ac}$ is non-vanishing only for particles inside of the compact support kernel $\varepsilon_{a}$ of $a$, which makes the calculation of terms like $\Omega_{a}$ computationally simpler and more exact (depending on the implementation). Or we could even choose $\hat{\phi}_{bc} \rightarrow \hat{\phi}_{bc}({\bf x}_{b} - {\bf x}_{c},\,\varepsilon_{0})$ for some constant $\varepsilon_{0}$ or $\varepsilon_{0}\rightarrow 0$ (the un-softened $\phi$), which would make $\Omega_{a} = 1$, both simplifying and (for $\varepsilon_{0}$ large) enforcing greater ``smoothness'' in the softening field $\varepsilon_{a}$ and terms $\boldsymbol{\Upsilon}$. 

Note that like the neighbor-based criteria in \S~\ref{sec:ex.rho}-\ref{sec:ex.n}, this method does not immediately meet all of the ``desireable'' criteria for a force softening scheme listed in \S~\ref{sec:intro}. Notably, it is not gauge-invariant. We will discuss the consequences of this further below.

\subsection{Example: Softenings Scaling with the Acceleration ${\bf a}$}
\label{sec:ex.a}

Building on \S~\ref{sec:ex.phi}, consider a softening which scales with the local gravitational acceleration scale, designed such that the maximum acceleration a particle can feel on a close $N$-body encounter with another particle ($\mathcal{O}(G\,m_{a}/\varepsilon_{a}^{2})$) is always some (small) fraction of the ``smooth'' or ``collective'' acceleration scale $|{\bf a}|$. Specifically consider
\begin{align}
\frac{G\,m_{a}}{\varepsilon_{a}^{2}} = \xi\,|\hat{\bf a} |_{a}
\end{align}
where $\hat{\bf a}_{a} \equiv \hat{\bf a}^{\rm grav,\,a} \equiv -\nabla \hat{\Phi}_{a}$, so ${\bf G}_{a} = \hat{\bf a}_{a}$, $\boldsymbol{\mathcal{G}}_{ac} = (m_{c}/2)\,\nabla \hat{\phi}_{ac}$, and  $\partial_{{\bf x}_{a}} {\boldsymbol{\mathcal{G}}_{bc,\,i}} \rightarrow (m_{c}/2)\,\partial_{i}\,\partial_{{\bf x}_{a}} \hat{\phi}_{bc}$ and $\partial_{{\bf G}_{b,\,i}} \mathcal{E}_{b} \rightarrow -\varepsilon_{b}/(2\,|\hat{\bf a}_{b}|^{2}) \, \hat{\bf a}_{b,\,i}$. Here we denote the vector components of ${\bf a}$ with the index $i$ which runs from $i=1$ to $i=D=3$. This gives us
\begin{align}
  \boldsymbol{\Upsilon}_{ab} &\rightarrow 
 \frac{\zeta_{a}\,\varepsilon_{a}}{\Omega_{a}\,4 | \hat{\bf a}_{a} |^{2}}\,\hat{\bf a}_{a}^{i}\,{\partial_{i}} \frac{\partial \hat{\phi}_{ab}}{\partial {\bf x}_{a}} 
\end{align}
\begin{align}
\Omega_{a} &\rightarrow 1 + \frac{1}{4} \frac{\varepsilon_{a}}{|\hat{\bf a}_{a}|^{2}} \, \hat{\bf a}_{a}^{i}  \sum_{c} m_{c}\,\partial_{i}\,  \left[\frac{\partial \hat{\phi}_{ac}}{\partial \varepsilon_{a}}  +  \frac{\partial \hat{\phi}_{ac}}{\partial \varepsilon_{c}} \mathcal{S}_{2,\,ac}   \right]
\end{align}
where we use an Einstein summation convention for the indices $i$, i.e.\ $\hat{\bf a}_{a}^{i} \partial_{i} \equiv \sum_{i=1}^{3} \hat{\bf a}_{a}^{i} \partial_{{\bf x},\,i} = \hat{\bf a}_{a} \cdot \nabla$. Note that as \S~\ref{sec:ex.phi}, we define this in terms of $\hat{\phi}_{bc}$ so $\hat{\bf a}$ does not necessarily have to be the exact same value as the numerically-calculated gravitational acceleration on particle $a$. Though in some implementations, it is most convenient if it is, i.e.\ if one sets $\hat{\phi}_{bc} = \tilde{\phi}_{bc}$, there are others where one might wish to use the un-symmetrized $\hat{\phi}_{bc}$ as discussed in \S~\ref{sec:ex.phi}.

As above, this method is also not gauge-invariant, and can produce some other unexpected behaviors, which we demonstrate below.

\subsection{Example: Softenings Scaling with the Tidal Tensor ${\bf T}$}
\label{sec:ex.T}

Going one step further we note that some estimator of the tidal tensor ${\bf T}\equiv -\nabla \otimes \nabla \hat{\Phi}$ (again in terms of $\hat{\phi}_{bc}$ per \S~\ref{sec:ex.phi}) has units of $\mathcal{O}(G\,\rho)$, so one can define a softening scaling as:\footnote{Note that given the tidal tensor ${\bf T}$, one can compute a variety of gauge invariant quantities from its eigenvalues $\lambda_{i}$, including the Frobenius norm $\|{\bf T}\|^{2} = \sum \lambda_{i}^{2}$, the trace ${\rm Tr}({\bf T}) = \sum \lambda_{i}$, determinant ${\rm Det}({\bf T}) = \prod \lambda_{i}$, minimum ${\rm min}(\lambda_{i})$ or ${\rm min}(|\lambda_{i}|)$, maximum ${\rm max}(\lambda_{i})$ or ${\rm max}(|\lambda_{i}|)$, and combinations thereof (any of which could be used dimensionally to construct a softening length with $G$ and $m$, e.g.\ $\epsilon \propto (G^{3}\,m_{a}^{3}/|{\rm Det}({\bf T})|)^{1/9}$). We adopt the Frobenius norm $\sum \lambda_{i}^{2}$ for several reasons. In highly-anisotropic potentials, $\|{\bf T}\|$ is well-behaved but the determinant ${\rm Det}({\bf T})$ and ${\rm min}(\lambda_{i})$-based scalings become ill-conditioned and can give unphysical (divergent or zero or imaginary) softenings (while being similar to our default $\|{\bf T}\|$-based scaling in isotropic potentials). The trace ${\rm Tr}({\bf T}) = -4\pi\,G\,\rho$ contains {\em only} the local density information, so using it to calculate $\epsilon$ becomes identical to the method in \S~\ref{sec:ex.rho}, and would eliminate the external tidal field information needed to define e.g.\ the tidal or Hill radius of interest here for long-range forces. A tidal/Hill-radius type argument as above suggests using either $\|{\bf T}\|$ or ${\rm max}(|\lambda_{i}|)$ (unless one generalizes to anisotropic softenings as in e.g.\ \citealt{pfenniger.friedli:1993.anisotropic.softening}, based on the principle axes of ${\bf T}$). In practice, these two are always similar, i.e.\ ${\rm max}(|\lambda_{i}|) \sim \|{\bf T}\|$  (they differ by at most a factor of $\sqrt{3}$, and since $\epsilon\propto \|{\bf T}\|^{-1/3}$ this means only at most a factor of $3^{1/6} \sim 1.2$ difference in the values for $\epsilon$ -- much smaller than the ambiguity in the coefficient $\xi$ which we will consider). But $\|{\bf T}\|$ can be computed directly from the components of $\nabla \otimes \nabla \Phi$ in a way that is easily differentiable (unlike ${\rm max}(|\lambda_{i}|)$ which involves functions like eigenvalue determination, absolute value, and ${\rm max}$, so does not lend itself to easily computing derivatives like $\partial_{{\bf G}_{b}} \mathcal{E}_{b}$, etc.). So we therefore default to the $\|{\bf T}\|$-based formulation here.} 
\begin{align}
\label{eqn:tidal.soft} 
\| \mathbf{T} \|_{a} &= \frac{G\,m_{a}}{(\varepsilon_{a}/\xi)^{3}}
\end{align}
or $\varepsilon_{a} = \xi\,({G\,m_{a}} /{\| \mathbf{T} \|_{a}} )^{1/3}$, where $\|\mathbf{T}\|^{2} \equiv {\bf T}:{\bf T} \equiv \lambda_{i}\lambda^{i} \equiv {\bf T}_{ij}\,{\bf T}^{ij}$ (using an Einstein summation convention for simplicity with indices $i$, $j$ running from $1$ to $D=3$, with $\lambda_{i}$ the eigenvalues of ${\bf T}$). 
So ${\bf G}_{a} = {\bf T}_{a}$ or ${\bf G}_{a,\,ij} = {\bf T}_{a,\,ij} = -\partial_{i}\,\partial_{j} \hat{\Phi}_{a}$, $\boldsymbol{\mathcal{G}}_{ac,\,ij} =  (m_{c}/2)\,\partial_{i}\,\partial_{j}\,\hat{\phi}_{ac}$. Therefore $\partial_{{\bf x}_{a}} {\boldsymbol{\mathcal{G}}_{bc,\,ij}} \rightarrow (m_{c}/2)\,\partial_{i}\partial_{j}\,\partial_{{\bf x}_{a}} \hat{\phi}_{bc}$ and $\partial_{{\bf G}_{b,\,ij}} \mathcal{E}_{b} \rightarrow -\varepsilon_{b}/(3\,\|{\bf T}_{b}\|^{2}) \, {\bf T}_{b,\,ij}$, and we have
\begin{align}
\label{eqn:tidal.corr} \boldsymbol{\Upsilon}_{ab} &\rightarrow 
\frac{\zeta_{a}\,\varepsilon_{a}}{\Omega_{a}\,6\| {\bf T}_{a} \|^{2}}\,{\bf T}_{a}^{ij}{\partial_{i}\partial_{j}} \frac{\partial \hat{\phi}_{ab}}{\partial {\bf x}_{a}} 
\end{align}
\begin{align}
\label{eqn:tidal.omega} \Omega_{a} &\rightarrow 1 + \frac{1}{6} \frac{\varepsilon_{a}}{\|{\bf T}_{a}\|^{2}} \, {\bf T}_{a}^{ij} \, \sum_{c} m_{c}\,\partial_{i}\,\partial_{j}\,  \left[\frac{\partial \hat{\phi}_{ac}}{\partial \varepsilon_{a}}  +  \frac{\partial \hat{\phi}_{ac}}{\partial \varepsilon_{c}} \mathcal{S}_{2,\,ac}  \right]
\end{align}

Physically, we can think of this as reflecting a generalized tidal (e.g.\ Hill or Roche or tidal radius) criterion, where in the limit of relevance for $N$-body treatments of collisionless fluids (where the mass of one particle $m_{a}$ is small compared to the collective background), this enforces that the softening $\varepsilon_{a}$ scales as a fixed multiple of the tidal radius of the mass represented by that $N$-body particle. In denser/less dense regions, this will be tidally compressed or sheared out approximately following Eq.~\ref{eqn:tidal.soft} (though of course the actual self-consistent evolution would require evolving $f_{s}$).

Numerically, this is designed to ensure both of the criteria in \S~\ref{sec:ex.phi}-\ref{sec:ex.a}, that the maximum velocity/energy deflection, and maximum acceleration, as well as the maximum tidal field strength and gravitational jerk, in a two-body encounter are always small compared to the smooth background component (see \S~\ref{sec:pros} \&\ \S~\ref{sec:tidal.norm} for more discussion). 

While the exact value of $\xi$ in Eq.~\ref{eqn:tidal.soft} has no effect on our subsequent equations, in  \S~\ref{sec:tidal.norm} we discuss the its normalization and provide some guidelines for reasonable choices for a given kernel.

\subsection{Illustration in an Example Problem}
\label{sec:ex.demo}

In Fig.~\ref{fig:plummer.h}, we illustrate the different softening rules proposed above in a simple problem. We initialize an isotropic, equilibrium Plummer sphere with $N=10^{5}$ equal-mass $N$-body collisionless particles, and evolve it for $\sim10$ dynamical times in the {\small GIZMO} code \citep{hopkins:gizmo}, with each softening rule in turn. Because the exact normalization of $\varepsilon$ is somewhat arbitrary, we have repeated this with each force law for different normalizations of $\varepsilon$ (e.g.\ $\xi$ values) and different softening kernels. First, we can immediately validate that all the models discussed here, when the $\boldsymbol{\Upsilon}$ terms are included, conserve momentum and energy to the desired integration accuracy (Appendix~\ref{sec:extra.tests}).\footnote{Given the antisymmetry of Eq.~\ref{eqn:eom}, momentum conservation would be machine-accurate as opposed to integration-accurate in our formulation if one ensured equal timesteps for all pairs $a$, $b$, and explicit pair-wise forces. However, standard tree and tree-PM methods using hierarchical timestepping, as adopted here, reduce this to integration accuracy \citep{springel:gadget}.} If we drop the $\boldsymbol{\Upsilon}$ terms, we see significant violations of energy conservation appear in a few dynamical times. This is expected and has been seen in many previous, much more detailed studies of these methods, to which we refer for details \citep[e.g.][]{price:2007.lagrangian.adaptive.softening,springel:arepo,iannuzzi:2011.collisionless.adaptive.softening.gadget,iannuzzi:2013.no.need.adaptive.softening.for.dm,hopkins:gizmo}. 

Comparing the actual values of $\varepsilon$, we see that most of the ``adaptive'' methods similarly produce $\varepsilon\sim$\,constant in the Plummer sphere ``core'' (where the mass density and therefore particle number density is also constant), with rising $\varepsilon$ at larger radii (where the density declines $\propto r^{-5}$), as desired. But as we will discuss in \S~\ref{sec:pros}, we see some ambiguities in the $\hat{\Phi}$ and $\hat{\bf a}$ methods in their normalization, where e.g.\ adopting a more naive $\xi \sim \mathcal{O}(1)$ for e.g.\ the $\hat{\Phi}$ method would produce unreasonable small softenings (a factor $\sim m_{\rm particle} / M_{\rm plummer}$ smaller than the inter-particle spacing), in a way which appears to require some manual resolution-dependent rescaling of $\xi$. For the $\hat{\bf a}$ method, we also see that $\varepsilon$ actually begins to {\rm rise} again in the central regions: this is because even though the density is constant, the acceleration vanishes (so formally $\varepsilon$ would diverge if we had a particle exactly at the origin in a well-sampled potential) as $r\rightarrow 0$. These issues in part (but not entirely) arise from the lack of gauge-invariance of said methods.

We also repeat the same for a \citet{hernquist:profile} profile (with the same particle number). The results are similar, though notably in the center, the profile features a cusp with $\rho \propto r^{-1}$, which means the $\tilde{\rho}$-based methods and tidal ${\bf T}$-based method continue to feature smaller $\varepsilon \propto r^{1/3}$ as $r\rightarrow 0$ (as we would desire for a steep/self-gravitating cusp), while the $\hat{\Phi}$ and $\hat{\bf a}$ methods have $\varepsilon \rightarrow $\,constant (making preserving such a cusp non-linearly more difficult), despite the accelerations at $r\rightarrow 0$ being dominated by the local enclosed mass.

Fig.~\ref{fig:plummer.err} plots a diagnostic of the errors (particle-by-particle, relative to analytic) in the potential and acceleration. Importantly, we see that there is not actually much difference between the different softening rules here (even the small differences that appear are largely noise-dominated and average out over time). For this reason it makes almost no visible difference if we plot the result at $t=0$ (the initial conditions, so identical particle positions for all models) or some later time (we show the latter since it provides an upper limit to the differences, but again the difference is very small).  If, however, we change the mass resolution (number of particles), we see the errors uniformly move up or down, approximately scaling as $\propto N^{-1/2}$ for acceleration (where $N$ is the number of particles). This confirms the well-known result that force softening actually has very weak effects on either the instantaneous acceleration or $L_{1}$-norm type errors (what is plotted here) in the acceleration/potential, or convergence properties of collisionless $N$-body integrators for {\em most} properties (some major exceptions discussed below), so long as a ``catastrophically bad'' (orders-of-magnitude too small or large) value of the softening is not used (for different detailed studies of this, we refer to e.g.\ \citealt{power:2003.nfw.models.convergence,2004MNRAS.353..624D,2009MNRAS.398L..21S,wheeler:dwarf.satellites,hopkins:fire2.methods,vandenbosch:2018.disruption.over.soft,vandenbosch:2018.oversoft.disruption.numerical.discreteness.noise}).\footnote{Formally, as discussed in \citet{ma:2022.discrete.df.estimator}, one can think of force softening as truncating a Coulomb logarithm that arises from integrating the contributions to discreteness noise from particle sampling of the potential on all scales. Hence the weak (logarithmic) effect of changing $\varepsilon$ on the instantaneous errors in ${\bf a}$.} This is important to remember in any application of these methods: mass resolution is usually far more important than force softening. That in turn means an adaptive force softening method which is too expensive will produce worse errors, in practice, than a faster method that enables one to run larger-particle-number simulations. But we stress that this does not mean force-softening is meaningless: {\em some} softening is required, and even the studies above found that seriously incorrect results could be obtained for much too-large or too-small $\varepsilon$. In a simple, single-species Plummer sphere problem here, it is easy to assign a ``reasonable'' $\varepsilon$, but this is much more difficult to define {\em a priori} in a high-dynamic range multi-physics simulation (e.g.\ galaxy formation simulations). And we will show below that there are differences in the results from such choices, they just do not appear so obviously in the classic ``error norm'' type tests of Fig.~\ref{fig:plummer.err}.

\section{Enforcing a Maximum or Minimum Softening}
\label{sec:minmax}

It is often desirable, even when adaptive softenings are used, to enforce some minimum (and sometimes maximum) softening length. This may be for purely numerical optimization reasons, or to reflect situations where other other physics (e.g.\ collapse to compact objects) would become unresolved. Usually, this is enforced ad-hoc by simply imposing  $\varepsilon_{a} = {\rm MAX}(\mathcal{E}_{a},\,\varepsilon_{\rm min})$ or $\varepsilon_{a} = {\rm MIN}(\mathcal{E}_{a},\,\varepsilon_{\rm max})$, but this is not differentiable and so can cause numerical errors and even instability when applied using the scalings above (even if one simply sets $\boldsymbol{\Upsilon}_{ab} = \mathbf{0}$ for sufficiently small $\varepsilon_{a}$). 

However it is easy to incorporate maxima or minima self-consistently in our formulation, by slightly modifying the definition of $\mathcal{E}_{a}$, as e.g.\ $\varepsilon_{a}^{n} = \varepsilon_{\rm min,max}^{n} + \mathcal{E}_{a}^{n}$ (where $n>0$ enforces a minimum, $n<0$ a maximum), or 
\begin{align}
\label{eqn:minmax} \mathcal{E}_{a} \rightarrow \left( (\varepsilon_{\rm min/max})^{n} + (\mathcal{E}_{a}^{0})^{n} \right)^{1/n}
\end{align}
where $\mathcal{E}_{a}^{0}$ is the function $\mathcal{E}_{a}$ we would have in the absence of the introduced maximum/minimum softening. Then everything above in our derivations is perfectly identical, we just need to slightly modify $\partial \mathcal{E}_{a} / \partial {\bf G}_{a}$ appropriately for the revised definition of $\mathcal{E}_{a}$. 

For example, if we adopt a tidal softening (Eq.~\ref{eqn:tidal.soft}) with a minimum $\varepsilon_{\rm min}$ and $n=1$, we have $\varepsilon_{a} = \mathcal{E}_{a} = \varepsilon_{\rm min} + \xi\,(G\,m_{a} / \| {\bf T}_{a} \|)^{1/3}$, and we simply replace $\varepsilon_{a} \rightarrow \varepsilon_{a} - \varepsilon_{\rm min}$ in Eq.~\ref{eqn:tidal.corr}-\ref{eqn:tidal.omega} for $\boldsymbol{\Upsilon}_{ab}$ and $\Omega_{a}$.

\begin{figure}
	\includegraphics[width=0.95\columnwidth]{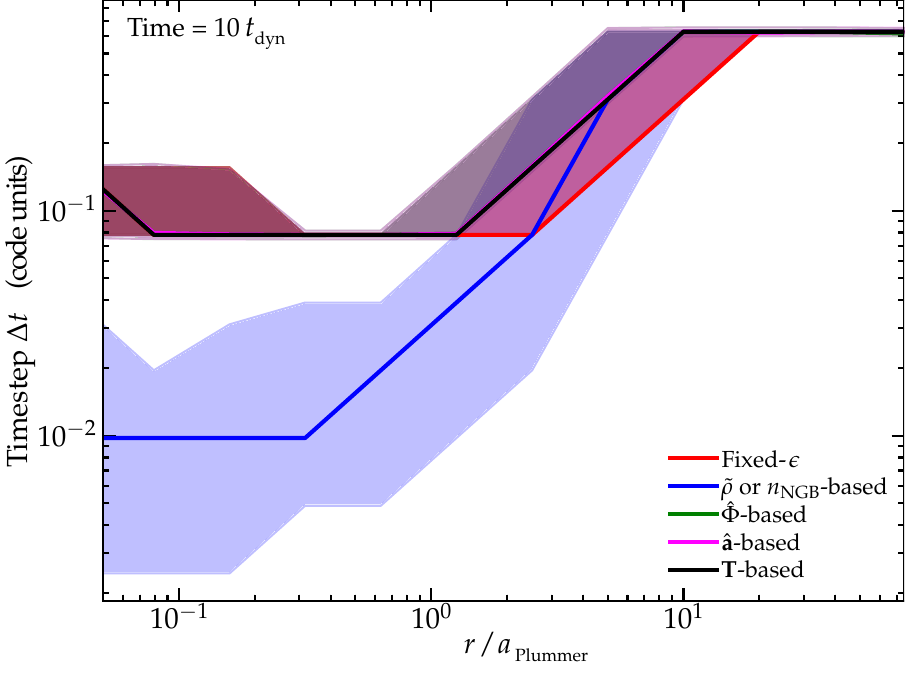}
	\vspace{-0.1cm}
	\caption{Numerical timesteps in the Plummer test from Fig.~\ref{fig:plummer.h}. Note the code uses a discretized power-of-two timestep hierarchy (the reason for discrete ``levels'') and adopts a standard universal timestep criterion for integrating gravity alongside the adaptive-softening-specific additional criterion in Eq.~\ref{eqn:courant} (with $C=0.25$). The fixed-$\varepsilon$ and potential/acceleration/tidal criteria have timesteps primarily determined by the standard gravity criterion (Eq.~\ref{eqn:courant} is already met and imposes no significant additional burden). The neighbor-based ($\tilde{\rho}$ and $n_{\rm NGB}$) models require much shorter timesteps to accurately integrate through collisionless particle-particle crossings owing to the sensitivity of $\varepsilon$ to the strictly-local particle distribution (see \S~\ref{sec:timesteps}). 
	\label{fig:plummer.dt}}
\end{figure}

\begin{figure}
	\hspace{0.8cm}\includegraphics[width=0.85\columnwidth]{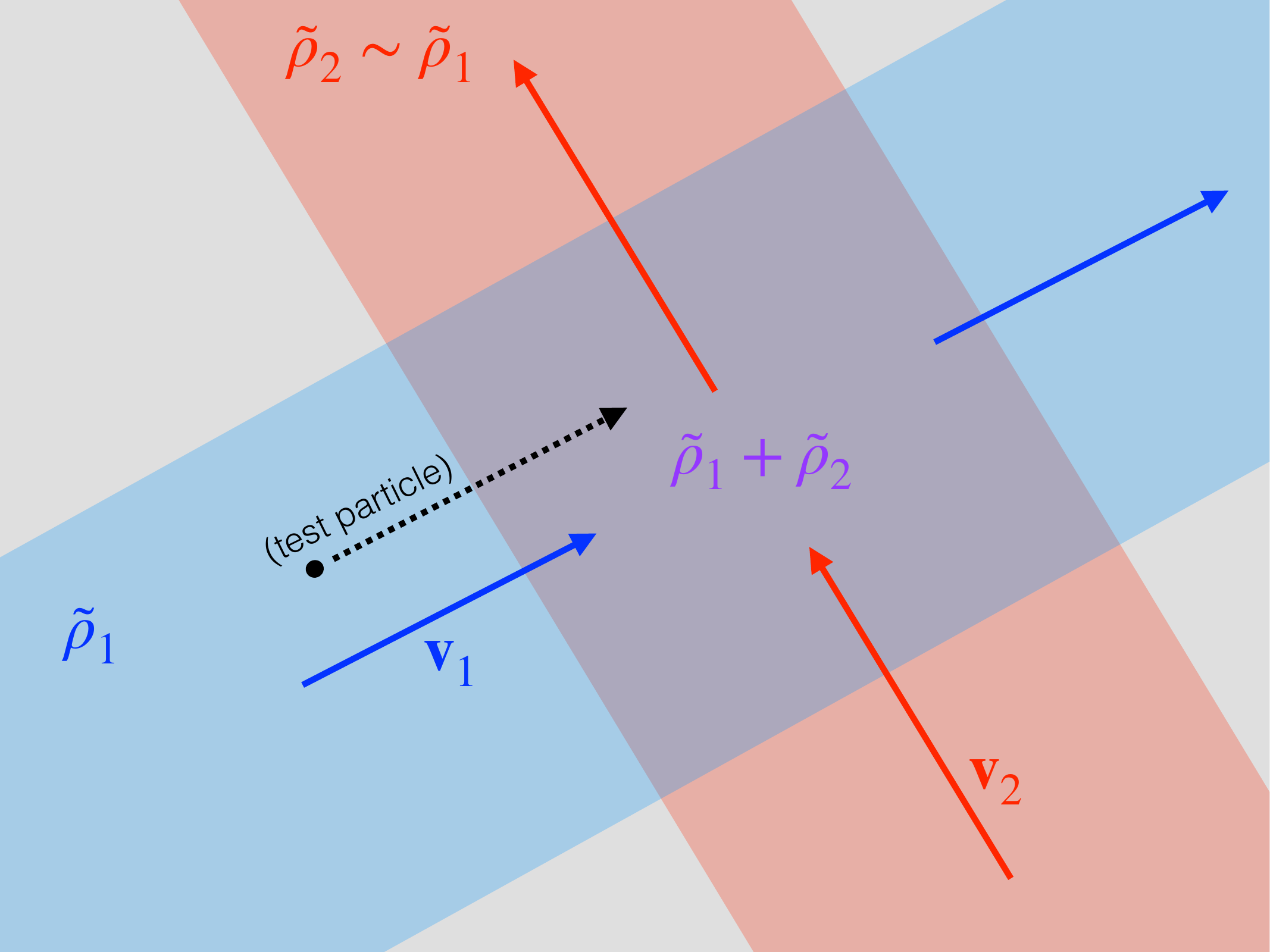}
	\includegraphics[width=0.95\columnwidth]{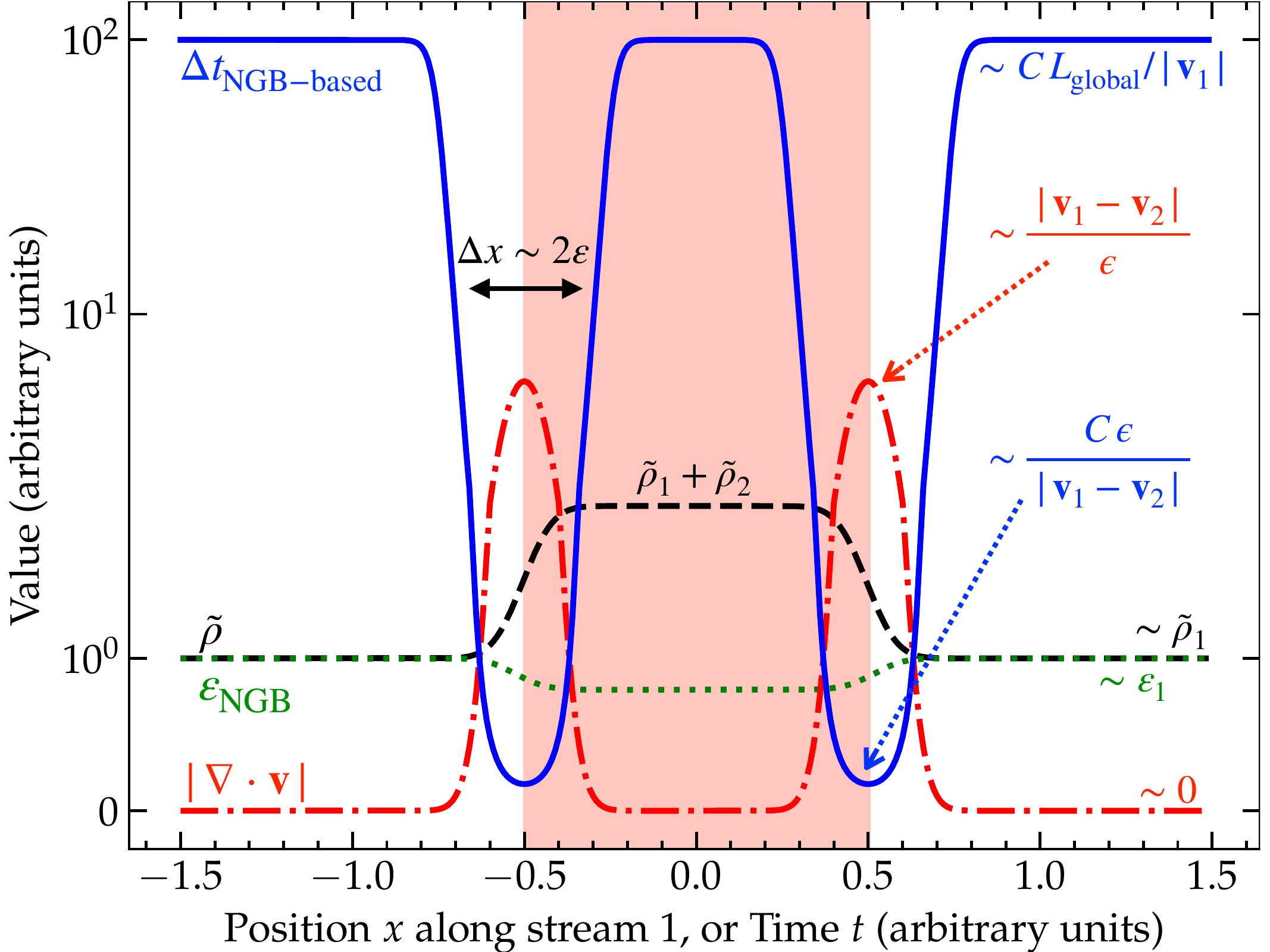}
	\vspace{-0.1cm}
	\caption{Toy illustration of the reason for the timestep difference in Fig.~\ref{fig:plummer.dt}, following discussion in \S~\ref{sec:timesteps}. 
	Consider two streams or sheets ($1$ \&\ $2$) of collisionless particles, intersecting at an arbitrary angle in an external potential (negligible self-gravity), with local densities $\tilde{\rho}_{1}$, $\tilde{\rho}_{2}$ (estimated as \S~\ref{sec:ex.rho}) and velocities ${\bf v}_{1}$, ${\bf v}_{2}$, respectively ({\em top}). 
	There are no local interactions, so this crossing has no effect on the dynamics, and in the fixed-$\varepsilon$ or gravity ($\hat{\Phi}$/$\hat{\bf a}$/${\bf T}$)-based schemes, negligible effect on the timesteps $\Delta t \approx \Delta t_{\rm global}$, which remain constant at their ``global'' value ($\sim C\,L_{\rm global}/|{\bf v}|$, where $L_{\rm global}$ is some global potential scale-length) set by the gravity integrator. 
	But in the neighbor ($\tilde{\rho}/n_{\rm NGB}$)-based schemes ({\em bottom}), the change in $\tilde{\rho}$/$n_{\rm NGB}$ (equivalently, the {\em local} particle/mass arrangement) as a test particle moving with stream $1$ intersects (or moves out of) stream $2$ leads by definition to a change in $\varepsilon$ over a scale length $\Delta x \sim \varepsilon$, or timescale $\delta t \sim \varepsilon/|{\bf v}| \sim 1/| \tilde{\nabla} \cdot {\bf v}|$ (where $\tilde{\nabla}$ is an ``SPH-like'' divergence estimator based on the in-kernel neighbors). So maintaining energy conservation in the $\tilde{\rho}/n_{\rm NGB}$ schemes necessarily requires a timestep $\Delta t \lesssim C\,\delta t$, a factor of $\sim (\epsilon/L_{\rm global})$ smaller than the orbital/gravitational dynamics timestep $\Delta t_{\rm global}$.
	\label{fig:h.crossing}}
\end{figure}

\section{Implications for the Timestep}
\label{sec:timesteps}

Any N-body method imposes various timestep criteria for the sake of stability, usually posed in terms of the acceleration ${\bf a}$ or tidal tensor ${\bf T}$ \citep[see e.g.][]{rasio:1991.stellar.merger.hydro,monaghan:1992.sph.review,kravtsov:1997.ART,truelove:1998.gmc.frag,aarseth:2003.nbody.review,power:2003.nfw.models.convergence,rodionov:2005.optimal.force.softening,dehnen:2011.nbody.review,grudic:2020.tidal.timestep.criterion}. Our derivations and the introduction of adaptive force softening do not fundamentally alter any of these. However, this does introduce a new variable $\varepsilon_{a}$ which can vary in time and appears in multiple places in the dynamics equations. It is therefore immediately obvious that we should also impose a restriction requiring that $\varepsilon_{a}$ cannot change too much in a timestep $\Delta t_{a}$, i.e.\ $|\Delta t_{a} \, {\rm d} \varepsilon_{a} / {\rm d} t| < C\,\varepsilon_{a}$ where ${\rm d} \varepsilon_{a} / {\rm d} t$ is the comoving derivative with particle $a$ and $C$ is some Courant-like factor. But expanding ${\rm d} \varepsilon_{a} / {\rm d} t$, it is easy to show that it is trivially related to the quantities we already derived above, giving: 
\begin{align}
\left| \frac{{\rm d}\varepsilon_{a}}{{\rm d}t}  \right| &= \left|  \sum_{b} \frac{m_{b}}{\zeta_{a}}\,\boldsymbol{\Upsilon}_{ab} \cdot \left({\bf v}_{b} - {\bf v}_{a} \right) \right| \\ 
\label{eqn:courant} \Delta t_{a} &< C\,\frac{\varepsilon_{a}}{| {\rm d}\varepsilon_{a} / {\rm d} t |} 
\end{align}
This can easily be computed alongside the forces defined above at no additional computational cost. 

Briefly, for the ``local neighbor'' based models for $\varepsilon_{a}$ (e.g.\ the $\tilde{\rho}$ and $n_{\rm NGB}$ models in \S~\ref{sec:ex.rho}-\ref{sec:ex.n}), inserting the definition of $\boldsymbol{\Upsilon}$ immediately shows that this is equivalent to $\Delta t_{a} \lesssim C / |\tilde{\nabla} \cdot {\bf v}|$, where $\tilde{\nabla} \cdot {\bf v}$ is a simple kernel estimator of the velocity divergence (identical to the estimator commonly used in SPH methods). This is itself roughly equivalent to a (comoving) Courant-like criterion, $\Delta t_{a} \lesssim C \, \varepsilon_{a} / v_{{\rm sig},\,a}$, where the signal velocity $v_{{\rm sig},\,a}$ is some appropriately-weighted measure of the maximum {\em relative} velocity |${\bf v}_{a} - {\bf v}_{b}|$ of all particles within the interaction kernel $W$. This is illustrated in Fig.~\ref{fig:plummer.dt}. As shown in \citet{hopkins:fire2.methods} (especially \S~2 \&\ 4 and Figs.~15-20 therein), where adaptive softening methods following both the $\tilde{\rho}$ and $n_{\rm NGB}$ methods are extensively tested with different timestep and integration schemes, it is essential to include this sort of timestep limiter in order to integrate these schemes stably, because of how the softening (and therefore energy conservation) is sensitive to the local particle order (by definition). There the authors show explicitly that such a restriction (with $C \lesssim 0.25$, for Eq.~\ref{eqn:courant}, and a similar signal-velocity based criterion to catch cases where $|\tilde{\nabla} \cdot {\bf v}|$ under-estimates the change about to occur), as well as a corresponding ``wakeup'' condition for adaptive timestepping as in \citet{saitoh.makino:2009.timestep.limiter}\footnote{Defined, like with hydrodynamics, such that any particle which has a potentially-interacting neighbor approach its kernel on a much smaller timestep must be ``awakened'' and moved to a lower corresponding timestep to prevent the faster-evolving neighbor from changing {\em its own} neighbor configuration too much in a timestep.} are all required not just to maintain accuracy, but also to actually ensure that the scheme conserves energy (the point of adding the correction terms in the first place) and to prevent a serious numerical instability that can occur which can cause artificial collapse of particle clusters to arbitrarily high densities (somewhat akin to hydrodynamic instabilities with $C \gg 1$, but with a pairing-instability-like runaway condition).

As discussed in \citet{hopkins:fire2.methods}, this can be simply understood by considering the ``test particle limit,'' illustrated in Fig.~\ref{fig:h.crossing}. Consider two nearby dynamically cold ``clouds'' or ``sheets'' or ``streams'' of particles (or even two particles in relative isolation), approaching one another with large relative velocity in a smooth, static background potential  -- an example chosen to be an (idealized) representation of a very common situation in the cosmological simulations shown in \S~\ref{sec:cosmo}. Since they are collisionless, the particles/structures should simply ``move through'' one another without feeling any perturbation, and most integrators with e.g.\ fixed softening will allow them to do so (correctly) in an arbitrarily large timestep. The ``gravity-based'' methods (the $\hat{\Phi}$, $\hat{\bf a}$, ${\bf T}$-based softening models in \S~\ref{sec:ex}) of course allow this (as they should), since the background potential is smooth, so $\varepsilon$ barely changes in such an intersection. But in the ``local neighbor'' based methods ($\tilde{\rho}$ and $n_{\rm NGB}$-based rules in \S~\ref{sec:ex}), the softening length by definition depends strictly on the {\em local} particle or mesh-generating-point configuration/density. This changes on a timescale $\sim1/|\tilde{\nabla} \cdot {\bf v}|$, deforming $\varepsilon$ and giving rise to significant changes in $\boldsymbol{\Upsilon}$ as the two groups ``move through'' each other's softening kernels (with the change especially rapid, i.e.\ $\sim1/|\tilde{\nabla} \cdot {\bf v}| \sim \varepsilon/|\Delta v|$, as their centers pass through the kernel ``core''). In other words, $\varepsilon$ fluctuates much more rapidly in response to local structure. As a result, integrating ``smoothly'' and accurately through this and maintaining energy conservation (since this is only ensured at integration-accuracy level in the methods here) necessarily requires a Courant-like condition where the particles are only allowed to move a small fraction of their own softening length per timestep, to spread the ``encounter'' over many timesteps.

In contrast, from the definition of $\varepsilon_{a}$, we immediately have for the potential/acceleration/tidal models $\varepsilon_{a} / |{\rm d}_{t}\varepsilon_{a}| \sim |\hat{\Phi}_{a}|/|{\rm d}_{t} \hat{\Phi}_{a}|$ or $\sim 2\,|\hat{\bf a}_{a}|/|{\rm d}_{t} |\hat{\bf a}_{a}| |$ or $\sim 3\,\|{\bf T}_{a}\|/|{\rm d}_{t} \| {\bf T}_{a}\| |$. In other words, the timestep criterion imposed here is simply that the potential/acceleration/tidal tensor seen by particle $a$ should not change by a large ($\mathcal{O}(1)$ or larger) factor within a single timestep. But any reasonable timestep criterion for accurate gravitational integration should {\em already} ensure this (see references above) -- otherwise the particle orbit could not possibly be integrated correctly because a particle could simply ``skip over'' a region with a strongly varying tidal tensor in a single timestep. So in practice, for these potential-based model choices for $\varepsilon_{a}$, we find that formally including the restriction of Eq.~\ref{eqn:courant} makes no significant difference compared to just using the ``normal'' timestepping, in terms of accuracy as well as energy conservation and numerical stability.

Some illustrations of this are shown in Fig.~\ref{fig:plummer.dt} for the Plummer sphere test of Fig.~\ref{fig:plummer.h} (\S~\ref{sec:ex.demo}), which plots the distribution of numerical timesteps at different radii in the Plummer sphere. Because we adopt a standard timestep criteria for integrating gravity (here following \citealt{dehnen:2011.nbody.review,grudic:2020.tidal.timestep.criterion}), this largely dominates over Eq.~\ref{eqn:courant} and means the timestep distribution is essentially identical for the fixed-$\varepsilon$, and potential/acceleration/tidal models. But for the ``local neighbor'' models we see in the Plummer sphere core (where particle-particle encounters are more common) that the median timestep is an order-of-magnitude smaller (with the larger scatter in timestep associated with different impact parameters for particle encounters). We see the same effect (with similar magnitude) in our tests with cosmological simulations discussed further below.

\section{Correctly Handing Mixed Interactions in Multi-Species Simulations}
\label{sec:multi}

Imagine we have different softening rules for different particle {\em species} (e.g.\ gas, stars, dark matter, black holes, planets, neutrinos, etc.), so $\mathcal{E}_{b} \rightarrow \mathcal{E}_{b}(... \, , \, s_{b})$ where $s_{b}$ indicates the species label of particle $b$. We will also allow the ``symmetrization rule'' or form of $\tilde{\phi}_{bc}$ to depend on $(s_{b},\,s_{c})$. This adds $s_{b}$ to the state vector ${\bf U}_{b}$ above, but does not fundamentally change our derivation because $s_{b}$ does not change in time or space under pure gravitational operations. 

We therefore have, for a given species and particle $a$, that $\zeta_{a}$ $\Omega_{a}$, and even $\boldsymbol{\Upsilon}_{ab}$ remain the same as we derived above, but e.g.\ $\boldsymbol{\Upsilon}_{ab}$, $\zeta_{a}$, etc. (all entries leading index ``$a$'') use the rule for particle $a$. Their ``rule'' $\mathcal{E}_{a}$ is what appears in these terms. And likewise for $b$. Mixed terms will still automatically give rise to antisymmetric forces, according to the definition of Eq.~\ref{eqn:eom}.

So for example, imagine type $s=0$ (e.g.\ a collisionless gas) uses the $\tilde{\rho}$ rule, based on a kernel $W$ that only includes particles of the same type; while type $s=1$ (e.g.\ dark matter) uses the tidal tensor ${\bf T}$; and type $s=2$ (e.g.\ sink particles) uses constant softenings. For each pair in the force sum $a$, $b$, if $s_{a}=s_{b}$, obviously we use the ``normal'' $\boldsymbol{\Upsilon}_{ab} - \boldsymbol{\Upsilon}_{ba}$ according to the rule for that species. In an interaction with $s_{a}=0$, $s_{b}=1$, we have $\boldsymbol{\Upsilon}_{ab} = (\zeta_{a}\,\varepsilon_{a}/\Omega_{a}\,3\,\tilde{\rho}_{a})\,\partial W_{ab}(\varepsilon_{a},\,{\bf U}_{a},\,{\bf U}_{b}) / \partial {\bf x}_{a} $ (the rule for $s_{a} = 0$), but this vanishes ($W_{ab}\rightarrow \mathbf{0}$), because the kernel function $W_{ab}$ (by definition) used to estimate $\varepsilon_{a}$ vanishes for any particles of a different type (the $s=0$ or ``gas'' particles only ``see''  particles of their same species when calculating their $\varepsilon_{a}$, in this example). But since $b$ with $s_{b}=1$ uses the tidal tensor to estimate its $\varepsilon_{b}$, which ``sees'' all particles that have non-zero mass, we have $\boldsymbol{\Upsilon}_{ba} = (\zeta_{b}\,\varepsilon_{b}/\Omega_{b}\,3\,\|{\bf T}_{b}\|^{2})\,{\bf T}_{b}^{ij}\,\partial_{i}\partial_{j} \partial_{{\bf x}_{b}} \hat{\phi}_{ab} \ne 0$ as usual. If $s_{a}=0$, $s_{b} = 2$, we have $\boldsymbol{\Upsilon}_{ba}=\mathbf{0}$, because this is the rule for $s_{b}=2$ (fixed softenings), and $\boldsymbol{\Upsilon}_{ab} = (\zeta_{a}\,\varepsilon_{a}/\Omega_{a}\,3\,\tilde{\rho}_{a})\,\partial W_{ab}(\varepsilon_{a},\,{\bf U}_{a},\,{\bf U}_{b}) / \partial {\bf x}_{a} = \mathbf{0}$, because again $b$ does not appear in the kernel for $a$. 

As detailed in \S~\ref{sec:gas}, these rules are consistent in mixed interactions with species such as gas in finite-volume fluid dynamics simulations, or ``true'' point-mass like particles, for which the ``correct'' force softenings can be rigorously derived.

\begin{figure}
	\includegraphics[width=0.95\columnwidth]{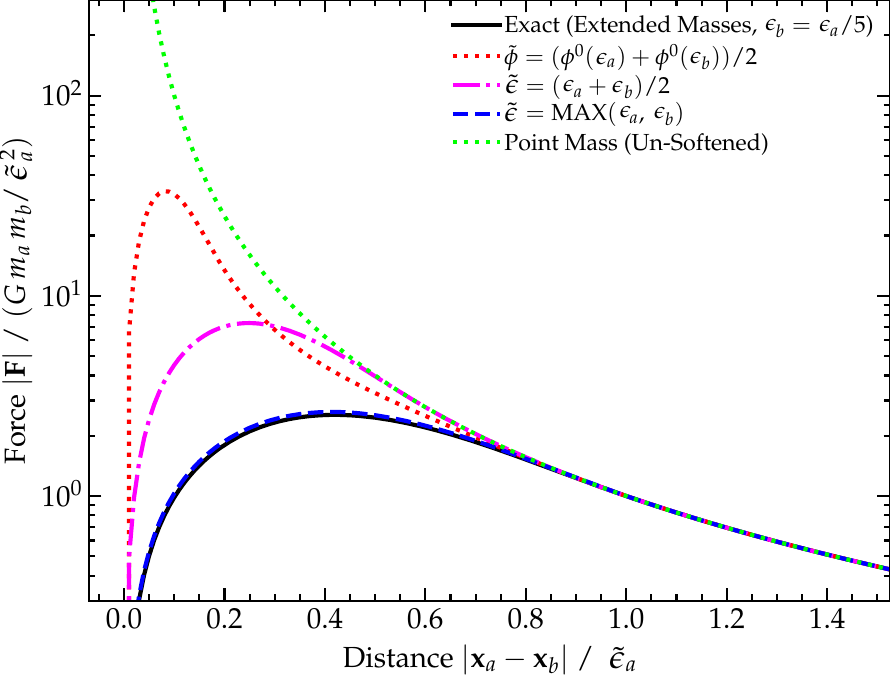}
	\includegraphics[width=0.95\columnwidth]{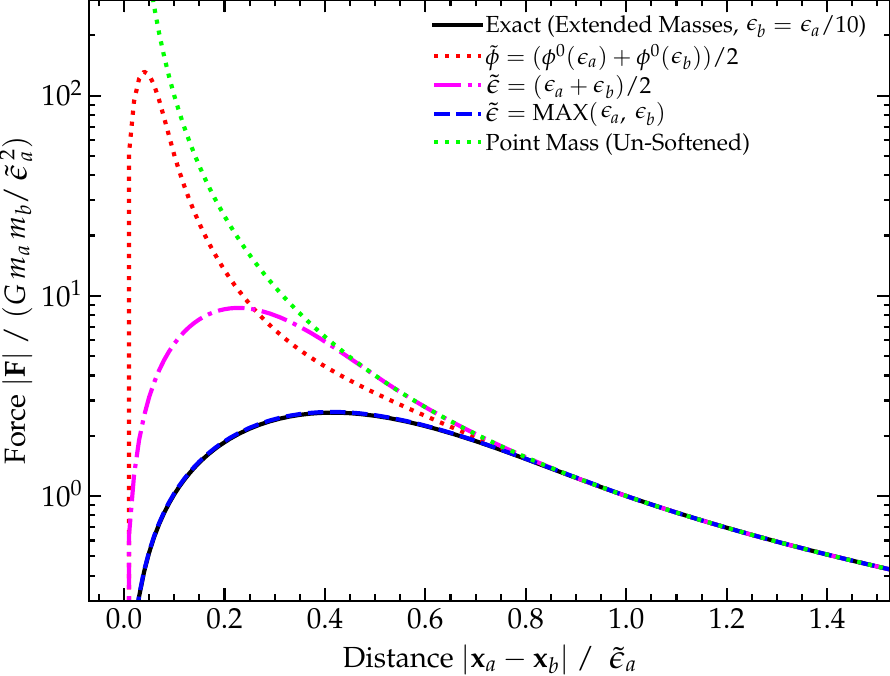}
	\vspace{-0.1cm}
	\caption{Behavior of the force in a close particle encounter following different symmetrization rules for $\tilde{\phi}_{bc}=\tilde{\phi}_{cb}$ discussed in \S~\ref{sec:symm}, as a function of the separation between the particles in units of the compact support radius $\tilde{\varepsilon}$ of the (non-Keplerian) softening kernel. We consider an encounter of two particles $a$, $b$ of masses $m_{a}$, $m_{b}$ with unequal softenings $\varepsilon_{a}\ne \varepsilon_{b}$ (ratio of $\varepsilon_{a}$ and $\varepsilon_{b}$ taken to be $=5$ or $=10$ at {\em top} and {\em bottom}), and plot the magnitude of the total force along the line connecting their centers-of-mass. We compare the un-softened limit (treating both as point masses) and the softened result (all adopt a cubic spline $K$, see \S~\ref{sec:kernels}) for the ``average the force'' rule ($\tilde{\phi} = (1/2)\,(\phi^{0}(\varepsilon_{a}) + \phi^{0}(\varepsilon_{b}))$, Eq.~\ref{eqn:phisymm.avg}), ``average the softening'' ($\tilde{\phi}=\phi^{0}(\tilde{\varepsilon})$ with $\tilde{\varepsilon}=(\varepsilon_{a}+\varepsilon_{b})/2$, Eq.~\ref{eqn:hsymm.avg}) and ``maximum softening'' ($\tilde{\varepsilon}={\rm MAX}(\varepsilon_{a},\,\varepsilon_{b})$) rules. We also compare the exact result if we treat each particle as an extended mass distribution with size parameter $\varepsilon_{a}$ obeying $\nabla^{2} \phi_{a}(\varepsilon_{a}) = 4\pi\,G\,\rho$ (\S~\ref{sec:kernels}). By definition all models become Keplerian at large distance. In close encounters, the ``${\rm MAX}$'' rule gives nearly identical results to the exact extended-mass calculation (and becomes exact for highly-unequal softenings), while the ``average $\tilde{\varepsilon}$'' rule deviates systematically and the ``average the force'' rule remains nearly-Keplerian (un-softened) until both particles are well inside the {\em smaller} of the two force softenings.
	\label{fig:force.symm.vs.exact}}
\end{figure}

\begin{figure}
	\includegraphics[width=0.95\columnwidth]{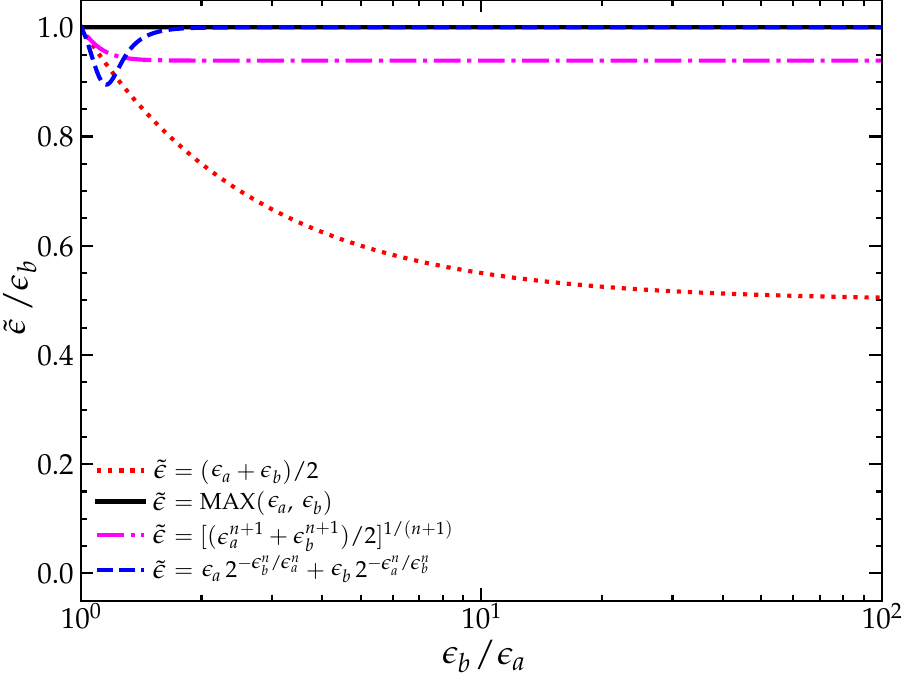}
	\vspace{-0.1cm}
	\caption{Illustration of the behavior of different symmetry rules for $\tilde{\varepsilon}$ discussed and proposed in \S~\ref{sec:symm}. For a particle pair $a$, $b$ we compare the symmetrized kernel size $\tilde{\varepsilon}$ (in units of $\varepsilon_{b}$) to the ratio $\varepsilon_{b}/\varepsilon_{a}$ (for values $\varepsilon_{b} \ge \varepsilon_{a}$; we can trivially swap $a$ and $b$ to see behavior for $\varepsilon_{b} \le \varepsilon_{a}$). The ${\rm MAX}$ rule gives $\tilde{\varepsilon}=\varepsilon_{b}$ trivially. The averaging rule leads to $\tilde{\varepsilon} \sim \varepsilon_{b}/2$ when $\varepsilon_{b} \gg \varepsilon_{a}$, the origin of the systematic deviation seen in Fig.~\ref{fig:force.symm.vs.exact}. Our new proposed rules Eq.~\ref{eqn:hsymm.weight} and Eq.~\ref{eqn:hsymm.weight.exp}, here both plotted for $n=10$, behave most similar to the ${\bf MAX}$ rule but remain differentiable for all $\varepsilon$ (though Eq.~\ref{eqn:hsymm.weight} produces a small systematic deviation for $\varepsilon_{b} \gg \varepsilon_{a}$, which is smaller for larger $n$). 
	\label{fig:force.symm.vs.rule}}
\end{figure}

\section{Choice of Symmetry Rule for the Softened Potential}
\label{sec:symm}

As discussed in \S~\ref{sec:deriv}, the function $\tilde{\phi}_{bc}=\tilde{\phi}_{bc}({\bf x}_{b}-{\bf x}_{c},\,\varepsilon_{b},\,\varepsilon_{c})$ must obey the symmetry $\tilde{\phi}_{bc} = \tilde{\phi}_{cb}$ for any physical system. The choice of how to ensure this is arbitrary as is the detailed functional form of $\tilde{\phi}$, but the overwhelming majority of the literature makes one of a couple choices to ensure this, which we review and generalize here. Recall, when the function $\tilde{\phi}_{bc}$ is Newtonian (outside the compact support radius of $\varepsilon_{b}$, $\varepsilon_{c}$), $\tilde{\phi}_{bc} = G/|{\bf x}_{b} - {\bf x}_{c}|$, so $\tilde{\phi}_{bc} = \tilde{\phi}_{cb}$ is trivially satisfied. But with two different softening parameters $\varepsilon_{b} \ne \varepsilon_{c}$, it is convenient to define $\tilde{\phi}(\varepsilon_{b},\,\varepsilon_{c})$ with respect to the ``single softening parameter'' function $\phi^{0}$, defined as 
\begin{align}
\phi_{ab}^{0} &\equiv \phi_{ab}^{0}(r_{ab} \equiv |{\bf x}_{a} - {\bf x}_{b}|\, ,\,\tilde{\varepsilon}) \ .
\end{align}
Many different symmetrization options have been proposed but the most widely-used include:
\begin{align}
\label{eqn:phisymm.avg} \tilde{\phi}_{ab} &\rightarrow \frac{1}{2} \left[ \phi^{0}_{ab}\left( r_{ab}\, , \, \tilde{\varepsilon}=\varepsilon_{a} \right) + \phi^{0}_{ab} \left( r_{ab} \, , \, \tilde{\varepsilon}=\varepsilon_{b} \right)  \right] \\
\label{eqn:hsymm.avg} \tilde{\phi}_{ab} &\rightarrow \phi^{0}_{ab}\left(r_{ab}\, , \, \tilde{\varepsilon}=\frac{\varepsilon_{a} + \varepsilon_{b}}{2} \right) \\
\label{eqn:hsymm.max} \tilde{\phi}_{ab} &\rightarrow \phi^{0}_{ab}\left(r_{ab}\, , \, \tilde{\varepsilon}={\rm MAX}\left[ \varepsilon_{a} \, , \, \varepsilon_{b} \right] \right) 
\end{align}
The first of these (Eq.~\ref{eqn:phisymm.avg}, averaging the forces) is by far most widely used for prior adaptive softening implementations \citep[e.g.][]{price:2007.lagrangian.adaptive.softening,springel:arepo,hopkins:gizmo,hubber:gandalf.gizmo.methods,sphenix:swift.mfm.implementation.and.sph.methods}, and the second (Eq.~\ref{eqn:hsymm.avg}, which amounts to averaging $\varepsilon$ before computing the forces) is also discussed in e.g.\ \citet{hernquist:1990.barnes.nbody.review,benz:1990.sph.review,price:2007.lagrangian.adaptive.softening}. The third (Eq.~\ref{eqn:hsymm.max}) is most widely used for fixed-softening simulations \citep[e.g.][]{barnes:1985.merger.sims,springel:gadget}. 

Most previous comparisons of these choices (e.g.\ \citealt{hernquist:1990.barnes.nbody.review,dehnen:2001.optimal.softening}) considered idealized test problems with a single collisionless species, such that the softening lengths of neighboring particles were generally similar. In that case, these and other subsequent studies largely concluded that the differences in accuracy were minimal (there could be some advantages to the averaging based schemes, but these are usually negligible compared to differences in the gravity integrator, timestepping scheme, and normalization of the softenings chosen, see e.g.\ \citealt{power:2003.nfw.models.convergence,grudic:2020.tidal.timestep.criterion}). However going back to \citet{barnes:1985.merger.sims}, and more recently in e.g.\ \citet{grudic:starforge.methods} and references therein, it is well-known that the ``averaging'' schemes (especially Eq.~\ref{eqn:phisymm.avg}) can produce catastrophically large errors and unphysical results in systems with highly unequal softenings. Consider, for example, two neighboring particles in a smooth background, one ``more extended'' (``$a$'') with a potential given by $m_{a}\,\phi^{0}_{a}(|{\bf x} - {\bf x}_{a}|\,,\,\varepsilon_{a})$, the other (``$b$'') with mass $m_{b}$ and softening $\varepsilon_{b} \ll \varepsilon_{a}$ which is very small (illustrated explicitly in Fig.~\ref{fig:force.symm.vs.exact}). This can arise quite easily in a multi-physics, multi-species simulation: for example a simulation of star formation or stellar or black hole or planetary dynamics where $a$ represents some smooth extended ``background'' collisionless medium (e.g.\ dark matter, or populations of low-mass stars on galactic scales), while $b$ represents a point-like mass (e.g.\ a sink particle or individual black hole/star/planet) which should have $\varepsilon_{b} \rightarrow 0$, effectively. Since these particles are collisionless, they can and should be able to ``pass through'' one another so imagine that particle $b$ approaches and moves through the center of particle $a$, i.e.\ ${\bf x}_{b} \rightarrow {\bf x}_{a}$ (there is no force that can or should ``restore particle order'' or prevent this). In any reasonable physical picture, the compact object $b$ should pass through a ``sea'' of background particles represented by the extended softening of $a$. Physically, it is trivial to show that in the limit $\varepsilon_{b}\rightarrow 0$ (a point mass $b$), the ``interaction potential'' (excluding self terms of $a$ and $b$ in isolation) is exactly $m_{a}\,m_{b}\,\tilde{\phi}^{0}_{a}(|{\bf x}_{b}-{\bf x}_{a}|\,,\,\varepsilon_{a})$, i.e.\ $\tilde{\varepsilon} = {\rm MAX}(\varepsilon_{a},\,\varepsilon_{b})$ as Eq.~\ref{eqn:hsymm.max}. In contrast, if we used Eq.~\ref{eqn:phisymm.avg}, then the interaction potential would unphysically diverge as ${\bf x}_{b} \rightarrow {\bf x}_{a}$, as $\sim (1/2)\,m_{a}\,m_{b}/|{\bf x}_{b} - {\bf x}_{a}|$ (i.e.\ it is only suppressed by a factor of $2$, relative to completely un-softened gravity). This can obviously produce extremely large $N$-body scattering events, and in \citealt{grudic:starforge.methods} the authors note this can easily lead to unphysical ejection of stars or gas, artificial dissolution of binaries and multiples, and related effects. It also defeats the entire point of introducing softened gravity in the first place. Eq.~\ref{eqn:hsymm.avg} avoids the divergence, but gets the interaction potential systematically wrong by a factor of $\sim 2$ in this regime. 

Other examples (e.g.\ interactions of two extended spheres) are discussed in e.g.\ \citet{dyer.ip:1993.softening.law}, and in complete generality solving for the interaction potentials between two arbitrary extended mass distributions with arbitrary kernel functions is highly non-trivial. But it is always the case that when the softenings are highly unequal ($\varepsilon_{b} \gg \varepsilon_{a}$ or $\varepsilon_{a}\gg \varepsilon_{b}$), the $\tilde{\varepsilon}={\rm MAX}(\varepsilon_{b},\,\varepsilon_{a})$ approach (Eq.~\ref{eqn:hsymm.max}) gives a result much closer to any real physical model (for the reasons above) and most importantly avoids the unphysical divergence of Eq.~\ref{eqn:phisymm.avg}; whereas when the softenings are similar, all these approaches give very similar results. However this ${\rm MAX}(\varepsilon_{a},\,\varepsilon_{b})$ approach naively has the same problem for our scheme as the enforcement of a minimum/maximum softening by simple application of a ${\rm MIN}$ or ${\rm MAX}$ function in \S~\ref{sec:minmax}, in that it is not differentiable so seems at first incompatible with our derivation. We therefore introduce the formulations 
\begin{align}
\label{eqn:hsymm.weight} \tilde{\phi}_{ab} &\rightarrow \phi^{0}_{ab}\left(r_{ab}\, , \, \tilde{\varepsilon}=\left[ \frac{\varepsilon_{a}^{n+1} + \varepsilon_{b}^{n+1}}{2} \right]^{\frac{1}{n+1}} \right) \\ 
\label{eqn:hsymm.weight.exp} \tilde{\phi}_{ab} &\rightarrow \phi^{0}_{ab}\left(r_{ab}\, , \, \tilde{\varepsilon}=\left[ \varepsilon_{a}\,2^{-\varepsilon^{n}_{b}/\varepsilon^{n}_{a}} + \varepsilon_{b}\,2^{-\varepsilon^{n}_{a}/\varepsilon^{n}_{b}} \right] \right) \ .
\end{align}
which allow one to interpolate as desired between a linear average like Eq.~\ref{eqn:hsymm.avg} ($n\rightarrow 0$) and ${\rm MAX}(\varepsilon_{a},\,\varepsilon_{b})$ ($n\rightarrow\infty$), but also allow for harmonic averages and other behaviors. A simple example is illustrated in Fig.~\ref{fig:force.symm.vs.rule}. In practice, using Eq.~\ref{eqn:hsymm.weight} or Eq.~\ref{eqn:hsymm.weight.exp} with any modest $n \gtrsim 10$ gives results nearly identical to the ${\rm MAX}$ formulation of Eq.~\ref{eqn:hsymm.max} while remaining differentiable (though Eq.~\ref{eqn:hsymm.weight} produces a small systematic deviation as $\varepsilon_{b}/\varepsilon_{a} \rightarrow \infty$ shown in Fig.~\ref{fig:force.symm.vs.rule}, this is not a large source of error in most applications). 

For any of the formulations where $\tilde{\varepsilon}=\tilde{\varepsilon}(\varepsilon_{a},\,\varepsilon_{b})$, then in e.g.\ $\zeta$ and $\Omega$ terms where one needs to calculate derivatives such as $\partial_{\varepsilon_{a}} \tilde{\phi}_{bc} = \partial \tilde{\phi}_{bc}/{\partial \varepsilon_{a}}$, it is trivial to evaluate these according to a simple chain rule 
$\partial_{\varepsilon_{a}} \tilde{\phi}_{bc} \rightarrow (\partial_{\tilde{\varepsilon}} \tilde{\phi}_{bc})\,(\partial_{\varepsilon_{a}} \tilde{\varepsilon})$, etc. As noted below there is essentially no detectable difference in computational cost per timestep between these different formulations.

\section{Choice of the Softening Function}
\label{sec:kernels}

Of course, for a real simulation, we must choose and define some $\tilde{\phi}$ by virtue of defining $\phi^{0}(r,\,\tilde{\varepsilon})$ in \S~\ref{sec:symm}. The easiest way to do so in a manner that will {automatically} ensure that the final function meets all of the additional criteria defined in \S~\ref{sec:deriv} is to build $\phi^{0}$ directly from a dimensionless kernel function $K$ which has the properties needed for good behavior in applications such as kernel density estimation, kernel-based volume partition or Voronoi approximation, or smoothed-particle hydrodynamics \citep[e.g.][]{schoenberg:1946.smoothing.kernels,liu:1995.reproducing.kernel.particle.methods,fulk.quinn:1996.sph.kernels,liu:2003.sph.kernels,hongbin.xin:05.sph.kernels,dehnen.aly:2012.sph.kernels,yang:2014.sph.kernels}.\footnote{For example, $K(u)$ is  positive-definite, finite, and sufficiently smooth within a domain of compact support with vanishing central and boundary derivatives, and is normalized so $\int K(u)\,{\rm d}^{3}{\bf u} = 1$. Note many of these conditions are similar to those considered ``optimal'' in \citet{dehnen:2001.optimal.softening}, but as noted in subsequent studies (including \citealt{dehnen:2011.nbody.review,dehnen.aly:2012.sph.kernels}) the ``negative density'' kernels proposed therein (equivalent to a non-positive-definite $K$ here) can produce not just numerical instabilities but a tidal tensor without negative definite eigenvalues or trace, which leads immediately to  failure of many higher-order numerical integration methods, timestepping schemes, and other multi-physics applications like sink formation/accretion criteria \citep[see references above and][]{hopkins:fire2.methods,guszejnov:2018.isothermal.nocutoff,grudic:2020.tidal.timestep.criterion,grudic:starforge.methods,grudic:2021.accelerating.hydro.with.adaptive.force.updates}. This leaves the ``optimal'' softening schemes studied therein (which considered accuracy of reproduction of forces in an instantaneous sense) akin to the spline models here.} This amounts to 
\begin{align}
\label{eqn:kernel.def} \phi^{0} \rightarrow -\frac{4\pi\,G}{\tilde{\varepsilon}}\left[ \frac{1}{u}\int_{0}^{u} K(u^{\prime})\,u^{\prime 2}\,{\rm d}u^{\prime} + \int_{u}^{\infty} K(u^{\prime})\,u^{\prime}\,{\rm d}u^{\prime} \right]
\end{align}
where $u\equiv r/\tilde{\varepsilon}$ \citep{binneytremaine}.

Consider, for example, the popular \citet{morris:1996.sph.stability} cubic B-spline in $D=3$: $K=(8/\pi)\,(1-6\,u^{2}\,(1-u))$ for $u\le 1/2$, $K=(16/\pi)\,(1-u)^{3}$ for $1/2<u\le 1$, and $K=0$ for $u>1$. Eq.~\ref{eqn:kernel.def} immediately gives $\phi^{0}= (2\,G/15\,\tilde{\varepsilon})\,(-21+40\,u^{2}-72\,u^{4}+48\,u^{5})$ for $u\le 1/2$, $\phi^{0}=(G/15\,r)\,(1-48\,u+160\,u^{3}-240\,u^{4}+144\,u^{5}-32\,u^{6})$ for $1/2<u\le 1$, and $\phi^{0}=-G/r$ for $u>1$. 

For completeness, in the public {\small GIZMO} code\footnote{\gizmourl} \citep{hopkins:gizmo,hopkins:mhd.gizmo,hopkins:gizmo.public.release}, the functions: $\phi^{0}$, $\partial_{\tilde{\varepsilon}} \phi^{0}$, $\partial_{{\bf x},i} \phi^{0}$, $\partial_{{\bf x},\,i}\partial_{{\bf x},\,j} \phi^{0}$, $\partial_{{\bf x},\,i}\partial_{{\bf x},\,j} \partial_{{\bf x},\,k} \phi^{0}$, $K$, $\partial_{u} K$ needed to construct any quantity defined in this paper, in $D=1$, $2$, and $3$ dimensions, are given for each of many different generating kernels $K$. The functions $K$ there include: the \citet{schoenberg:1946.smoothing.kernels} or \citet{morris:1996.sph.stability} cubic, quartic, and quintic B-splines ($b_{4}$, $b_{5}$, $b_{6}$), the Wendland $C^{2}$, $C^{4}$, and $C^{6}$ functions \citep{dehnen.aly:2012.sph.kernels}, and the quadratic 2-step and ``peak'' kernels and linear ramp kernels widely used in image processing. 

Our derivations are agnostic to this choice, but we have experimented with a variety of kernels using the different $\varepsilon_{a}$ estimators here, in a variety of applications including cosmological, galaxy-formation, star-formation, accretion disk, and stellar and planetary dynamics simulations. We find (consistent with most previous studies) no appreciable systematic increase in accuracy using more complicated higher-order kernels compared to the cubic spline or $C^{2}$. This is quite different from e.g.\ the situation with some kernel-based hydrodynamics solvers such as SPH (where high-order kernels are often necessary; see e.g.\ \citealt{agertz:2007.sph.grid.mixing,read:2010.sph.mixing.optimization,dehnen.aly:2012.sph.kernels}), for several reasons: most obviously, in any situation where adaptive softening is needed the forces should be dominated by collective long-range effects, not by the immediate neighbors; moreover there are no ``intercell fluxes'' or pressure forces across some effective face where higher-order kernels are needed to minimize E0 and other errors in SPH related to the ``closure'' of the faces \citep{abel:2011.sph.pressure.gradient.est,hopkins:lagrangian.pressure.sph,hopkins:cg.mhd.gizmo,hopkins:gizmo.diffusion}; in addition the softening kernel $\phi^{0}$ is already automatically two orders ``more smooth'' than $K$, if defined as above. On top of this, to achieve similar accuracy, the higher-order kernels ($C^{>2}$, $b_{>4}$) require larger radii of compact support, which substantially increases the computational expense (by factors of several).\footnote{With the higher-order kernels one can also potentially degrade the effective resolution if the scaling of the radius of compact support relative to kernel size/shape is not well-chosen. In comparing different kernels, $\varepsilon$ from \S~\ref{sec:ex} should represent something like the ``core size'' of the kernel, for which we follow \citet{dehnen.aly:2012.sph.kernels} by setting $\varepsilon=2\,\sigma$ with $\sigma$ the kernel standard deviation (Eq.~8 therein). This means that the ratio of $\varepsilon$ to the kernel radius of compact support or effective ``neighbor number'' for e.g.\ the neighbor-based methods varies following Table~1 therein. We refer to the public code for implementation details.} 
So we generally favor the simpler kernels in our practical applications in {\small GIZMO} multi-physics simulations.

\section{Computational Expense}
\label{sec:cpu}

It is always difficult to compare computational/CPU costs of different methods, because it is highly problem and implementation-dependent. Comparing  adaptive softening methods  to  ``fixed softening'' methods is especially fraught. Of course, the added operations here add some CPU cost per timestep if all else (e.g.\ $\varepsilon$) were exactly equal. But if the problem has a high dynamic range of densities (e.g.\ a cosmological simulation, as in \S~\ref{sec:cosmo}), then in fixed-softening methods with the softening chosen to match some mean inter-particle spacing the dense regions will have huge numbers of particles inside the softening kernel which becomes very expensive (shared-softening generally requiring explicit $N^{2}$ operations). One could adopt fixed softening with  small $\varepsilon$, but this would give large force errors, and could become expensive if the timestep criterion depends on $\varepsilon$ \citep[as commonly adopted, e.g.][]{power:2003.nfw.models.convergence}.\footnote{For the cosmological simulations in \S~\ref{sec:cosmo}, for example, we discuss the relative CPU cost to run different simulations to $z=0$. The simulations with constant softening set to the smallest values therein are least expensive, followed by the tidal softening models ($\sim 25-50\%$ slower) then the simulations with larger fixed softening (several times more expensive), and finally the local density $\tilde{\rho}/n_{\rm NGB}$ based models are the most expensive by another factor of several (cost increasing with $N_{\rm NGB}$).} And CPU cost ``per timestep'' or ``per problem'' is not the most helpful metric -- if different methods enable different effective resolution or accuracy, CPU cost ``to fixed accuracy of solution'' is more important. Moreover, adaptive and fixed-softening methods often develop non-linearly different small-scale structures, which lead to different CPU costs that usually swamp any ``all else equal'' cost comparison \citep[see][and simulation examples below]{hopkins:fire2.methods}. 

It is slightly more well-posed to compare the adaptive methods here (\S~\ref{sec:ex}) to one another. All of the method variants discussed here fundamentally require the same number of ``loops'' of the neighbor/force trees.\footnote{Though of course some loops -- e.g.\ those over just neighbors as compared to over all long-range forces -- are less expensive, in our implementation each requires an independent $\mathcal{O}(N\,\log{N})$ tree walk, and so the cost difference is swamped by the timestepping difference noted above.} This is a modest cost and at least one such loop is already required for fixed softenings. Moreover, modern higher-order integrators already require multiple tree-walks (e.g.\ the fourth-order Hermite integrator in \citealt{grudic:starforge.methods}, implemented in our tests in {\small GIZMO}, or any other integrator built on \citealt{makino.aarseth.1992:hermite.integrator}, or those in \citealt{aarseth:2003.nbody.review}), so all computations can be done alongside existing operations.\footnote{Ideally in such a higher-order integrator one would re-compute the softenings for each force computation within a single timestep. We have experimented with this in the tests using the higher-order integration schemes presented in Appendix~\ref{sec:extra.tests:energy}, considering (1) complete re-computation (requiring additional loops); (2) drifting the softening according to the time-derivative as outlined in \S~\ref{sec:time.variant}, but only within a single timestep (i.e.\ re-computing at the beginning of the timestep then drifting within the timestep); and (3) simply using a static softening for each particle across its single timestep (re-computed at the beginning of each timestep, allowing it to vary at timestep-to-timestep as desired). We find in practice the differences are negligible compared to other variations discussed therein.}
 The ``local neighbor'' type criteria (e.g.\ the $\tilde{\rho}$ and $n_{\rm NGB}$-based rules) may require more loops, since the equation defining $\varepsilon$ is usually solved implicitly via an iterative bisection method \citep{springel:entropy}, but these iterations can be made efficient (see the public {\small GIZMO} code for our specific implementation, or {\small GADGET}-4 for another). If desired, we actually find in most of our experiments that one can use a suitably drifted value of quantities like $\varepsilon$, ${\bf T}$, $\zeta$ from the previous timestep, allowing one to collapse all operations into a single loop,\footnote{This is discussed further in Appendices~\ref{sec:extra.tests} \&\ \ref{sec:time.variant} where we describe both technical details and show the effects on energy conservation from such an approximation.} without much loss of accuracy. On a per-loop basis, if $\varepsilon$ were identical, some of the softening rules for $\varepsilon$ in \S~\ref{sec:ex}, such as the tidal tensor ${\bf T}$ based method, appear to involve more computations, but that difference is largely negligible:  the cost of a few additional floating-point operations inside the loop is minuscule compared to the tree-walk and communication/imbalance costs, and many codes (e.g.\ those above, or following \citealt{dehnen:2011.nbody.review,grudic:2020.tidal.timestep.criterion} for timestepping, or those with sink formation/accretion prescriptions as reviewed in e.g.\ \citealt{federrath:2010.sink.particles,grudic:starforge.methods,hopkins:fire3.methods}) already compute quantities like ${\bf T}$. While it is true that -- for fixed $\varepsilon$ -- some neighbors can be ``skipped'' for computing terms in $\boldsymbol{\Upsilon}$ in the $\tilde{\rho}$ or $n_{\rm NGB}$ methods as compared to the gravity-based methods, this has small effects on performance because the same number of neighbors must still be used for the direct (softened) versus unsoftened (or tree/multipole) calculations, so it again simply adds or removes a few floating-point operations at the end of each walk. Comparing different softening kernels, the cost scales primarily with the radius of compact support as discussed in \S~\ref{sec:kernels} -- adding ``more neighbors'' (larger radius of compact support) increases the cost as expected, but there is little difference between the cost of different kernel function floating-point evaluations for fixed neighbor number. Properly-optimized, the CPU cost difference between different symmetry rules (\S~\ref{sec:symm}) is negligible as expected. And the cost difference ``all else equal'' of enforcing a minimum/maximum softening is also negligible (\S~\ref{sec:minmax}), although this can be useful in some specific types of problems to prevent very small timesteps from small $\varepsilon$ or to prevent large load imbalances from large $\varepsilon$.

The biggest difference in cost between the methods here for a given softening is largely between the ``local neighbor'' type criteria  and the ``gravity'' type criteria (e.g.\ the $\hat{\Phi}$, $\hat{\bf a}$, and $\hat{\bf T}$-based rules) for $\varepsilon$, where the ``local neighbor'' criteria tend to be much more expensive (by factors of two to an order of magnitude, in e.g.\ a dark-matter only cosmological simulation), owing primarily to the Courant-like timestep condition it imposes (\S~\ref{sec:timesteps}) -- when two particles or groups/streams/sheets of dynamically cold particles are near each other in the ``test particle'' limit, the ``local neighbor'' methods require much smaller timesteps compared to the gravity-based methods. Over the course of a simulation with highly inhomogeneous density fields, this is a much larger effect compared to the others discussed above.

Of course, non-linear differences either in the problem evolution, or distribution of $\varepsilon$ values and how this interacts with load balancing and timestepping can in many problems dominate. And in multi-physics simulations such as galaxy/star/planet/black hole formation, the relative cost differences between any force-softening methods for collisionless species are usually small compared to the cost of the collisional parts of the integration.

\begin{figure*}
	\includegraphics[width=0.33\textwidth]{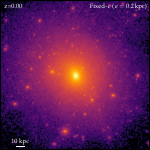}
	\includegraphics[width=0.33\textwidth]{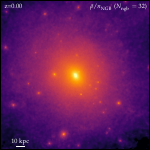}
	\includegraphics[width=0.33\textwidth]{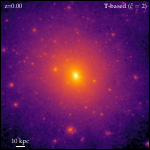} \\
	\includegraphics[width=0.33\textwidth]{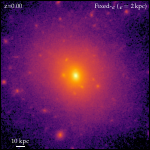}
	\includegraphics[width=0.33\textwidth]{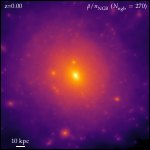} 
	\includegraphics[width=0.33\textwidth]{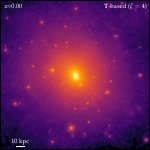} \\
	\vspace{-0.1cm}
	\caption{Application of different force-softening rules (from \S~\ref{sec:ex}, labeled) to a cosmological dark-matter-only simulation evolved to present-day (redshift $z=0$), zooming in on the $\sim 100$ kpc around a $\sim 10^{11}\,{\rm M_{\odot}}$ dark matter halo realized with a few million equal-mass $N$-body particles. We plot the projected surface mass density map, after ``smearing'' each $N$-body particle into the corresponding extended mass density distribution $K$ for the cubic spline kernel adopted (with its value of softening $\varepsilon$). 
	{\em Left:} The fixed-$\varepsilon$ (\S~\ref{sec:ex.fixed}) simulation with small $\varepsilon=0.2\,$kpc shows the most substructure but also visible noise in the ``density'' field and many of the smallest visually-apparent substructures consist of just $\sim 1-10$ particles. Making $\varepsilon=2\,$kpc much larger wipes out most of the substructure, with only a modest decrease in the visible noise at large radii. 
	{\em Middle:} The $\tilde{\rho}$ or $n_{\rm NGB}$ (\S~\ref{sec:ex.rho}-\ref{sec:ex.n}) simulations exhibit a much less noisy distribution but the scheme appears to suppress a considerable amount of substructure, even for $N_{\rm ngb}=32$ (the smallest neighbor number for this softening kernel where the kernel core radius would include more than a couple particles). This scheme also requires shorter timesteps which greatly increase the CPU cost (\S~\ref{sec:cpu}). 
	{\em Right:} The tidal or ${\bf T}$-based (\S~\ref{sec:ex.T}) simulations are similarly smooth but appear to retain all but the smallest (marginally-resolved) substructures even with the larger $\xi=4$ and essentially all with $\xi=2$. Note $\xi=2$ ($=4$) would produce and $\varepsilon$ equivalent to $N_{\rm ngb}=32$ ($=270$) in an isotropic, homogeneous medium, so this is not simply an effect of $\varepsilon$ being larger or smaller on average.
	\label{fig:cosmo.image}}
\end{figure*}

\begin{figure}
	\includegraphics[width=0.95\columnwidth]{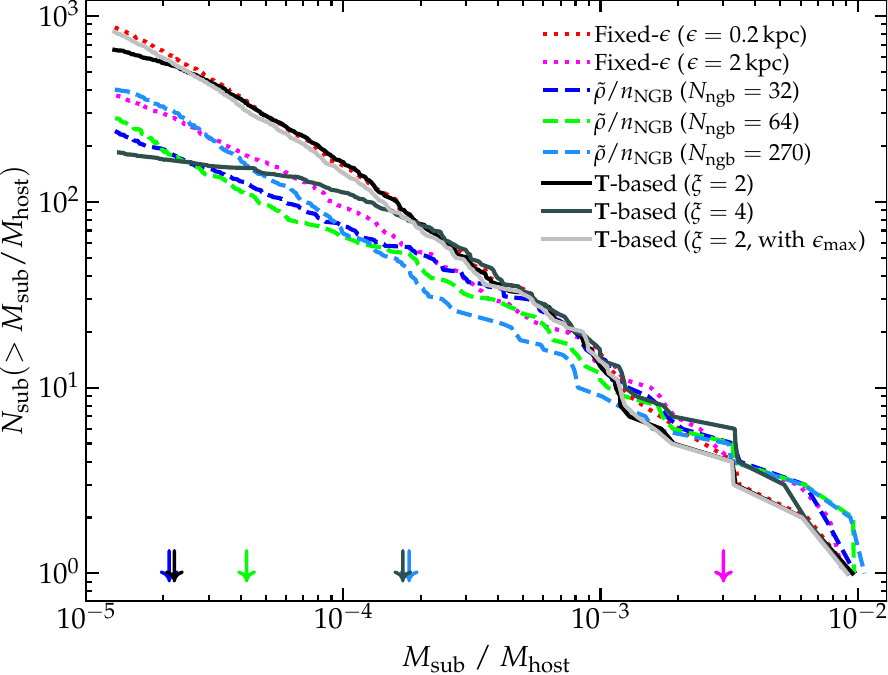}
	\includegraphics[width=0.95\columnwidth]{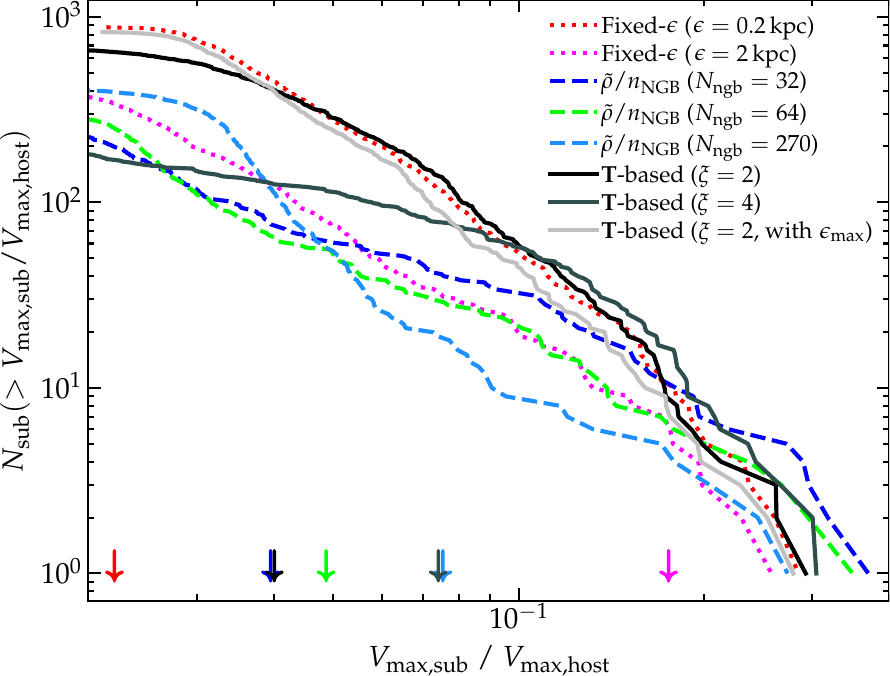}
	\vspace{-0.1cm}
	\caption{{\em Top:} Subhalo mass functions (MFs; cumulative number above some mass relative to the host, at $z=0$, at radii $<2\,R_{\rm vir}$) for different softening models from Fig.~\ref{fig:cosmo.image}. We also consider a tidal model with a maximum $\varepsilon_{\rm max}=0.2$\,kpc ($n=-1$ in Eq.~\ref{eqn:minmax}) instead of a minimum (as assumed otherwise). 
	{\em Bottom:} Same for $V_{\rm max}$. 
	In both, we label (arrows) the approximate virial mass or $V_{\rm max}$ where an NFW halo {\em in isolation} (with concentration $\sim 10$) would have a softening kernel with a radius of compact support $\gtrsim 1.6\,R_{s}$ (the NFW scale-radius) and thus its concentration could not be resolved. 
	The tidal models can suppress the subhalo MF at sufficiently small $M_{\rm sub}$, but only below the ``threshold'' mass where $\varepsilon \gtrsim R_{s}$, as expected. 
	The neighbor-based models, on the other hand, suppress the subhalo MF even at masses much larger than this threshold.
	\label{fig:subhalo.mf}}
\end{figure}

\begin{figure}
	\includegraphics[width=0.95\columnwidth]{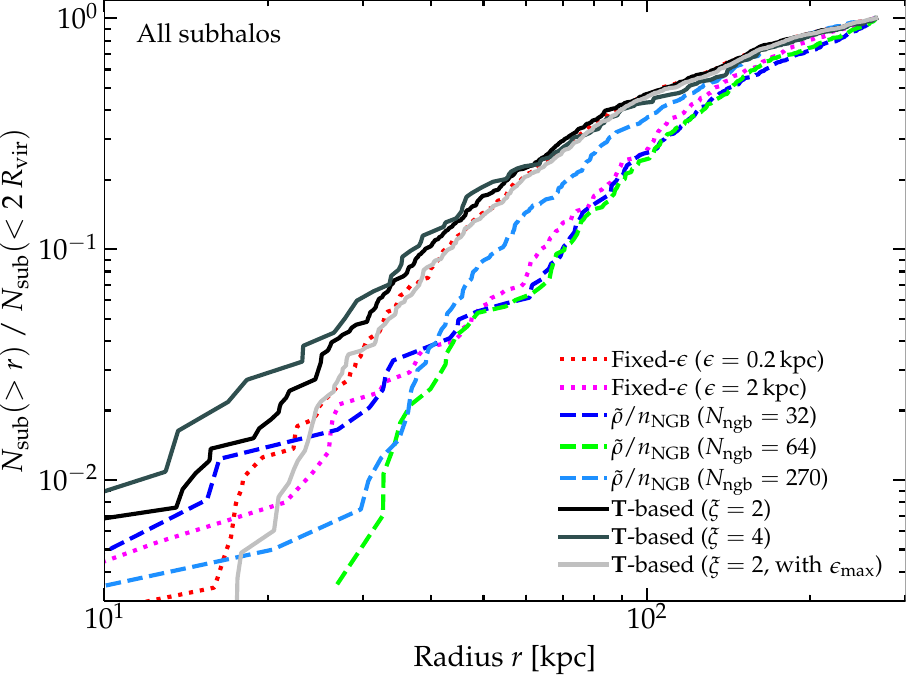}
	\includegraphics[width=0.95\columnwidth]{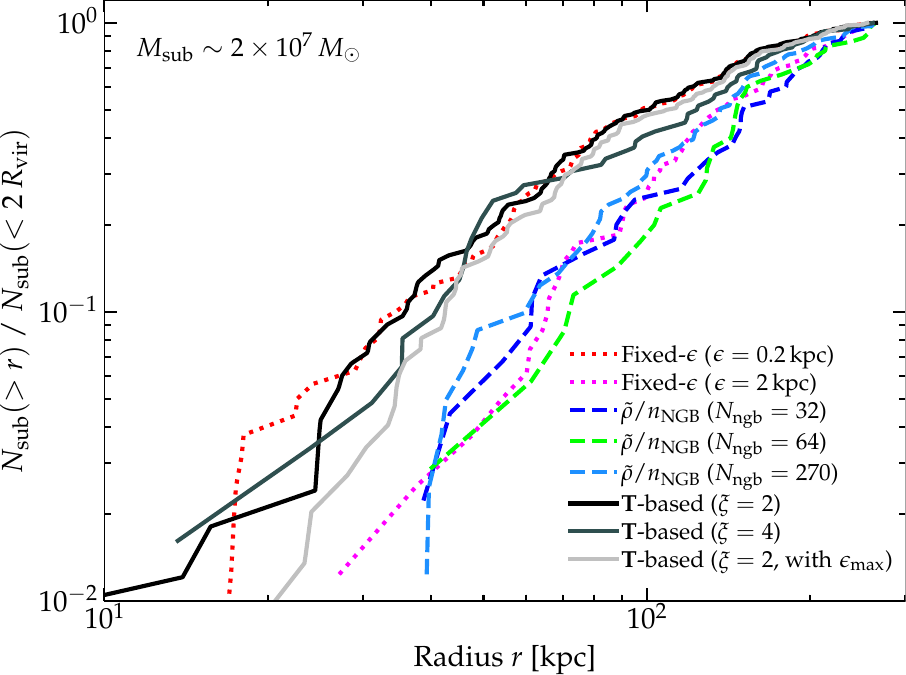}
	\vspace{-0.1cm}
	\caption{Radial distribution of the subhalos in Fig.~\ref{fig:subhalo.mf}. We plot the cumulative number of subhalos versus halo-centric radius, normalized to the total number at $<2\,R_{\rm vir}$, for each softening model. 
	{\em Top:} Including all subhalos in Fig.~\ref{fig:subhalo.mf}. 
	{\em Bottom:} Focusing on a narrow range (factor $<3$) in subhalo mass (where the larger-$\xi$ tidal softening is more complete). The results are qualitatively independent of the mass range we consider: in all cases the tidal-softening models produce similarly concentrated profiles to the small fixed-$\varepsilon$ model. This means the suppression of the subhalo MF for $\xi=4$ is to first order a pure mass/resolution effect, independent of tidal environment. The ``neighbor-based'' models and large fixed-$\varepsilon$ model produce less-concentrated radial distributions: subhalos are preferentially artificially disrupted in stronger tidal environments.
	\label{fig:subhalo.r}}
\end{figure}

\begin{figure}
	\includegraphics[width=0.95\columnwidth]{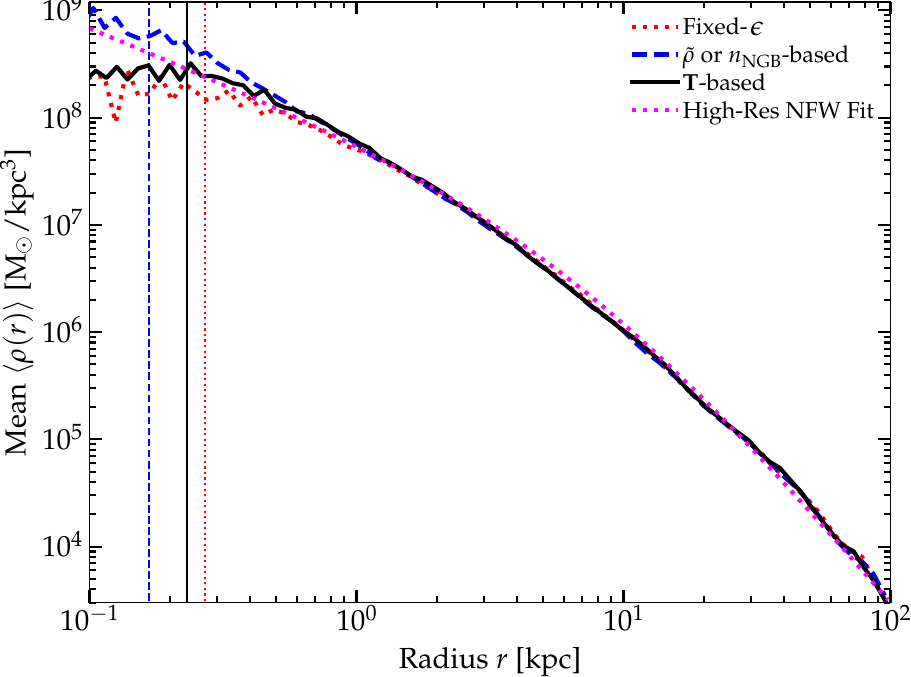}
	\includegraphics[width=0.95\columnwidth]{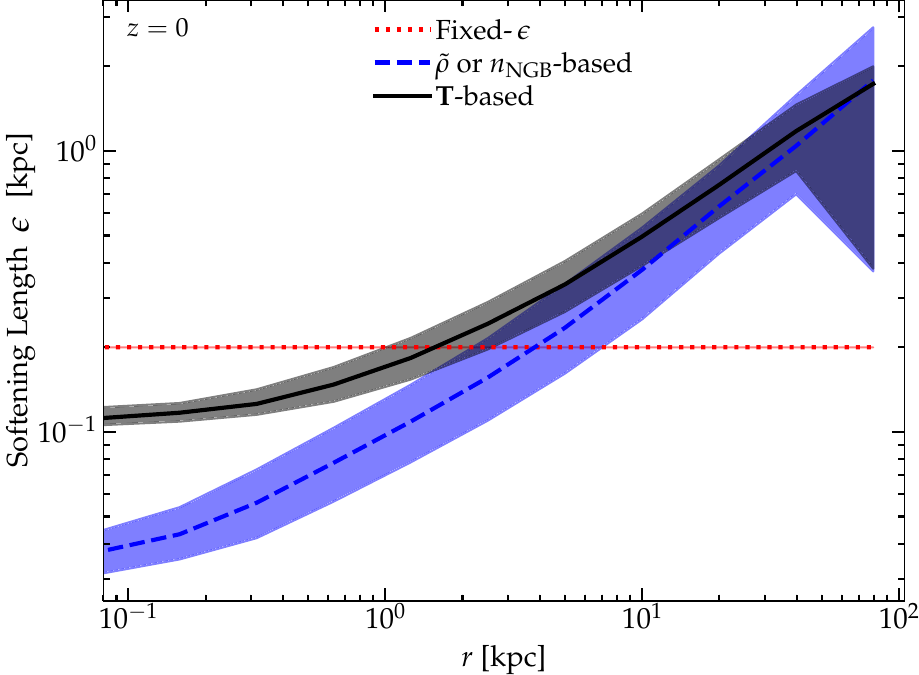}
	\vspace{-0.1cm}
	\caption{Radial profiles of the spherically-averaged dark matter mass density ({\em top}), and softening length $\varepsilon$ ({\em bottom}, as Fig.~\ref{fig:plummer.h}) in a subsample of the cosmological simulations from Fig.~\ref{fig:cosmo.image} (for the full set, see \S~\ref{sec:extra.tests}). The $\varepsilon$ profiles show a similar qualitative behavior to the idealized Plummer test, as expected (note the scatter at large $r$ arises from dense subhalos in which $\varepsilon$ is smaller). In the density profiles, we compare the best-fit NFW profile from a much higher-resolution fixed-$\varepsilon$ simulation (from \citet{hopkins:fire2.methods}), as well as the radii interior to which the profile encloses $<200$ particles (vertical lines) -- an approximate resolution criterion from \citet{hopkins:fire2.methods}. We see excellent convergence in the main-halo mass profile at larger radii. The ${\bf T}$-based model may (tentatively) show {\em slightly} better convergence by a factor of $\sim 2$ in spatial scale. The neighbor-based ($\tilde{\rho}$ and $n_{\rm NGB}$) methods produce a slight excess at smaller radii, for reasons discussed in \citet{hopkins:fire2.methods}. The mass profile and other properties of the ``primary'' (extremely well-resolved) halo are not especially sensitive to details of the softening scheme (nor to the value of $\varepsilon$, $N_{\rm ngb}$, $\xi$).
	\label{fig:cosmo.profiles}}
\end{figure}

\section{Some Advantages and Disadvantages of Different Methods}
\label{sec:pros}

Our primary goal here is to present the derived equations-of-motion for a flexible family of energy and momentum-conserving schemes for adaptive force softening for collisionless fluids. Given a choice of schemes, of course the ``optimal'' method will always be problem-dependent. Nevertheless, it is useful to comment on some advantages and disadvantages of different choices for the many options discussed here, in common astrophysical simulation contexts. 

\subsection{Choice of Softening Rule}

Comparing the choices of the scheme used to determine $\varepsilon$ in \S~\ref{sec:ex}, we generally find that the ``local neighbor'' based methods ($\tilde{\rho}$ and $n_{\rm NGB}$ in \S~\ref{sec:ex.rho}-\ref{sec:ex.n}) have some significant disadvantages in a variety of problems. First, as discussed in \S~\ref{sec:timesteps} \&\ \ref{sec:cpu}, these methods impose a serious CPU cost in the form of their much stricter timestep condition. They also can introduce undesired behaviors, because they ``respond'' and their and their represented potential changes anytime local neighbors move with respect to one another, even in the collisionless N-body test particle limit (where the only robust prediction of the simulation is the ensemble/smooth components of the potential). For the sake of illustration, consider two collisionless $N$-body particles $a$ and $b$ which ``pass through'' one another (close to the kernel center) in a smooth, static, dominant background field. This is not uncommon in non-linear simulations like the cosmological examples below. As detailed in \S~\ref{sec:timesteps}, because $\varepsilon$ depends only on the local neighbor/mesh-generating-point configuration, this not only imposes short timesteps, but the values of $\varepsilon$ and therefore properties of the potential deform as if the individual particles were ``compressing'' or ``expanding.'' This deformation is desired when the particles represent a collisional hydrodynamic fluid, like in SPH for which the ``neighbor-based'' methods were first derived in \citet{price:2007.lagrangian.adaptive.softening}, because it represents fluid flow actually compressing/deforming neighbor cells. But physically, in this situation for a {\em collisionless} system, there should be no such deformation.\footnote{Another way of saying this is that the method will ``over-reconstruct'' the density distribution generating the potential in these neighbor-based methods, by representing it implicitly at the particle-scale.} Finally, while it often works reasonably well in sufficiently smooth, homogeneous systems, there is no actual guarantee that softening with something like the ``inter-particle separation'' actually ensures that the two-body deflection or acceleration from an individual close $N$-body encounter is actually small compared to the characteristic background gravitational velocity or acceleration, because there is (by construction) no information about those background (long-range) velocities or accelerations in the softening criterion.\footnote{One can of course reduce two-body deflections by choosing a larger neighbor number for neighbor-based softenings, and if the problem is roughly homogeneous then the usual inter-neighbor separation criterion works well as we discuss above, but the point is that in a non-linear problem, there is no obvious {\em a priori} choice that can ensure the deflections for this purely-local choice of $\varepsilon$ will be small compared to the long-range/smooth component of the forces, which do not enter $\varepsilon$.}

The ``gravity based'' $\hat{\Phi}$ (\S~\ref{sec:ex.phi}) and $\hat{\bf a}$ (\S~\ref{sec:ex.a}) models avoid most of these disadvantages, but have some ambiguities. The $\hat{\Phi}$ model, in particular, is not gauge-invariant and requires a definition of the zero-point of $\hat{\Phi}$ ($\Phi_{0}$), which is arbitrary and difficult to define at all in e.g.\ a cosmological  or periodic box simulation. Independently, in regions like the center of some massive structure where the potential becomes $\sim$\,constant, there is no clear way to choose the normalization parameter $\xi$ except to require $\xi \gtrsim 1$, but seemingly sensible choices can lead to problematic values of $\varepsilon$ (for example, in the Plummer sphere test, we have to choose $\xi \sim 10^{4}$ to obtain a value of $\varepsilon_{a}$ which is order-of-magnitude comparable to the other softening schemes). It also introduces a quite strong dependence of softening on particle mass, in a manner which does not strictly ensure that the single-particle contribution to gradients of $\Phi$ (e.g.\ the acceleration) are small (e.g.\ for fixed $\xi$, changing the particle number by a factor of $\sim 10^{3}$, which would naively translate to changing the linear spatial resolution by a factor of $\sim10$, produces a factor of $\sim 1000$ change in $\varepsilon$). For the $\hat{\bf a}$ model there are similar challenges. Most notably, in a locally isotropic potential in the usual ``lab/simulation frame'' gauge which is implicitly chosen in most simulations, $\hat{\bf a}$ should vanish, so $\varepsilon \rightarrow \infty$. Because the model is not gauge-invariant this could technically be transformed away by suitable choice of gauge (boosting to a frame with large uniform mean ${\bf a}$), but that would invariably lead to gauge-dependent results (always undesireable), produce very weak variation of $\varepsilon$ with position (defeating the purpose of the adaptive scheme), and potentially cause additional problems with common gauge-invariance-violating terms which depend on $|{\bf a}|$ in e.g.\ time-integration schemes \citep{power:2003.nfw.models.convergence} or tree opening criteria \citep{vogelsberger:2011.arepo.vs.gadget.cosmo}. The $\hat{\bf a}$ criterion also cannot always ensure the jerk and orbital perturbation are small in a two-body interaction. 

On the other hand, the tidal-tensor (${\bf T}$)-based scheme (\S~\ref{sec:ex.T}) appears to avoid all of these disadvantages. Per \S~\ref{sec:timesteps} \&\ \ref{sec:cpu} it is slightly more computationally complex but as cheap or cheaper than other schemes above. Its physical behaviors are generally well-motivated (\S~\ref{sec:ex.T}). It pairs naturally with adaptive timestepping criteria one would already use for gravitational dynamics, such as that in \citet{dehnen:2011.nbody.review,grudic:2020.tidal.timestep.criterion}. It depends only on $\|{\bf T}\|$ which is a true invariant of what is often considered the only ``real'' gravitational object: it is coordinate and frame-independent and gauge-invariant and ensures all of the desired symmetries (e.g.\ translation and Galilean invariance and the equivalence principle) are respected. And (as discussed in \S~\ref{sec:ex.T} \&\ \ref{sec:tidal.norm}) unlike the other softening schemes above, in some background gravitational field with characteristic total mass $M_{\rm enc}$ and length scale $\ell$ (i.e.\ acceleration $\mathcal{O}( G\,M_{\rm enc}/\ell^{2} )$), in the only limit where $N$-body collisionless dynamics should be valid (particle mass $m_{a} \ll M_{\rm enc}$), it is easy to see immediately from Eq.~\ref{eqn:tidal.soft} that the maximum two-body tidal field ($\mathcal{O}(G\,m_{a}/\varepsilon_{a}^{3})$), acceleration ($\mathcal{O}(G\,m_{a}/\varepsilon_{a}^{2})$) and energy change/work done ($\mathcal{O}(G\,m_{a}/\varepsilon_{a})$) will all be small compared to the background tidal field/acceleration/potential, respectively. In other words, the knowledge of the long-range forces helps to ensure that the two-body deflection or acceleration remains small compared to the characteristic background gravitational velocities or accelerations.

\subsection{The Symmetry Rule and Other Choices}

For practical reasons, it is almost always useful to define a maximum and minimum softening as in \S~\ref{sec:minmax}. We see no advantage or disadvantage to choosing $n\gg 1$ instead of simply $n=1$ (which makes the equations simpler) in Eq.~\ref{eqn:minmax}. For the reasons given in \S~\ref{sec:kernels}, there is also no obvious advantage (while there is an obvious computational cost) to using higher-order kernels, as compared to a relatively simple, well-behaved, and widely-used kernel such as the cubic-spline or Wendland $C^{2}$ for $K$ used to define $\phi^{0}$ in Eq.~\ref{eqn:kernel.def}. For the choice of how to ``symmetrize'' $\tilde{\phi}(\phi^{0})$, the examples given in \S~\ref{sec:symm} and \S~\ref{sec:gas} strongly favor a ${\rm MAX}(\varepsilon_{a},\,\varepsilon_{b})$-type choice (Eq.~\ref{eqn:hsymm.max}, \ref{eqn:hsymm.weight}, or \ref{eqn:hsymm.weight.exp}), while the ``$\varepsilon$-averaging'' Eqs.~\ref{eqn:hsymm.avg} can produce systematically incorrect results and ``force-averaging'' Eq.~\ref{eqn:phisymm.avg} can produce spurious divergences and severe $N$-body scattering as described therein. Eq.~\ref{eqn:hsymm.weight.exp} with modest $n\sim 10$ has the advantages of differentiability and reducing exactly to the correct answer independent of $n$ in the $\varepsilon_{a} \ll \varepsilon_{b}$ or $\varepsilon_{b} \ll \varepsilon_{a}$ limits, but in practice the differentiability concern appears minor and we have not seen much difference between Eq.~\ref{eqn:hsymm.max} and Eq.~\ref{eqn:hsymm.weight.exp} in real simulations.

\section{Example Application in a Cosmological Simulation: The Importance for Substructure}
\label{sec:cosmo}

\subsection{Setup and Overview}

Figs.~\ref{fig:cosmo.image}-\ref{fig:cosmo.profiles} consider these choices from \S~\ref{sec:pros} above, comparing the ${\bf T}$-based $\varepsilon$ rule to the neighbor-based or fixed-$\varepsilon$ schemes, in a slightly more realistic application, namely a cosmological simulation of collisionless dark matter forming collapsed dark matter halos. Specifically, we use {\small GIZMO} to run a dark-matter-only cosmological zoom-in simulation of the halo {\bf m11i} from the Feedback In Realistic Environments (FIRE) project \citep{hopkins:2013.fire},\footnote{\FIREurl} using the methodology studied extensively in \citet{hopkins:fire2.methods,hopkins:fire3.methods} but updated with the force softening options proposed herein. The simulation begins from cosmological initial conditions (small fluctuations in the linear regime) at redshift $z>100$ and is evolved to redshift $z=0$. The halo of interest (around which the high-resolution region is constructed) has a virial mass of $\sim 10^{11}\,{\rm M_{\odot}}$ at $z=0$ and a collisionless particle mass here of $\sim 4\times10^{4}\,{\rm M_{\odot}}$ (though since this is a dark-matter only simulation and structure formation is approximately scale-free, the number of particles is much more important than the actual mass scale of either). For completeness, we adopt the cubic-spline $K$ kernel, with $\xi$ or $N_{\rm ngb}$ similar to Fig.~\ref{fig:plummer.h}, and Eq.~\ref{eqn:hsymm.weight.exp} for the symmetry rule. We have experimented with these choices: re-running with different splines with the same $\varepsilon$ gives little difference in the results, consistent with Fig.~\ref{fig:plummer.err}. Here the symmetry rule also produces little effect since as we show below there is little dispersion in $\varepsilon$ at a given halo-centric radius -- we expect the effects to be much larger in multi-physics simulations including e.g.\ sink particles, for the reasons given in \S~\ref{sec:symm}. 

We focus here on the softening scheme: we have tested simulations (a) with fixed-$\varepsilon$ set to $\varepsilon=(0.01,\,0.05,\,0.2,\,2)$\,kpc; (b) using the ``neighbor-based'' schemes ($\tilde{\rho}/n_{\rm NGB}$, both are identical since the particle masses are equal here) with effective neighbor number $N_{\rm ngb}=(32,\,48,\,64,\,270)$ (corresponding to $\varepsilon \approx (1,\,1.1,\,1.2,\,2)\,\langle \Delta x \rangle_{\rm NGB}$; see \S~\ref{sec:ex.n}); (c) the tidal or ${\bf T}$-based schemes with $\xi=(1,\,2,\,4)$ (corresponding to $\varepsilon \approx (0.5,\,1,\,2)\,\langle \Delta x \rangle_{\rm NGB}$ in a homogenous, isotropic medium; see \S~\ref{sec:tidal.norm}), with a minimum softening $\varepsilon_{\rm min}=(0.01,\,0.03,\,0.03)\,{\rm kpc}$ (enforced with Eq.~\ref{eqn:minmax} with $n=1$, but varying this has little effect) and one run with $\xi=2$ and instead a maximum $\varepsilon_{\rm max}=0.2\,$kpc (Eq.~\ref{eqn:minmax} with $n=-1$).

\subsection{Subhalo Mass Functions \&\ Radial Distributions}

Fig.~\ref{fig:cosmo.image} shows images of the projected surface-mass-density distribution at $z=0$. This demonstrates that the force softening algorithm clearly makes a difference for small substructures in the halo. Specifically, the constant-softening model with a small $\varepsilon$ ($=0.2$\,kpc), while visibly noisy, retains the most substructure (down to subhalos resolved via a few particles). Increasing the softening to a much larger value, $=2$\,kpc wipes out the majority of the substructure, but has surprisingly little effect on the visible noise level at large radii. 
The neighbor-based ($\tilde{\rho}$/$n_{\rm NGB}$) models exhibit reduced noise but wipe out almost all the substructure as well, even when we make the neighbor number essentially ``as small as possible'' while retaining a smoothing length comparable to the inter-neighbor separation ($N_{\rm ngb}=32$), with surprisingly little difference even at much larger $N_{\rm ngb} = 270$. Both of these also impose the timestep penalty discussed in \S~\ref{sec:pros} and their total CPU cost is several times larger than the ${\bf T}$-based models; following particles in practice we see that it is usually situations directly analogous to the idealized example in Fig.~\ref{fig:h.crossing} which produce rapid local fluctuations in $\varepsilon$ on collisionless stream crossing/intersection which drive this. 
For the ${\bf T}$-based models, we see relatively little noise at large radii, but significantly more substructure: with $\xi = 2$ ($\varepsilon \approx 1\,\langle \Delta x \rangle_{\rm NGB}$ in a homogeneous, isotropic medium) we see essentially all the substructure of the small $\varepsilon=0.2\,$kpc run (far more than the $N_{\rm ngb}=32$ nearest-neighbor ``equivalent'' run), and even for $\xi=4$ ($\varepsilon \approx 2\,\langle \Delta x \rangle_{\rm NGB}$ in a homogeneous, isotropic medium) we still see more substructure than any $\tilde{\rho}$/$n_{\rm NGB}$-based or the large fixed-$\varepsilon=2\,$kpc run (with only the smallest substructures missing).

Note that the runs with fixed-$\varepsilon=0.01$\,kpc or tidal with $\xi=1$ or $\xi=2$ and $\varepsilon_{\rm max}=0.2\,$kpc are visually identical to the $\varepsilon=0.2\,$kpc and $\xi=2$ runs, respectively.

Fig.~\ref{fig:subhalo.mf} makes this more quantitative, showing the actual subhalo mass and $V_{\rm max}$ distributions,\footnote{Subhalo properties are computed using the {\small ROCKSTAR} halo finder \citep{behroozi:2013.rockstar.halo.finder}. Systematically varying the parameters used for (sub)halo identification produces systematic effects on the resulting mass functions and other statistics, as expected, but does not have any effect on our {\em relative} comparison of different softening methods.} with some of the additional runs from the survey above. 
The ``smaller-$\varepsilon$'' group of both fixed-$\varepsilon$ and ${\bf T}$-based models, specifically the fixed-$\varepsilon=0.01$\,kpc, $\varepsilon=0.2$\,kpc, ${\bf T}$-based with a maximum $\varepsilon_{\rm max}$, and ${\bf T}$-based with $\xi=1$ or $\xi=2$ models are all very similar to one another. For all of these, the subhalo MF exhibits essentially identical normalization within the shot noise and a slope close to the approximate expected value $d n /dM_{\rm sub} \propto M_{\rm sub}^{-2}$ down to $M_{\rm sub} \sim 10^{-5}\,M_{\rm host}$, corresponding to substructures with $\lesssim 20$ particles. 

Fig.~\ref{fig:subhalo.r} plots the radial distribution of subhalos, both across the entire mass range and in a narrow mass interval (our comparison is not sensitive to which mass range we choose). Again, the ``smaller-$\varepsilon$'' group of both fixed-$\varepsilon$ and ${\bf T}$-based models are quite similar. And in \S~\ref{sec:extra.tests} we consider a number of other (sub)halo properties including axis ratios, scale radii, and velocity dispersions; but all of these are consistent with the same trends we see in Figs.~\ref{fig:subhalo.mf}-\ref{fig:subhalo.r}. 

For reference, Fig.~\ref{fig:cosmo.profiles} plots radial profiles of the (spherically-averaged) mass density of the primary halo and range of softening lengths $\varepsilon$ for a subset of the runs (the full set is given in \S~\ref{sec:extra.tests}). As expected the primary halo is close to NFW, so has a stronger tidal field at smaller $r$ which we will discuss below. The qualitative behavior of the median $\varepsilon$ and its dispersion with radius (which is generally quite small except within dense substructures) is very similar to the idealized Plummer test in Fig.~\ref{fig:plummer.h} and our simple analytic expectations.

\subsubsection{Fixed-$\varepsilon$ Models}

In the fixed-$\varepsilon$ models, we see the results are essentially independent of $\varepsilon$ for sufficiently small $\varepsilon \lesssim 0.2\,$kpc, but substantial suppression of the subhalo MF up to $\sim 10^{-3}\,M_{\rm host}$ for the large $\varepsilon=2$\,kpc model. We see a more dramatic effect in the $V_{\rm max}$ distribution where even the most massive subhalos (that do survive) appear to have slightly suppressed $V_{\rm max}$ for $\varepsilon=2\,$kpc. This is consistent with the additional diagnostics in \S~\ref{sec:extra.tests} that show that at essentially all masses probed here, the $\varepsilon=2\,$kpc model (sub)halos are directly influenced by softening (exhibiting e.g.\ inflated scale radii and suppressed binding energy and velocity dispersions). In the radial distribution, we see a clear bias where subhalos at small $r$ (in a stronger tidal field) are preferentially suppressed at large $\varepsilon$.

This is consistent with a number of recent studies that have argued that {\em over}-softening can be a serious problem and significantly suppress substructure in collisionless $N$-body simulations. These studies -- which variously considered idealized (non-cosmological, analytic host-halo) satellite simulations \citep{vandenbosch:2018.disruption.over.soft,vandenbosch:2018.oversoft.disruption.numerical.discreteness.noise} or more exotic idealized configurations \citep{shen:2022.axion.disruption}, semi-analytic models \citep{jiang:2021.satgen,2021MNRAS.503.4075G,2022MNRAS.tmp.2560B}, zoom-in simulations like those here \citep{nadler:2022.symphony}, or large-volume cosmological simulations \citep{2019MNRAS.488.3663L,mansfield.avestruz:2021.halo.bias.over.softening,2021MNRAS.501.5051J,2021arXiv211002097M}, generally with only collisionless dark matter and fixed-$\varepsilon$ models (varying $\varepsilon$) -- argued that unlike most of the properties of the {\em primary} halo, satellite or subhalo disruption is quite sensitive to force softening. In particular, these studies concluded that some common force softening prescriptions tend to {\em over-soften} in tidal encounters; even a subhalo using a softening which was ``optimal'' in an isolated environment would be over-softened in a strong tidal field, resulting in easier tidal disruption in peri-centric passages, and causing a runaway effect where this leads to even easier subsequent unbinding. 

Based on these arguments we can crudely estimate where we might expect over-softening to become problematic, by calculating the halo mass (or $V_{\rm max}$) at which a halo {\em in isolation} (assuming an NFW profile and a concentration $c=10$) would have its scale radius $R_{s}$ appreciably softened. For specificity, we define this to be when the acceleration at $R_{s}$ would deviate by more than $\sim 20\%$ from its Newtonian value (though our comparisons are not especially sensitive to this choice), which corresponds (for our adopted kernel) to roughly $\varepsilon \gtrsim 0.8\,R_{s}$ or the radius of compact support of the kernel $\gtrsim 1.6\,R_{s}$. For the fixed-$\varepsilon$ models, this is reasonably consistent with where we see deviations in the subhalo MF, though we do see some propagation of errors to even larger halos in the $\varepsilon=2\,$kpc model, consistent with the more detailed studies in the references above.

\subsubsection{Neighbor-Based Models}

In the neighbor-based models, we see more problematic behavior. Even for the smallest $N_{\rm ngb}$, where the median softening $\varepsilon$ is $\ll 2\,$kpc at most radii, we see suppression of the subhalo MF and $V_{\rm max}$ distribution up to much larger masses -- much larger, in particular, than the halo mass where we would naively expect isolated halos to be ``over-softened'' (where $\varepsilon \gtrsim 0.8\,R_{s}$, calculated now appropriate for this model for $\varepsilon$ given $N_{\rm ngb}$ and our mass resolution). There is still a clear trend that larger $\varepsilon$ (larger $N_{\rm ngb}$ here) causes deviations at larger (sub)halo masses, but the suppression at lower masses even for the smallest $N_{\rm ngb}$ can be significantly larger than even the largest fixed-$\varepsilon=2\,$kpc model, and is surprisingly weakly sensitive to $N_{\rm ngb}$ (and even non-monotonic). Moreover, at essentially all masses -- even masses where the subhalo MF appears ``complete'' in Fig.~\ref{fig:subhalo.mf} for $N_{\rm ngb}=32$, we see a radial bias in the subhalo mass distribution akin to the $\varepsilon=2\,$kpc model (subhalos preferentially suppressed at small-$r$). 

While the studies focused on ``over-softening'' referred to above focused on fixed-$\varepsilon$ models, it is worth noting that \citet{vandenbosch:2018.oversoft.disruption.numerical.discreteness.noise} specifically suggested that traditional adaptive softening schemes -- essentially all based on the ``local neighbor'' approximation ($\tilde{\rho}$ or $n_{\rm NGB}$) -- would make the ``over-softening problem'' {\em worse}, not better. This is because of how $\varepsilon$ expands in the test particle limit as a subhalo unbinds even if the ``core'' of the halo should retain a constant density as it is tidally stripped or undergoes tidally-induced mass loss (it should indeed lose mass, but retain the dense, tidally resistant core). In short, $\varepsilon$ tends to respond in the opposite manner to that ``desired'' during tidal deformation.

\subsubsection{Tidal Models}

In the tidal models, the $\xi=1$ and $\xi=2$ with $\varepsilon_{\rm max}=0.2\,$kpc models resemble closely the ``small fixed-$\varepsilon$'' models in essentially every respect we examine. There is a very slight suppression of the very smallest substructures (with $\lesssim 30-40$ particles) in the $\xi=2$ model (without a maximum $\varepsilon$) and more notable suppression for $\xi=4$ extending up to $M_{\rm sub} \sim 10^{-4}\,M_{\rm host}$ or $V_{\rm max,\,sub} \sim 0.1\,V_{\rm max,\,host}$) ($\lesssim 200$ particles). This also defines the mass scale where deviations appear in the additional (sub)halo properties in \S~\ref{sec:extra.tests}.

However, three things are noteworthy about this suppression even in the large-$\xi$ simulation:
\begin{enumerate}

\item{At masses above where the (sub)halos become spatially unresolved owing to the force softening, the agreement is nearly perfect. In other words, there appears to be no visible ``upwards propagation'' of softening errors to larger scales or instability/runaway effect.} 

\item{There is no radial bias in the subhalo MF or $V_{\rm max}$ distribution at a given mass -- i.e.\ even if small halos are uniformly suppressed by large softening (as they should be because their scale radii cannot be resolved), this is not any {\em worse} in dense tidal environments.}

\item{The deviations from the smaller-softening runs appear almost exactly where we would naively predict -- i.e.\ where the softening becomes larger than the halo scale radii. Thus the convergence properties appear remarkably well-defined (even moreso than for constant-softening models).}

\end{enumerate}

We stress that these desirable behaviors do not simply stem from ``smaller softenings on average.'' For example, at radii $\gtrsim 100\,$kpc where half the subhalos in Fig.~\ref{fig:subhalo.r} reside, we see (e.g.\ Fig.~\ref{fig:cosmo.profiles}) that the median softening in the ${\bf T}$-based models with $\xi=2-4$ is comparable to $\sim2\,$kpc, and yet we see nothing like the suppression in the fixed-$\varepsilon=2\,$kpc model. Moreover, the models with $\xi=2-4$ feature comparable or larger median $\varepsilon$ at essentially {\em all} radii compared to the ``neighbor-based'' models with $N_{\rm ngb} \lesssim 64$, and yet again show none of the systematic biases those models exhibit. But this is because we are comparing median softenings across all particles at a given radius, which as we see in the mass profile in Fig.~\ref{fig:cosmo.profiles} is dominated by mass {\em not} contained in dense subhalos. Since the ${\bf T}$ and $\tilde{\rho}$ or $n_{\rm NGB}$ models here are calibrated to give broadly similar softening in an isotropic, locally-constant-density medium (see \S~\ref{sec:tidal.norm}), they have similar values (as they should), but since ${\bf T}$ has an additional contribution from the external tidal field while $\tilde{\rho}$ or $n_{\rm NGB}$ can see {\em only} the local neighbor density, $\varepsilon$ is able to drop more rapidly within a dense subhalo in the tidal models than in the local neighbor models, which we confirm by direct examination.

In the context of the ``over-softening'' errors, \citet{vandenbosch:2018.oversoft.disruption.numerical.discreteness.noise} went so far as to state that their simulation results were ``extremely sensitive to the softening length used, whose optimal value appears to depend on the strength of the tidal field.'' This, of course, is exactly what our tidal model seeks to address. It appears that for reasonable choices of $\xi$, these models are indeed able to avoid over-softening while still minimizing noise and $N$-body scattering and retaining a softening length comparable to the inter-neighbor separation in diffuse regions.

\subsection{Convergence of the Primary (Isolated) Halo}

Returning to Fig.~\ref{fig:cosmo.profiles}, we note that the mass density profiles of the primary halo show remarkably good agreement at almost all radii outside $\gtrsim 0.5\,$kpc (approximately radii enclosing $\sim 2000$ particles, suggested as a more conservative convergence criterion by e.g.\ \citealt{power:2003.nfw.models.convergence}). The same is true for the larger set of models shown in \S~\ref{sec:extra.tests}. Clearly, the different substructures seen in Fig.~\ref{fig:cosmo.image} do not appreciably influence the average mass profiles on large scales. It is only on smaller scales, enclosing a few hundred particles, i.e.\ radii just a $\sim5-6$ times the mean inter-particle spacing (or smaller), that we start to see appreciable differences from force softening. Indeed, the only model which shows deviations outside these radii is the fixed-$\varepsilon=2\,$kpc model, which (unsurprisingly) modifies the profile out to $\sim2\,$kpc.

Note that differences in the {\em isolated} halo mass functions at low masses are qualitatively broadly similar to what we discussed at length above for subhalo MFs, but exhibit a weaker dependence on the softening, consistent with previous studies in e.g.\ \citet{springel:2008.aquarius,2008ApJ...688..709T,2013MNRAS.431.1866R,2016MNRAS.458.2870V,2017MNRAS.472..657J}. Hence our focus on substructure above.

This is yet another example (as discussed in \S~\ref{sec:ex.demo}) of the fact that mass resolution or particle number has a much larger effect on formal convergence of mass profiles, velocity distribution functions, and related quantities of the ``primary halo,'' compared to the actual force softening algorithm. Since this has been explored rather extensively in previous studies, including those with adaptive softening, we do not explore it further here but refer to \citet{iannuzzi:2011.collisionless.adaptive.softening.gadget,iannuzzi:2013.no.need.adaptive.softening.for.dm}. On these very small scales, we do see some tentative evidence for slightly better convergence with our ${\bf T}$-based scheme compared to a fixed-$\varepsilon$ scheme, but more study would be needed to know if this is systematic. The neighbor-based schemes produce a small excess density at small radii, related to ``capture'' that can occur in close passages when the effective densities of particle neighbors become high (see Fig.~\ref{fig:h.crossing}), as discussed and shown extensively in \citet{hopkins:fire2.methods}. 

\subsection{Small Fixed-$\varepsilon$ Softening, or Tidal Softening?}

The comparisons above clearly favor use of either fixed-$\varepsilon$ models with ``sufficiently small'' $\varepsilon$, or the tidal softening models with some reasonable $\xi \sim \mathcal{O}(1)$. They do not alone strongly argue for one or the other. For the sake of computational simplicity and speed, it will still be preferable in many circumstances to adopt a simple fixed-$\varepsilon$ model with an ``appropriately chosen'' value of $\varepsilon$ (following e.g.\ \citealt{2019MNRAS.488.3663L}). 

However, it is easy to imagine circumstances where the tidal ${\bf T}$-based models here would have advantages. For example: 
\begin{itemize}

\item{In many simulations, it is not obvious {\em a priori} what a ``good'' value of $\varepsilon$ will be. Prescriptions like those in \citet{2019MNRAS.488.3663L} or \citet{mansfield.avestruz:2021.halo.bias.over.softening} are specific to cosmological simulations of NFW-like cold dark matter halos and their substructure -- and even then, were not actually predicted {\em a priori} but are empirically derived criteria from many high-resolution simulations run with varied $\varepsilon$. In simulating different regimes and parameter spaces, non-cosmological simulations, new physics (e.g.\ alternative dark-matter), or multi-physics simulations which can produce non-linearly different types of substructure it is not obvious if these prescriptions provide useful guidance.}

\item{As the dynamic range of any simulation increases, it becomes more and more problematic to invoke a single fixed $\varepsilon$. Consider for example the simple case of dark matter halos: if we ran a zoom-in of an NFW halo (with $c=10$) using $\sim 10^{10}$ particles (near the limit of what is possible today), and applied the precise criteria above (designed to ensure against ``over-softening'' and suppressing convergence of the halo profile or substructure), we would require $\varepsilon \lesssim 10^{-5}\,R_{\rm vir}$. But this implies in the diffuse IGM that $\varepsilon$ would be approximately $1000$ times smaller than the typical inter-particle spacing in the simulation! In short, the trade-off between noise/artificial scattering and suppression of substructure becomes more severe as one moves to higher resolution.\footnote{This is because the maximum softening $\varepsilon_{\rm max}$ needed to avoid artificially suppressing substructure and/or the central cusp structure in a $\rho \propto r^{-1}$ density profile scales as $\varepsilon_{\rm max} \propto N^{-1/2}$ \citep{2019MNRAS.488.3663L,mansfield.avestruz:2021.halo.bias.over.softening}, while the inter-particle spacing at a given density scales as $\propto N^{-1/3}$.}}

\item{In simulations with different particle masses (whether multi-species or simply particles with a range of masses representing the same collisionless species), the ``light'' particles can be more strongly heated and their mass profiles strongly perturbed by $N$-body heating where it would not appear in the ``heavy'' particles heating themselves \citep{2019MNRAS.488L.123L}. \citet{binney:2002.two.body.relaxation.cosmo.sims} showed that this effect is strongest in fixed-$\varepsilon$ simulations with small $\varepsilon$, and is greatly ameliorated by adopting an adaptive softening where the softening scales with some measure of the local density (which our tidal scheme automatically ensures).}

\item{It is well-known that dynamically cold structures are much more sensitive to $N$-body heating (see discussion and references in e.g.\ \citealt{tothostriker:disk.heating,hopkins:disk.heating}). Recently \cite{ludlow:2021.disk.heating.by.dm.particles,2022arXiv220807623W} pointed out that this could significantly heat cold structures such as thin disks embedded in a dynamically hot collisionless medium (although see also \citealt{hopkins:fire2.methods} for examples where the effect is weaker than might be expected). The small fixed-$\varepsilon$ models represent the ``worst case scenario'' for this: if the background collisionless medium has $\varepsilon$ much smaller than the inter-particle spacing for that background medium, then each particle acts like a bound ``point mass'' structure; in a cosmological simulation with a thin gas/stellar disk for example, setting a fixed, small $\varepsilon$ for dark matter means that every dark matter particle acts as if it were a very dense subhalo with mass equal to the particle mass when it encounters the disk. However it is expected from simple analytic considerations in tidal heating/shock calculations \citep{spitzer:tidal.disruption.impact.approx}, and confirmed in direct numerical experiments \citep[e.g.][]{ludlow:2021.disk.heating.by.dm.particles} that this effect is greatly reduced if the collisionless background is appropriately softened.}

\end{itemize}

Note that the latter two cases above are also cases where the choice of symmetry rule (\S~\ref{sec:symm}) can be potentially important, and in both cases a ${\rm MAX}$-like rule (e.g.\ our Eq.~\ref{eqn:hsymm.weight.exp}) will further minimize errors.

Thus, for situations like these, the tidal scheme proposed here could provide a substantial improvement in accuracy and enable more general applications to a broad range of problems and physics, without imposing either the computational/timestep costs or suppression of substructure imposed by the more traditional ``neighbor-based'' adaptive softening schemes.

\section{Conclusions}
\label{sec:conclusions}

We have discussed and derived generalized, flexible schemes for adaptive softening of collisionless systems in $N$-body methods, conserving energy and momentum. Specifically:

\begin{itemize}

\item We derive a  fully-generalized version of the energy-conserving equations for adaptive gravitational softening for an arbitrary choice of softening ``rule'' (\S~\ref{sec:deriv}) used to determine a variable softening.

\item We then provide several new examples of such softening ``rules'' (\S~\ref{sec:ex}) including softenings based on the potential (\S~\ref{sec:ex.phi}), acceleration (\S~\ref{sec:ex.a}) and tidal tensor (\S~\ref{sec:ex.T}). Some of these can have unique advantages in many applications (\S~\ref{sec:pros}). 

\item We derive a novel, simple mechanism to enforce maximum or minimum softenings, needed in many contexts, while maintaining the conservation properties of the method (\S~\ref{sec:minmax}). 

\item We provide a general timestep constraint condition applicable to any of these which must be enforced to maintain stability, accuracy, and conservation (\S~\ref{sec:timesteps}). 

\item We detail how to correctly deal with interaction terms between different particle species which use {\em different} softening {\em rules}, without sacrificing conservation (\S~\ref{sec:multi}). This is necessary in almost any applications with multi-physics, where e.g.\ some particles might have fixed softening (e.g.\ sinks, individual stars/black holes/planets), some consistent softenings set by finite-volume meshes (e.g.\ gas/fluids), some adaptive set by arbitrary criteria (e.g.\ dark matter or dust or stellar populations). 

\item We outline several different schemes for symmetrizing the potential (in order to ensure momentum conservation), including new schemes which capture advantages of older, non-differentiable schemes largely used for fixed-softening simulations in multi-fluid simulations in particular (\S~\ref{sec:symm}). 

\item We provide relevant scalings for many different softening kernel functions, in the public {\small GIZMO} code (\S~\ref{sec:kernels}), though our derivations are agnostic to this choice.

\item We also discuss computational costs (\S~\ref{sec:cpu}), advantages of different choices (\S~\ref{sec:pros}), and subtleties of cross-application of methods here to cases where a self-consistent softening can be defined (where the full phase-space distribution function is evolved; \S~\ref{sec:gas}) and of the ``self'' potential (\S~\ref{sec:self}), as well as guidance on normalization choices for softening (\S~\ref{sec:tidal.norm}).

\item We compare the tidal-softening scheme here to historically-adopted fixed-softening and neighbor-based schemes in cosmological dark matter simulations (\S~\ref{sec:cosmo}). We show that neighbor-based schemes exacerbate the ``oversoftening'' problems seen in fixed-$\varepsilon$ models when too-large a value of $\varepsilon$ is used, which leads to excessive disruption of substructures and artificially suppressed substructure in dense environments. The tidal-softening schemes however automatically adapt to this and agree well with fixed-$\varepsilon$ schemes with ``well-chosen'' (and sufficiently small) fixed softenings, but without needing to specify the softening in an ad-hoc manner or with foreknowledge of the problem, and while simultaneously reducing noise and $N$-body scattering (keeping the softening close to the inter-neighbor separation in diffuse environments).

\end{itemize}

For the sake of reproducibility and providing the complete details of the numerical implementations of the methods herein, we provide the full implementations of the numerical methods described here, for modular choices of different symmetry rules, softening rules, kernel functions, etc., in the public {\small GIZMO} code.

\acknowledgments{We thank our referee, Walter Dehnen, for many detailed checks and suggestions. Support for PFH was provided by NSF Research Grants 1911233, 20009234, 2108318, NSF CAREER grant 1455342, NASA grants 80NSSC18K0562, HST-AR-15800. Numerical calculations were run on the Caltech compute cluster ``Wheeler,'' allocations AST21010 and AST20016 supported by the NSF and TACC, and NASA HEC SMD-16-7592.}

\datastatement{The data supporting this article are available on reasonable request to the corresponding author. The public code with additional numerical implementation details is available at \gizmourl.} 

\bibliography{ms_extracted}

\begin{appendix}

\section{Application to and Interactions with Collisional Fluid or N-Body Dynamics}
\label{sec:gas}

We have been careful throughout to emphasize that our derivations apply to collisionless (or weakly-collisional) self-gravitating systems in methods (like standard $N$-body approaches) that do not evolve the 6-dimensional phase space distribution function $f_{s}({\bf x},\,{\bf v},\,t)$. Of course, if one did evolve the full $f_{s}$, then one can derive a unique and {\em consistent} energy-conserving adaptive force softening. By far the most common examples of this are for ``collisional'' systems: either collisional or ``direct'' $N$-body methods, or fluid dynamics. 

First consider the simple case of collisional/direct $N$-body methods, where the masses really are treated as point-mass-like relative to the simulation resolution. Then $f_{s} = \sum_{a} \delta_{a}({\bf x}-{\bf x}_{a}\, , \, {\bf v}-{\bf v}_{a} \, ,\, t)$ and we have a system of (un-softened) point masses, as we started from in \S~\ref{sec:deriv}. In this case, the correct equations of motion are just those in \S~\ref{sec:deriv} for a constant/fixed softening $\varepsilon_{a}$ ($\boldsymbol{\Upsilon}\rightarrow 0$; \S~\ref{sec:ex.fixed}) with $\varepsilon_{a} \rightarrow 0$ set to some arbitrarily small value (it can be non-zero to prevent divergences and represent e.g.\ compact object collisions or accretion). While of course more sophisticated methods are used in some contexts to treat phenomena outside the scope of our paper here (e.g.\ hard binaries and stellar/remnant mergers, in some star cluster simulations), methods like those here are still widely used \citep{wang:2015.nbody6.starcluster.dynamics,banerjee:2020.nbody7.stellar.models}, especially in multi-species/multi-component simulations, such as star formation (with point-like sinks/stars/remnants; \citealt{lee:2018.stellar.imf.larson.core,hennebelle:2020.stellar.radiative.feedback,guszejnov:2020.starforge.jets,guszejnov:2022.starforge.cluster.assembly,guszejnov:starforge.environment.multiplicity,grudic:2022.sf.fullstarforge.imf}), simulations of interacting supermassive black holes (SMBHs) in galactic nuclei \citep{rantala:2017.ketju.dynamics.smbh,mannerkoski:2021.smbh.multiple.galaxy.nucleus,neureiter:2023.smbh.interactions.models} or SMBH-loss cone dynamics (especially those modeling tidal disruptions and other events; \citealt{rantala:2022.bifrost.bh.integration.in.galaxies,liao:2022.ketju.with.gas}).

In most modern fluid dynamics methods where one is evolving {\em collisional} fluids like finite-volume/element methods (e.g.\ applications continuum fluids with pressure, hydrodynamics, magnetohydrodynamics, where the mean free-path and/or gyro radii are very small compared to characteristic lengths in the system), one does evolve the full distribution $f_{s}$ in each cell, typically by decomposition into $f_{s} = \rho({\bf x})\,\mathcal{U}_{v}({\bf v}-\langle {\bf v} \rangle \, , \, \langle {\bf v} \rangle)$, where $\rho({\bf x})$ is the {\em continuously-defined} $\rho$ (usually defined by virtue of the cell-centered values $\rho({\bf x}_{a})$ and some gradient-based reconstruction which depends on the details and order of the method), $\langle {\bf v} \rangle$ is the mean velocity of the fluid in the cell, and the velocity distribution function $\mathcal{U}_{v}$ is something like a Maxwellian (with the necessary moments of the distribution evolved via variables like the thermodynamic temperature).\footnote{The implicit assumptions of collisionality (mean free paths small compared to resolution) and molecular chaos effectively enforce these forms and close the Vlasov moments hierarchy. This of course means the fluid obeys some set of equations which, by definition, are distinct from the collisionless Vlasov-Poisson equation applicable to collisionless N-body systems.} Integrating over ${\rm d}^{3}{\bf v}$ at a given position, one simply obtains the usual Euler or Navier-Stokes or MHD equations. In finite-volume/Godunov-type methods, these can then be integrated over volume ${\rm d}^{3}{\bf x}$ appropriate for the method to define the integrated forces on a fluid element or ``cell'' at the desired integration accuracy for the method and cell shape. For some finite-volume examples: for fixed/static mesh codes,\footnote{For non-moving but adaptive-mesh refinement (AMR) methods, this is inconsistent with the actual refinement step. Just like for the fluid dynamics, the self-gravity method necessarily reverts to zeroth-order accuracy in a refinement step.} this gives gravitational force equations again equivalent to the ``fixed softening'' models ($\boldsymbol{\Upsilon}\rightarrow 0$;  \S~\ref{sec:ex.fixed}), with $\varepsilon_{a}$ directly related to the cell size $\Delta x_{a}$; for the quasi-Lagrangian MFM/MFV-type finite-volume methods, the consistent $\boldsymbol{\Upsilon}$ terms are essentially identical to those we derive for the $n_{\rm NGB}$ method (\S~\ref{sec:ex.n}) for the same kernel function ($K=W$) and radius of compact support (and therefore $\varepsilon_{a}$) used in the method to define the effective hydrodynamic mesh faces, as given in \citet[][Appendix~H]{hopkins:gizmo}; for Voronoi tesselation-based moving mesh methods, the $\boldsymbol{\Upsilon}$ terms are similar but given by a slightly different expression reflecting how the mesh depends on the neighboring mesh-generating point configuration, given in \S~5 of \citet{springel:arepo}.\footnote{Specifically, the $m_{a}\,{\bf a}_{a}^{\rm soft}$ terms in Eq.~107 therein are equivalent to our $m_{a}\,m_{b}\,(\boldsymbol{\Upsilon}_{ba}-\boldsymbol{\Upsilon}_{ab})$ term in Eq.~\ref{eqn:eom}.} Of course in 6-dimensional finite-volume Vlasov-integrator type methods, we could attempt to define a consistent softening for collisionless fluids in the same way, but that is not the subject of our study here.

This means that one can easily treat these ``consistent'' limits in any code alongside collisionless fluids where softening is chosen according to some rule like those in the main text, by simply noting the equivalence of these methods to different ``rules'' for $\varepsilon_{a}$ with correspondingly different $\boldsymbol{\Upsilon}$ terms. This immediately makes it clear how to handle interactions between these species and collisionless fluids using any of the ``rules'' from \S~\ref{sec:ex}, following the procedure in \S~\ref{sec:multi}, based on using the correct $\boldsymbol{\Upsilon}_{ab}$ based on whether ``$a$ sees $b$'' in $a$'s rule for $\varepsilon_{a}$ and vice versa. The only (relatively modest) ambiguity arises in the symmetrization (\S~\ref{sec:symm}), where symmetrizing a consistently-defined $f_{s}$ element (e.g.\ gas) with a more arbitrarily chosen function used to define a single collisionless $\varepsilon_{a}$ element will, of course, break the strict consistency of the former. However, one case where we can consider a ``mixed'' multi-species interaction with an exactly-defined $f_{s}$ is suggestive. Consider an interaction between a fluid-dynamic gas cell and a ``true'' point mass (direct-$N$-body) moving through it; it is easy to show the correct symmetrization rule in the mutual interaction is exactly the ${\rm MAX}$ rule (or our proposed Eq.~\ref{eqn:hsymm.weight}-\ref{eqn:hsymm.weight.exp} with large $n$), as argued in \S~\ref{sec:symm}. Finally, while for point-mass systems the self-potential term is not defined, if desired it can be defined consistently for fluid dynamics methods, where its effect {\em internally} causing a gas cell to collapse (resisted by e.g.\ pressure), as discussed in \S~\ref{sec:self}, can be self-consistently treated via sub-grid sink formation/accretion prescriptions.

\section{On the ``Self'' Contribution}
\label{sec:self}

In \S~\ref{sec:deriv}, we noted that the potential is defined only between masses, $\sum_{b} \sum_{c>b} \tilde{\phi}_{bc}$, i.e.\ there is no ``self'' contribution. This is important because, although some previous discussions of softening have argued that it can provide some definition of the self-potential (e.g.\ $m_{b}^{2}\tilde{\phi}_{bb} \sim G\,m_{b}^{2}/\varepsilon_{b} \ne 0$), including such a term (i.e.\ allowing $\tilde{\phi}_{bb} \ne 0$) would introduce both numerical and physical inconsistencies. Numerically, there are many subtle issues, but we will note two dramatic examples. First, in our derivation (\S~\ref{sec:deriv}), if we do not use the symmetry properties of $\tilde{\phi}_{ab}$ to simplify the derivation but retain all terms which should vanish under symmetry, then insert the symmetric form of $\tilde{\phi}_{ab}$ with respect to $\phi^{0}_{ab}$ at some later step of our derivation (as in e.g.\ \citealt{price:2007.lagrangian.adaptive.softening}), then depending on which symmetrized form of $\tilde{\phi}_{ab}$ we adopt and specifically where in the derivation we insert its definition, we can arrive at {\em different} number of countings of the ``self potential'' term (i.e.\ $\tilde{\phi}_{bb}$ appears with different pre-factors in terms like $\zeta$) in our final equations-of-motion, while all other terms are identical (as they should be). Thus, the only way to define a consistent equation of motion is to enforce $\tilde{\phi}_{bb}=0$. Second, consider an isolated particle $a$ in a dominant, smooth, static background potential. If we include ``self'' potential, then the ``self'' contribution can, for sufficiently small $\varepsilon_{a}$, dominate the potential $\Phi_{a}$ (and tidal tensor ${\bf T}_{a}$ and contribution to the kernel sums $\tilde{\rho}_{a}$, $n_{{\rm NGB},\,a}$). But per \S~\ref{sec:ex}, this is what is used to calculate $\varepsilon_{a}$ in the first place. Thus it is easy to show that there is no {\em unique} solution for $\varepsilon_{a}$ under these conditions, which can cause discontinuous ``jumps'' in $\varepsilon_{a}$.

Physically, there are a number of additional arguments one can make. Briefly, consider two common heuristic ``justifications'' of force softening for collisionless populations in $N$-body simulations often seen in the literature (stressing that these are simply intended to provide some intuitive guidance, as they lead to identical discrete equations-of-motion and there is no ``unique'' or rigorously-derived model here). The first is to view softening as a ``smoothing'' or ``averaging'' operation \citep[see e.g.][]{hernquistkatz:treesph,merritt:1996.optimal.softening,dehnen:2001.optimal.softening,barnes:2012.softening.is.smoothing}. In this picture, one ``begins'' from the $N$-body Lagrangian for a set of point masses ($\tilde{\phi}_{bc}=G/|{\bf x}_{b}-{\bf x}_{c}|$, for which the self-potential does not exist and the energy is given only by interaction terms), and then introduces softening to remove or ``smooth'' the divergences. In this case there is no ``self-potential'' to modify. The second is to view $N$-body particles as approximating an extended mass distribution (of many micro-physical collisionless particles) distributed with some characteristic spatial extent $\varepsilon_{a}$ (e.g.\ according to some density function like $K$ in \S~\ref{sec:kernels}; \citealt{dyer.ip:1993.softening.law,romeo.1998:optimal.softening,athanassoula:2000.optimal.force.softening.collisionless.sims,rodionov:2005.optimal.force.softening}). Naively this might appear (and has sometimes been claimed) to produce a consistent self-potential. However consider a particle in a static homogenous background with vanishing net external force: if we assign it a finite gravitational self-potential then in the absence of any balancing ``internal self energy'' or artificially introduced forces, $\varepsilon$ should decrease (``collapse''). This is of course unphysical, and for any reasonable softening law $\varepsilon$ will remain constant, but it illustrates that incorporating the self-potential in the Lagrangian would generate some artificial forces unless we further introduced some ``compensating'' terms. More formally, we can say that for fixed softening, the self-potential would be constant so have no effect on the dynamics, while for adaptive softening, its non-constancy necessarily generates a force when we write the energy-conserving equations (due to the dependence of the self-potential on $\varepsilon$, and hence on the position of the particle and all those on which its $\varepsilon$ depends), which is neither necessary not has a clear physical motivation.

In any case, all of these arguments require $\tilde{\phi}_{bb}$ vanish for consistency, for any standard $N$-body method integrating softened {\em collisionless} particles. One can, of course, define a meaningful self term for a population $f_{s}$ if one actually evolves the entire distribution function (as is done for collisional gases in e.g.\ finite-volume hydrodynamic methods; see \S~\ref{sec:gas}, as well as some Vlasov solvers which integrate the entire 6-dimensional $f_{s}$ for collisionless species in e.g.\ \citealt{filbet:2001.conservative.vlasov,yoshikawa:2013.6d.vlasov.direct,colombi:2014.waterbags,sousbie:2016.vlasov.poisson.tesselation,mocz:2017.integer.lattice.vlasov,tanaka:2017.6d.phase.space.solvers.finite.volume}, but these are not the methods of interest in our study here). 
This is important, because even though in the equations of motion $\nabla_{a} \tilde{\phi}_{aa}$ (defined in \S~\ref{sec:deriv} as the gradient {\em at a fixed $\varepsilon$}) would vanish for any of the definitions of $\phi^{0}$ in \S~\ref{sec:kernels} even if $\tilde{\phi}_{aa} \ne 0$, $\tilde{\phi}_{aa} \ne 0$ would still appear in the ``grad-$\varepsilon$'' or $\boldsymbol{\Upsilon}$ terms, not to mention any energy accounting of the system and terms like $\zeta$, $\Phi$, and ${\bf T}$ wherever these are used for determining $\varepsilon$ or timesteps or star formation or other rules in the simulation.

\section{Choosing a Reasonable Normalization for Tidal Softening}
\label{sec:tidal.norm}

In Eq.~\ref{eqn:tidal.soft} we propose a rule for the softening of the form $\varepsilon_{a} = \xi\,(G\,m_{a}/\| {\bf T}_{a} \|)^{1/3}$, and note that the choice of the normalization constant $\xi$ is somewhat arbitrary and not essential for our discussion. That said it is useful to give some estimates for ``reasonable'' values of $\xi$ (note the appropriate normalization constants $N_{\rm ngb}$ or $M_{0}$ for the $n_{\rm NGB}$ and $\tilde{\rho}$ methods are discussed already in detail in the literature cited in \S~\ref{sec:ex.rho}-\ref{sec:ex.n}). To do so, is helpful to consider two limiting cases for the matter distribution/potential.

First, consider the case of test particles of mass $m$ in some large-scale (i.e.\ smooth) external tidal field, generated collectively by the mass at some large distance $\sim \ell$. For specificity consider a Keplerian $\sim G\,M_{\rm enc}/\ell$ potential (a ``maximally anisotropic'' case, in terms of some characteristic mass $M_{\rm enc}$ and scale/distance from the mass $\ell$), for which we immediately obtain $\|{\bf T}\| = \sqrt{6}\,G\,M_{\rm enc}/\ell^{3}$. We then have the usual tidal/Hill/Roche radius around a mass $m$, $r_{\rm tidal} \approx (m/M_{\rm enc})^{1/3}\,\ell$ (with the exact normalization depending on the definition adopted), and from Eq.~\ref{eqn:tidal.soft}, we have $\varepsilon = \xi\,(G\,m/\| {\bf T} \|)^{1/3}$ so $\varepsilon/r_{\rm tidal} \approx 0.74\,\xi$. Intuitively, $\xi$ should be $\mathcal{O}(1)$ in this situation, but the desired value of $\varepsilon$ depends, of course, on how it is defined in a particular softening kernel function. For the sake of specificity let us take the softening kernel (\S~\ref{sec:kernels}) generated by $K$ equal to the cubic spline ($b_{4}$ kernel) in $D=3$ dimensions, with $\varepsilon = \varepsilon_{b_{4}}$ defined to be the usual ``core'' radius (close to the RMS or FWHM) equal to $=1/2$ the radius of compact support \citep{price:2012.sph.review}. Then requiring that the tidal radius be similar to the radius containing most of the kernel power suggests some $\xi \sim 1-$\,a few as a reasonable choice. Applying the same rule and kernel definitions, we can calculate  the maximum possible or worst-case ``two body'' acceleration experienced by such a particle passing directly through the center of another, $|{\bf a}|_{\rm max}^{a} \approx 1.2\,(m/M_{\rm enc})^{1/3}\,\xi^{-2}\,|{\bf a}_{\rm background}|$ where $|{\bf a}_{\rm background}| = G\,M_{\rm enc}/\ell^{2}$ is the dominant smooth collective acceleration. Since $m \ll M_{\rm enc}$ by definition in this situation, we are ensured $|{\bf a}|_{\rm max}^{a} \ll |{\bf a}_{\rm background}|$ for reasonable choice of $\xi$, but can minimize these errors by choosing a somewhat larger value of $\xi$, but this will come with a tradeoff in e.g.\ resolving substructure as discussed in the main text. The default implementation of this module in {\small GIZMO}, for example, adopts a value $\xi \approx 2$ (ensuring $|{\bf a}|_{\rm max}^{a} \ll |{\bf a}_{\rm background}|$ even as $m \rightarrow M_{\rm enc}$, while keeping $\varepsilon/r_{\rm tidal} \sim 1$). It is more cumbersome to calculate the rms velocity deflection and effective ``$N$-body heating rate'' for a given method \citep{ma:2021.seed.sink.inefficient.fire,ma:2022.discrete.df.estimator}, and this is only weakly influenced by softening (which suppresses only large-angle deflections), but a simple estimate of the ``worst-case scenario'' for a two-body encounter (marginalizing over particle trajectories and velocities) gives a fractional deflection (relative to Keplerian) of $\sim 0.01\,(m/M_{0})^{2/3}$ for $\xi \sim 2$ to $4$ and the same kernel (though a typical encounter will produce an effect orders-of-magnitude weaker, assuming the collisionless particles have relative velocities of order the Keplerian speed, see \citealt{hopkins:fire2.methods} for discussion).  

With this in mind, consider an opposite extreme, of particles (of mass $m$) at the center of a steeply-peaked isotropic mass distribution with central/local mass density $\rho$ (so the net acceleration from the surrounding large-scale mass distribution is weak). In this limit the norm of the tidal tensor approaches $\| {\bf T} \| \rightarrow (4\pi/\sqrt{3})\,G\,\rho$, so we obtain $\varepsilon \rightarrow 0.52\,\xi\,(m/\rho)^{1/3}$. Inserting $\xi \sim 2$ from above, we have $\varepsilon \sim (m/\rho)^{1/3}$, but this is just the mean local inter-particle spacing $\langle \Delta x \rangle$. So in the limit where local forces do dominate, this criterion just reduces (as it should) to the ``optimal'' softening for such a situation discussed in many numerical studies cited in \S~\ref{sec:intro} \&\ \ref{sec:ex}. As usual, smaller values will produce excess noise in the potential, while larger values will degrade the effective resolution (the accuracy is qualitatively similar to that of density estimation, see e.g.\ \citealt{dehnen.aly:2012.sph.kernels}, though note that there is more freedom in the problem here since there is no pairing instability).

\section{Additional Comparisons}
\label{sec:extra.tests}

Here we include some additional tests and comparisons from the cosmological simulations already presented in \S~\ref{sec:pros}, which inform the discussion in the main text. 

\subsection{Energy Conservation Tests}
\label{sec:extra.tests:energy}

\begin{figure}
	\includegraphics[width=0.95\columnwidth]{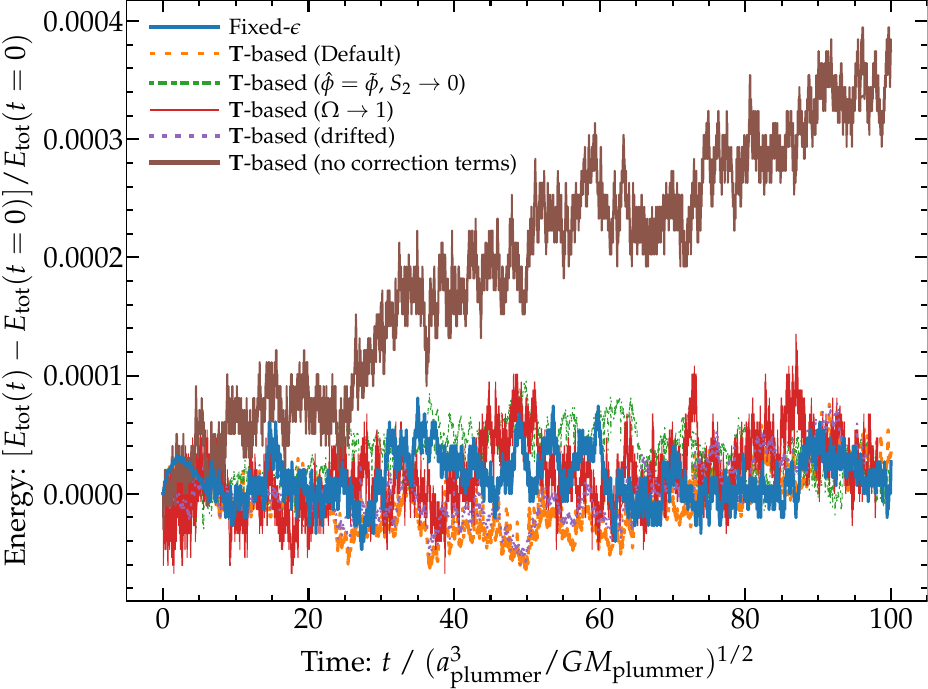}
	\hspace{0.2cm}
	\includegraphics[width=0.97\columnwidth]{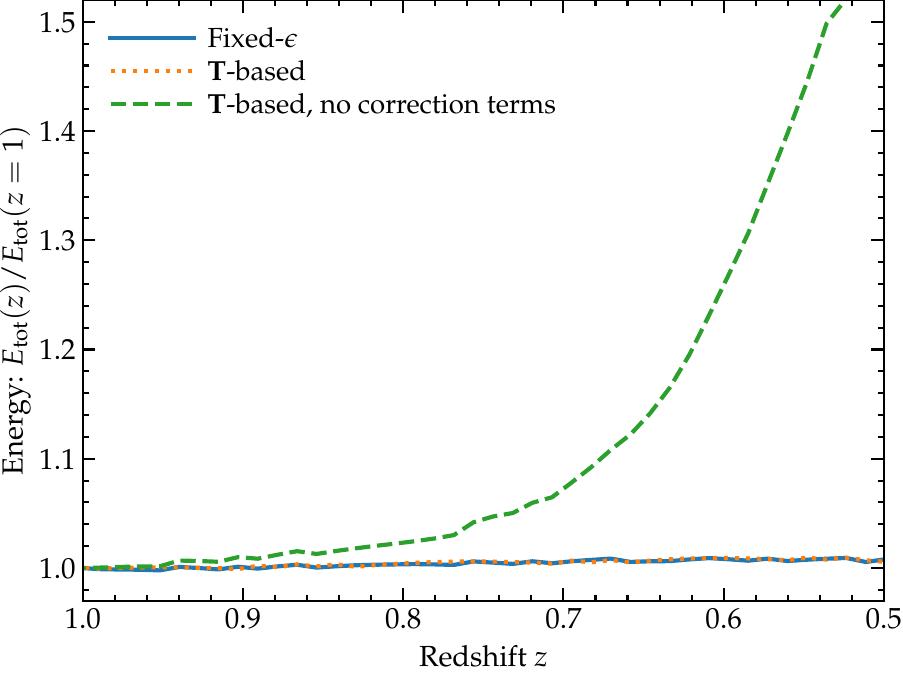}
	\vspace{-0.1cm}
	\caption{Tests of energy conservation in the methods from the text (\S~\ref{sec:extra.tests:energy}).
	{\em Top:} Change in total (potential plus kinetic) energy, relative to initial, in the Plummer sphere test from \S~\ref{sec:ex.demo}, run to $\sim 100$ global dynamical times. We compare several variant implementations of the tidal (${\bf T}$-based) models (\S~\ref{sec:ex.T}), including our default implementation, one using the ``drifted'' scheme from \S~\ref{sec:time.variant} to reduce the number of loops, and variants neglecting the $\mathcal{S}_{2}$ terms ($\mathcal{S}_{2} \rightarrow 0$) or even more simply setting $\Omega\rightarrow 1$ in Eq.~\ref{eqn:omega} even when these should be non-zero, and one entirely neglecting the conservation/correction terms $\boldsymbol{\Upsilon}$. All of the variant tidal implementations including the $\boldsymbol{\Upsilon}$ terms conserve energy at a similar level to the fixed-$\varepsilon$ models. Neglecting the $\boldsymbol{\Upsilon}$ terms produces a notable drift in the total energy.
	{\em Bottom:} As above, but for a ``worst case'' example, testing energy conservation in our cosmological simulations from \S~\ref{sec:cosmo}, restarted at $z=1$ after non-linear structure has formed with a low-order accuracy integration scheme and individual particle timesteps and softenings $\varepsilon$ (with large dynamic range), and featuring interactions between species using different softening rules. Despite this, the tidal scheme produces very similar energy conservation to the fixed-$\varepsilon$ scheme. In this more challenging problem, neglecting the  $\boldsymbol{\Upsilon}$ terms produces catastrophic $\mathcal{O}(1)$ energy-conservation errors.
	\label{fig:egycon}}
\end{figure}

Fig.~\ref{fig:egycon} validates the expected energy conservation properties of the method(s) discussed. First, we consider the simple Plummer sphere test from \S~\ref{sec:ex.demo}. We evolve the systems to many dynamical times and plot the total energy (potential plus kinetic, defined exactly as in \S~\ref{sec:deriv}, summed over all particles), relative to its initial value in each run (because the absolute total energy should be slightly different between models, as different softening leads to slightly different interaction potentials), over time. For both the fixed-$\varepsilon$ model and our default tidal model (see e.g.\ \S~\ref{sec:deriv.details:eliminate}), we see energy is conserved to integration error level as expected. We also compare several variant implementations of the tidal scheme. In one, we set $\hat{\phi}=\tilde{\phi}$ (with the ${\rm MAX}$ symmetrization scheme), which implies $\mathcal{S}_{2} \ne 0$ exactly, but we ignore this term by setting $\mathcal{S}_{2} \rightarrow 0$ anyways (ignoring higher-order corrections to $\Omega$) as described in \S~\ref{sec:deriv.details:ignore}. In another variant, we more radically simply take $\Omega \rightarrow 1$ always (ignoring all the higher-order correction terms within $\Omega$ in Eq.~\ref{eqn:omega}). As discussed below (\S~\ref{sec:deriv.details}), these correction terms ($|\Omega - 1|$) are small, so in either case, we see very little effect on the energy conservation properties of the system. In yet another experiment, we adopt the ``drifted'' scheme from \S~\ref{sec:time.variant} for calculating ${\bf T}$ and related terms; this again exhibits similar conservation properties. But if we omit the energy-conservation correction terms $\boldsymbol{\Upsilon}$ entirely, we immediately see much less accurate energy conservation, as expected.

We have re-run this test varying the force softening (choice of fixed $\varepsilon$, or $\xi$ in the tidal models); switching the timestepping criterion between the tidal-tensor-based \citealt{dehnen:2011.nbody.review,grudic:2020.tidal.timestep.criterion} or the acceleration-based \citet{power:2003.nfw.models.convergence} criteria in addition to the required criteria from \S~\ref{sec:timesteps}; using adaptive hierarchical timestepping (as in \citealt{springel:gadget}) versus uniform timestepping; switching between our higher-order Hermite integrator \citep{grudic:starforge.methods} and the standard second-order leapfrog scheme \citep{springel:gadget} in {\small GIZMO}; varying the standard tree opening and integration or force accuracy criteria \citep[see e.g.][]{hernquist:1990.barnes.nbody.review}; or varying the number of particles sampling the Plummer sphere. These choices systematically change the integration errors (as expected), at a level generally much larger than any difference between our (default) tidal model versus the fixed-softening model. More importantly the {\em relative} conclusion in Fig.~\ref{fig:egycon} remains the same: the various tidal and fixed-$\varepsilon$ model implementations exhibit similar energy conservation properties while the tidal method omitting correction terms entirely exhibits notably worse energy conservation, as expected.

But this is a highly-idealized experiment which has relatively small errors even when the $\boldsymbol{\Upsilon}$ terms are neglected entirely. To more realistically illustrate a ``worst case'' behavior in a real problem, in Fig.~\ref{fig:egycon} we also compare the results for some of our cosmological zoom-in simulations from \S~\ref{sec:cosmo}. Specifically, to have a controlled experiment, we select one of the halos run with the fixed-$\varepsilon$ model (with smaller softenings), and restart it from a snapshot at $z=1$, so the initial conditions are identical and feature well-developed highly-nonlinear structure already in place. We run to $z=0.5$, so a substantial fraction of the Hubble time will pass. We further adopt parameters designed to maximally ``stress test'' the implementation here: we allow arbitrarily-adaptive hierarchical timestepping (each particle only obeying its local timestep criterion, so there is a dynamic range of factor up to $\sim 10^{3}$ in timesteps between particles); we adopt the least-restrictive (and therefore also least-accurate) timestepping and force/integration/tree opening accuracy criteria recommended in the survey from \citet{power:2003.nfw.models.convergence} -- a common choice, it should be noted, for many state-of-the-art cosmological simulations (including some of those in e.g.\ \citealt{mansfield.avestruz:2021.halo.bias.over.softening,2021MNRAS.501.5051J,2021arXiv211002097M,nadler:2022.symphony} discussed in the text), of course with the additional criteria imposed as required by \S~\ref{sec:timesteps}; and use the lower-order leapfrog integrator (and, since this only requires one loop of the gravity calculation, adopt the ``drifted'' implementation from \S~\ref{sec:time.variant}). Like all of our ``zoom'' simulations, the initial ($z>100$) conditions feature particles with optional different ``type'' or species labels corresponding to different nested mass resolution (e.g.\ ``high'' resolution inside the zoom-in region, ``low'' resolution outside, and a nested level of ``buffer'' particles with intermediate resolution in between). To test interactions of different species with different softening laws, in our restart we adopt fixed-$\varepsilon$ softenings for the ``intermediate''/``buffer'' particles, and adaptive tidal softening but with different values of $\xi$ for particles of the ``high'' and ``low'' resolution tags, to ensure we have interactions between particle species using distinct softening laws, as well as a very wide range of $\varepsilon$. Finally note some care is needed discussing energy conservation in a cosmological simulation: we define the zero point of the energy at each redshift as the energy computed numerically using the standard definitions from \S~\ref{sec:deriv} of a uniform density periodic box moving with the Hubble flow (obtained by rescaling our cosmological initial conditions). 

Despite all these issues, we see that the energy conservation is nearly identical and quite accurate in the fixed-$\varepsilon$ and tidal schemes. Both feature a small drift of order $\sim 1\%$ per Hubble time, which is a known effect that owes primarily to the adaptive hierarchical timestepping scheme, rather than to the choice of softening. Using the same softening law for all the particle species gives similar results. But removing the correction terms $\boldsymbol{\Upsilon}$, we see the energy error runs away to $\mathcal{O}(1)$-level in a Hubble time. Moreover, in the ``no correction term'' (no $\boldsymbol{\Upsilon}$ terms) runs, on small scales an error appears which is very similar to that described in \citet{hopkins:fire2.methods} for test simulations which adopted the neighbor-based $\tilde{\rho}$/$n_{\rm NGB}$-based models but did {\em not} include an appropriate timestep restriction as discussed in \S~\ref{sec:timesteps} -- not surprising, since that also necessarily leads to large violations of energy conservation. In these cases, large local violations of energy conservation lead to the formation of spurious bound resolution-scale ``clumps'' (often consisting of tens of particles) which lose energy and so collapse down to whatever minimum softening length one imposes numerically. Such events are rare (a few per parent halo) but clearly unphysical.

\subsection{Additional Cosmological Halo Properties}
\label{sec:extra.tests:halos}

\begin{figure}
	\includegraphics[width=0.95\columnwidth]{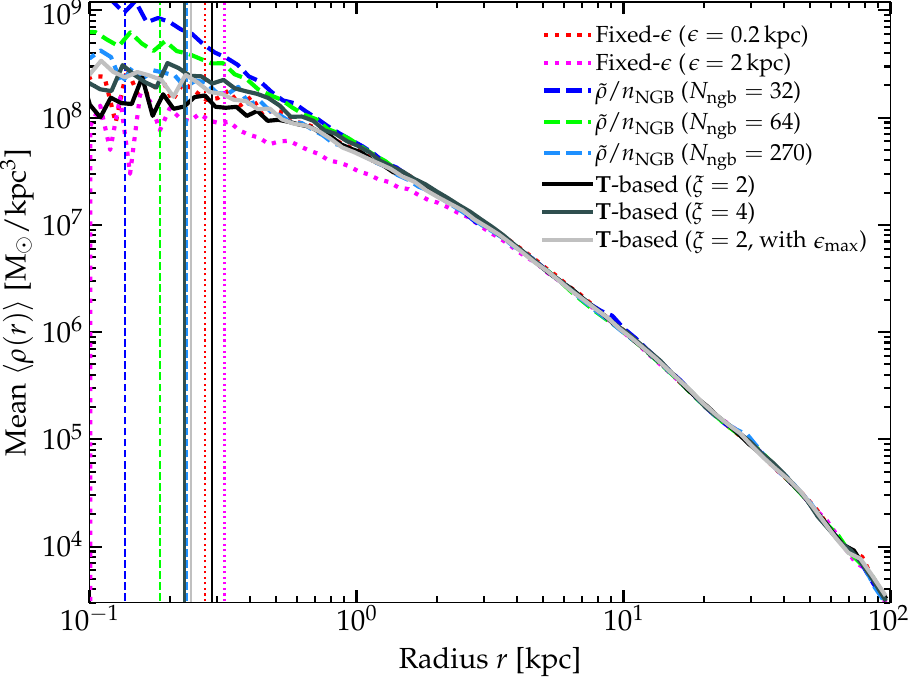}
	\includegraphics[width=0.95\columnwidth]{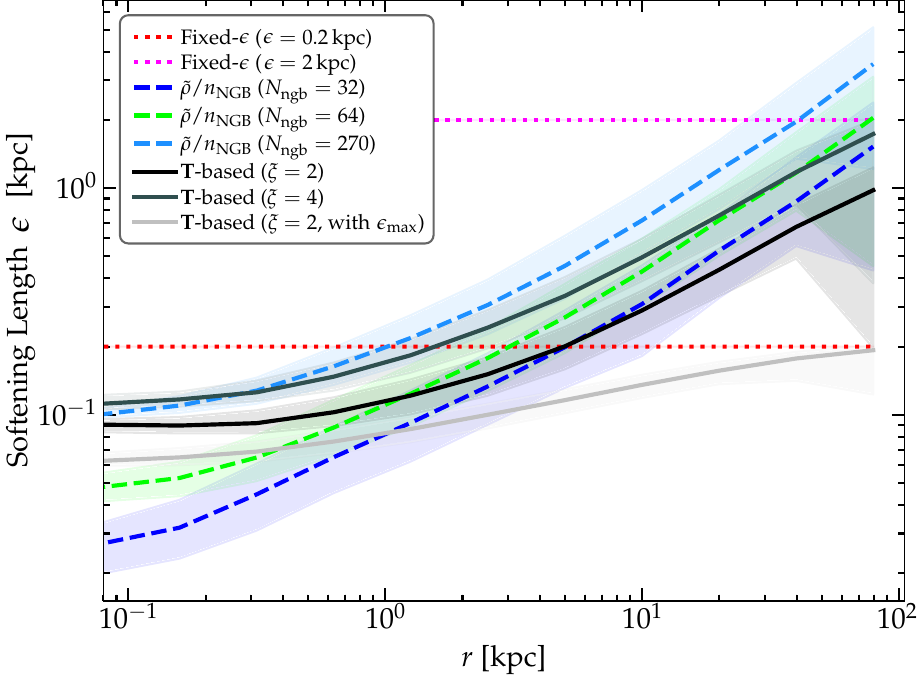}
	\vspace{-0.1cm}
	\caption{{\em Top:} Additional radial density profiles of the primary halo, as Fig.~\ref{fig:cosmo.profiles}, for the additional softening variants run. For each softening ``rule'' (e.g.\ fixed-$\varepsilon$, tidal, or neighbor-based), the behaviors are quite similar to the examples in Fig.~\ref{fig:cosmo.profiles}, and all show excellent agreement at radii well outside the expected convergence radii in this particular diagnostic (except for the largest fixed-softening run with $\varepsilon=2\,$kpc, which as expected deviates from the converged profile within a couple softening lengths).
	{\em Bottom:} Values of $\varepsilon$ versus radius $r$ for the same models.
	\label{fig:cosmo.profiles.extra}}
\end{figure}

Fig.~\ref{fig:cosmo.profiles.extra} repeats Fig.~\ref{fig:cosmo.profiles} from the main text, showing the primary halo mass density profile and median softening lengths, for a wider range of simulations. We did not show all of these for clarity in the main text because the different ``groups'' of simulations (using different models for $\varepsilon$) behave essentially identically.

\begin{figure}
	\includegraphics[width=0.95\columnwidth]{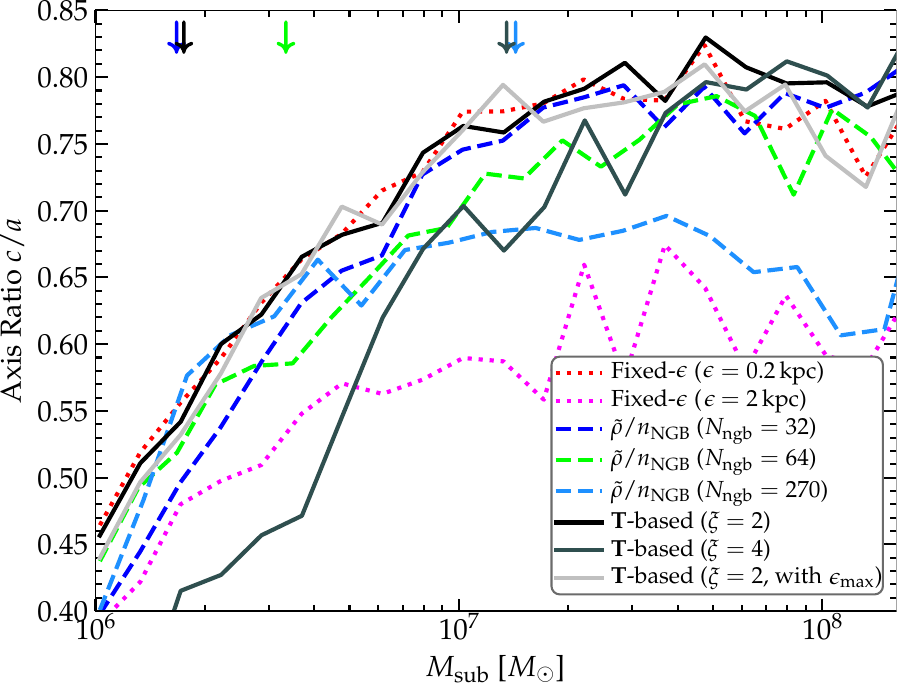}
	\includegraphics[width=0.95\columnwidth]{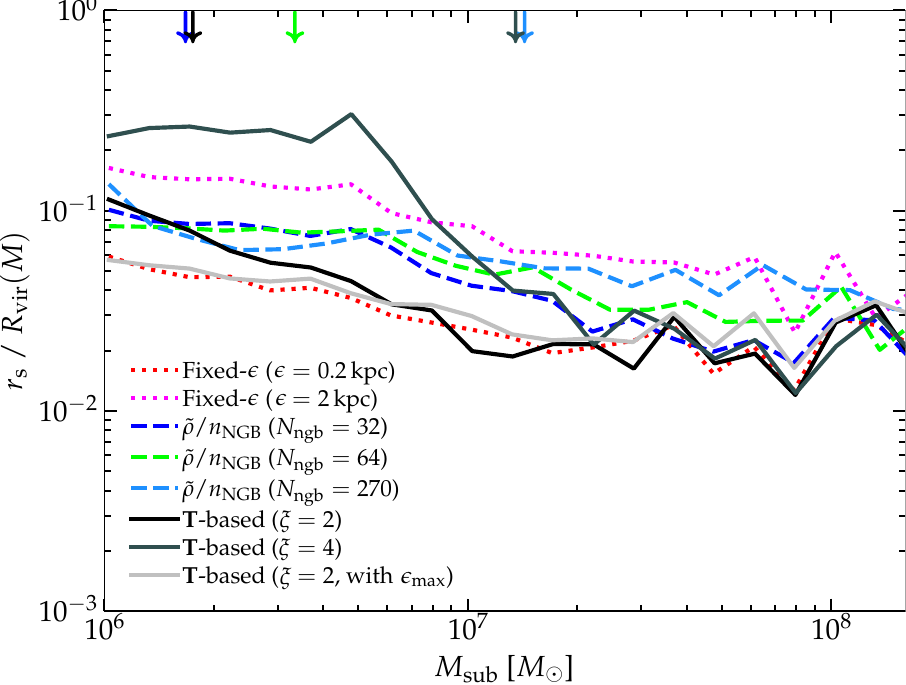}
	\includegraphics[width=0.95\columnwidth]{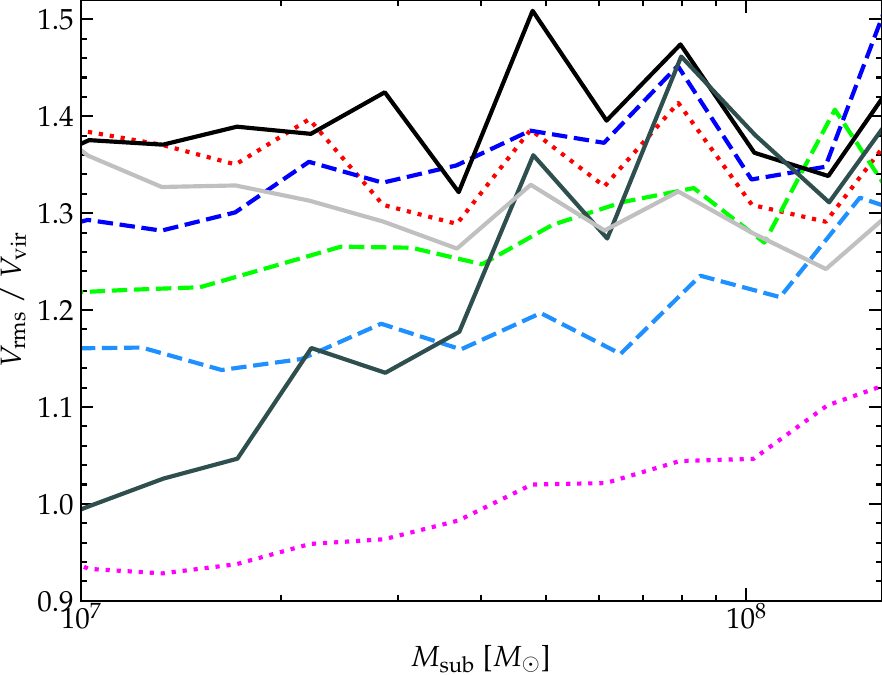}
	\vspace{-0.1cm}
	\caption{Internal properties of subhalos and isolated halos (qualitative trends are the same for both) in the specified mass range as a function of mass, for the simulations in Fig.~\ref{fig:subhalo.mf}.
	{\em Top:} Smallest-to-largest axis ratio $c/a$ from the moments of the halo. 
	{\em Middle:} Scale radius $r_{s}$, normalized to the expected virial radius for an isolated halo of the same virial mass as the instantaneous halo/subhalo mass. 
	{\em Bottom:} Root-mean-square (rms) velocity dispersion relative to the expected virial velocity for an isolated halo of the same virial mass as the instantaneous halo/subhalo mass.
	For each, we plot the median value for all (sub)halos in narrow mass intervals.
	Arrows label the mass where we expect the scale radii of isolated halos to become unresolved for a given softening model, per Fig.~\ref{fig:subhalo.mf}. 
	We see trends consistent with Fig.~\ref{fig:subhalo.mf}: at masses where a given model suppresses the subhalo MF, it tends to over-estimate $r_{s}/R_{\rm vir}$ (i.e.\ artificially ``puff up'' the halo) and under-estimate $c/a$ and $V_{\rm rms}/V_{\rm vir}$.
	\label{fig:cosmo.halo.props}}
\end{figure}

Fig.~\ref{fig:cosmo.halo.props} shows some additional (sub)halo properties versus mass as diagnostics of their internal structure and how it is influenced by softening. In each case we see what we would expect from e.g.\ Figs.~\ref{fig:cosmo.image}-\ref{fig:subhalo.r} in the main text, where the larger fixed-$\varepsilon$ and neighbor-based softening models exhibit systematic biases (with the sense of halos being artificially ``less bound'' or ``puffed up'') at the same mass scales where the subhalo MF was suppressed. 

\section{Integration Scheme Variants with Reduced Numbers of Loops}
\label{sec:time.variant}

In \S~\ref{sec:cpu} we noted that if desired, one could use values of quantities like $\varepsilon$, ${\bf T}$, $\zeta$ etc.\ from a previous timestep to reduce the number of loops needed in a single timestep. Of course, evaluating the gravitational force ${\rm d}{\bf p}_{a}/{\rm d}t$ (Eq.~\ref{eqn:eom}) requires at least one loop of the gravity force evaluation for all $a$. In such a loop, the first term in Eq.~\ref{eqn:eom} ($\propto \nabla_{a}\,\tilde{\phi}_{ab}$) requires knowledge of the particle positions, masses, and force softenings $\varepsilon_{a}$ but nothing else. Evaluation of $\boldsymbol{\Upsilon}_{ab}$ alongside this requires knowledge of $\zeta_{a}/\Omega_{a}$ and ${\bf G}_{a}$ (defined in \S~\ref{sec:ex}). All of these can be computed in prior loops in a timestep (\S~\ref{sec:cpu}). 

However, since the smoothing length $\varepsilon_{a}$ is some function of ${\bf G}_{a}$, one could use ${\bf G}_{a}$ computed in the previous timestep $n-1$ or half timestep $n-1/2$ to calculate $\varepsilon$ for the current timestep $n$: $\varepsilon_{a,\,n} = \mathcal{E}_{a}({\bf G}_{a,\,n-1},\,...)$. Then  calculate ${\bf G}_{a,\,n}$ alongside $({\rm d}{\bf p}_{a}/{\rm d} t)_{n}$ in the loop for step $n$. Given this approximation one can also use $(\zeta_{a}/\Omega_{a})_{n-1}$ and then compute the new $(\zeta_{a}/\Omega_{a})_{n}$. This makes it possible to evaluate all of the relevant quantities within only a single loop of the gravity force computation, which can (in some applications) lead to savings in CPU time and/or complexity.

Like with other integrators of this form, one can improve the accuracy of the method further by drifting the values to some drifted or ``predicted'' ${\bf G}_{a,\,(n)}^{\rm pred}$  (where here $(n)$ denotes the predicted value), e.g.\ ${\bf G}_{a,\,(n)}^{\rm pred} = {\bf G}_{a,\,n-1} + \dot{\bf G}_{a,\,n-1}\,(t_{n}-t_{n-1})$ defined in terms of some estimator $\dot{\bf G}_{a,\,n-1}$ of the time derivative of ${\bf G}_{a}$ evaluated at time $n-1$. But we already have such an estimator defined in \S~\ref{sec:timesteps}, by using ${\rm d}\varepsilon/{\rm d}t \sim \partial_{{\bf G}} \mathcal{E}\,\partial {\bf G}/\partial t$ (Eq.~\ref{eqn:courant}), which we already record and evolve, giving $\dot{\bf G}_{a,\,n-1} \approx ({\rm d}\varepsilon/{\rm d}t)_{a,\,n-1}/( \partial_{{\bf G}} \mathcal{E})_{a,\,n-1}$. For $\zeta/\Omega$, we can define an analogous drift estimate, e.g.\ $\dot{\zeta}_{b} \approx \sum_{c} m_{c}\,\partial_{\varepsilon_{b}} \partial_{\bf x_{a}} \tilde{\phi}_{bc} \cdot ({\bf v}_{b} - {\bf v}_{a})$ (for $\Omega$, since we show below that the $\Omega$ terms are very close to unity and have little effect on the conservation even if we simply ignore them, it seems safe to simply take $\Omega_{b,\,n-1}$). 

Regardless of the drift estimation, by using values from a previous timestep, such a scheme will inevitably break time symmetry and hence introduce long-term error drift in energy. While this is not ideal of course, in practice this may not be a large source of error (depending on the simulation), because $N$-body integration schemes usually introduce such drifts anyways in other parts of the operation (certainly the schemes tested in this paper do so), and this is only comparable to integration error. We show this explicitly in \S~\ref{sec:extra.tests}.

\section{Derivation and Methods of Treating the ``$\mathcal{S}_{2}$'' Terms}
\label{sec:deriv.details}

Recall from Eq.~\ref{eqn:presum} in the text, we have
\begin{align}
\label{eqn:presum.appendix} \frac{{\rm d} \varepsilon_{b}}{{\rm d} {\bf x}_{a}} &= \frac{1}{\Omega^{\prime}_{b}} \frac{\partial \mathcal{E}_{b}}{\partial \mathbf{G}_{b}} \cdot \sum_{c} \left( \frac{\partial \boldsymbol{\mathcal{G}}_{bc}}{\partial {\bf x}_{a}} + \frac{\partial \boldsymbol{\mathcal{G}}_{bc}}{\partial \varepsilon_{c}}  \frac{{\rm d} \varepsilon_{c}}{{\rm d} {\bf x}_{a}}
 \right) 
\end{align}
Inserting this equation into itself (with $b\rightarrow c$, $c\rightarrow d$, etc.) to replace ${\rm d}\varepsilon_{c}/{\rm d}{\bf x}_{a}$, and iterating, we see that this can be expanded into an infinite recursive series, which can be written as:
\begin{align}
\label{eqn:powerseries}\frac{{\rm d} \varepsilon_{k_{1}}}{{\rm d} {\bf x}_{k_{0}}} &= \frac{1}{\Omega^{\prime}_{k_{1}}} \frac{\partial \mathcal{E}_{k_{1}}}{\partial \mathbf{G}_{k_{1}}} \cdot \sum_{k_{2}}  \frac{\partial \boldsymbol{\mathcal{G}}_{k_{1}k_{2}}}{\partial {\bf x}_{k_{0}}} 
+ \frac{1}{\Omega^{\prime}_{k_{1}}} \frac{\partial \mathcal{E}_{k_{1}}}{\partial \mathbf{G}_{k_{1}}} \cdot \\
\nonumber & \sum_{i=3}^{\infty} 
\left [ 
\left \{ 
\prod_{j = 2}^{i-1}
\left (
\sum_{k_{j}=1}^{N}
\frac{1}{\Omega^{\prime}_{k_{j}}} \frac{\partial \mathcal{E}_{k_{j}}}{\partial \mathbf{G}_{k_{j}}}
\cdot
\frac{\partial \boldsymbol{\mathcal{G}}_{k_{j-1} k_{j}}}{\partial \varepsilon_{k_{j}}}
\cdot 
\right )
\right \}
\sum_{k_{i}=1}^{N}
\frac{\partial \boldsymbol{\mathcal{G}}_{k_{i-1} k_{i}}}{\partial {\bf x}_{k_{0}}}
\right ]
\end{align}
where we replace $ \{ a, b, c, d, ...\} \rightarrow \{ k_{0}, k_{1}, k_{2}, k_{3}, ...\} $ for bookkeeping in Eq.~\ref{eqn:powerseries}. 

Noting that this has the form of a series expansion with the leading term $(1/\Omega_{b}^{\prime})\,(\partial \mathcal{E}_{b}/\partial \mathbf{G}_{b}) \cdot \sum_{c} \partial \boldsymbol{\mathcal{G}}_{bc}/\partial {\bf x}_{a}$ which is straightforward to evaluate (discussed in detail in the text), and recalling from the main text that the multiplicative pre-factor $\Omega^{\prime} $ also behaves like a series expansion to leading order,\footnote{In fact, one can show that $\Omega^{\prime}_{b}$, with the form $1-\psi_{b}$ for $\psi_{b}\equiv (\partial \mathcal{E}_{b}/\partial {\bf G}_{b}) \cdot \sum_{c} (\partial \boldsymbol{\mathcal{G}}_{bc}/\partial \varepsilon_{b})$ can be derived from Eq.~\ref{eqn:h.deriv.chain} (albeit more tediously) by recursively replacing terms in that equation giving rise to a series expansion of the form $\sum \psi_{b}^{n} = 1/(1-\psi_{b}) = 1/\Omega^{\prime}_{b}$.} we find it helpful for the sake of presentation to introduce the notation from the text Eqs.~\ref{eqn:w.regular.omega}-\ref{eqn:omega}:
\begin{align}
\label{eqn:depsdx.appendix} \frac{{\rm d} \varepsilon_{b}}{{\rm d} {\bf x}_{a}} &\approx \frac{1}{\Omega_{b}} \frac{\partial \mathcal{E}_{b}}{\partial \mathbf{G}_{b}} \cdot \sum_{c}  \frac{\partial \boldsymbol{\mathcal{G}}_{bc}}{\partial {\bf x}_{a}} \\ 
\nonumber \Omega_{b} &\equiv \Omega^{\prime}_{b} -  \frac{\partial \mathcal{E}_{b}}{\partial \mathbf{G}_{b}} \cdot \sum_{c} \frac{\partial \boldsymbol{\mathcal{G}}_{bc}}{\partial \varepsilon_{c}} \, \mathcal{S}_{2,\,bc} \\
\label{eqn:omega.appendix}  &= 1 - \frac{\partial \mathcal{E}_{b}}{\partial \varepsilon_{b}} - \frac{\partial \mathcal{E}_{b}}{\partial \mathbf{G}_{b}} \cdot \sum_{c}\left[  \frac{\partial \boldsymbol{\mathcal{G}}_{bc}}{\partial \varepsilon_{b}}  +  \frac{\partial \boldsymbol{\mathcal{G}}_{bc}}{\partial \varepsilon_{c}}  \, \mathcal{S}_{2,\,bc} \right] \ .
\end{align}
We can in principle represent the dimensionless term $\mathcal{S}_{2}$ as an infinite tensor sum/power series for each component, but this does not aid  our intuition nor is it explicitly used in any of the actual implementations discussed below -- for now it is here to simply serve as a bookkeeping ``shorthand.'' Important for that purpose, we have ensured $\mathcal{S}_{2}$ is accompanied by the prefactor $\partial\boldsymbol{\mathcal{G}}_{bc}/\partial\varepsilon_{c}$, just like the original implicit term preceding ${\rm d}\varepsilon_{c}/{\rm d}{\bf x}_{a}$ in Eq.~\ref{eqn:presum.appendix}. If this term vanishes, then Eq.~\ref{eqn:depsdx.appendix} becomes exact (equivalent to Eq.~\ref{eqn:presum.appendix}), i.e.\ we are ensured that the all terms in $\mathcal{S}_{2}$ do not contribute in any equations in the text (we can simply set $\mathcal{S}_{2}=0$ and have exact expressions), and similarly when these higher-order expansion terms in Eq.~\ref{eqn:powerseries} are small, $\| \mathcal{S}_{2} \| \ll 1$ is also small. 

Now Eqs.~\ref{eqn:powerseries}-\ref{eqn:omega.appendix} provide an explicit expression for ${\rm d}\varepsilon_{b}/{\rm d}{\bf x}_{a}$ instead of the implicit Eq.~\ref{eqn:presum.appendix}, but at the cost of an infinite series of ever-larger product sets of sums over the particles. The latter obviously cannot realistically be evaluated by ``brute force'' in a computationally reasonable time. However, there are three different ways to handle this ``$\mathcal{S}_{2}$ term'', which we discuss in more detail here.

\subsection{Eliminate It}
\label{sec:deriv.details:eliminate}

The simplest solution is to ensure these terms vanish identically. As stated above, this is easiest to ensure if we simply define our scheme such that  $\partial\boldsymbol{\mathcal{G}}_{bc}/\partial\varepsilon_{c} \rightarrow \mathbf{0}$. And that is easily accomplished for all of the proposed softening schemes presented in \S~\ref{sec:ex}. 

For fixed softening (\S~\ref{sec:ex.fixed}) this always vanishes trivially along with $\boldsymbol{\Upsilon}$ itself. For the ``neighbor-based'' schemes, the default version of the schemes presented in \S~\ref{sec:ex.rho}-\ref{sec:ex.n} have $\boldsymbol{\mathcal{G}}_{bc} \equiv m_{c}^{0,1}\,W_{bc}({\bf x}_{b}-{\bf x}_{c},\,\varepsilon_{b})$, so $\boldsymbol{\mathcal{G}}_{bc}$ depends just on $\varepsilon_{b}$ and this term vanishes except for $c=b$, which makes the terms immediately straightforward to evaluate (see e.g.\ \citealt{price:2007.lagrangian.adaptive.softening} for details) and this is already included exactly in the presented form of $\boldsymbol{\mathcal{\Upsilon}}$ and $\Omega$ (note there are no $\mathcal{S}_{2}$ terms appearing for these schemes in \S~\ref{sec:ex.rho}-\ref{sec:ex.n}). 

For the ``gravity based'' schemes, we have $\boldsymbol{\mathcal{G}}_{bc} \equiv -(m_{c}/2)\,\hat{\phi}_{bc}$ (\S~\ref{sec:ex.phi}), $\boldsymbol{\mathcal{G}}_{bc,\,\alpha} \equiv (m_{c}/2)\,\partial_{\alpha} \hat{\phi}_{bc}$ (\S~\ref{sec:ex.a}),\footnote{Note that we use the notation $\partial_{\alpha}$, $\partial_{\beta}$ in the expressions in this Appendix in place of the main text notation $\partial_{i}$, $\partial_{j}$, to avoid confusion with the $k_{i}$, $k_{j}$ indexing in Eq.~\ref{eqn:powerseries}.} 
$\boldsymbol{\mathcal{G}}_{bc,\,\alpha\beta} \equiv (m_{c}/2)\,\partial_{\alpha} \partial_{\beta} \hat{\phi}_{bc}$ (\S~\ref{sec:ex.T}), so (recalling $\hat{\phi}_{cc}=0$) $\partial\boldsymbol{\mathcal{G}}_{bc}/\partial\varepsilon_{c} \rightarrow \mathbf{0}$ is ensured in all three cases if $\hat{\phi}_{bc}$ does not explicitly depend on $\varepsilon_{c}$. For our favored methods used to symmetrize the forces, $\tilde{\phi}_{bc}=\tilde{\phi}_{bc}({\bf x}_{b}-{\bf x}_{c},\,\varepsilon_{b},\,\varepsilon_{c})$ would indeed depend on both $\varepsilon_{b}$ and $\varepsilon_{c}$. But this is why we emphasized in \S~\ref{sec:ex.phi}-\ref{sec:ex.T} that $\hat{\phi}$ does not have to be the same as $\tilde{\phi}$. As noted therein, one can simply choose $\hat{\phi}_{bc} \rightarrow \hat{\phi}_{bc}({\bf x}_{b}-{\bf x}_{c},\,\varepsilon_{b})$ to be the {\em un-symmetrized} $\hat{\phi}$, which immediately gives $\partial\boldsymbol{\mathcal{G}}_{bc}/\partial\varepsilon_{c} \rightarrow \mathbf{0}$ and ensures all of these series expansion and $\mathcal{S}_{2}$ terms vanish. This is our default choice in our numerical tests in the text. Another reasonable option, also discussed in \S~\ref{sec:ex.phi}, is to choose some $\hat{\phi}_{bc}\rightarrow \hat{\phi}_{bc}({\bf x}_{b}-{\bf x}_{c},\,\varepsilon_{0})$ in terms of some constant $\varepsilon_{0}$, or any $\varepsilon$ which depends on other properties of the system which are not explicitly dependent on $\varepsilon_{c}$, or even $\varepsilon\rightarrow 0$ (the un-softened potential), which does not introduce any serious numerical problems here because it is only used to help obtain an estimate of the actual $\varepsilon$ to be used to ensure finite softening in the forces. 

\subsection{Integrate It}
\label{sec:deriv.details:integrate}

Imagine that one does wish to adopt $\hat{\phi} = \tilde{\phi}$ for a softening like our proposed tidal scheme (\S~\ref{sec:ex.T}), using a symmetrization scheme (\S~\ref{sec:symm}) which makes $\hat{\phi}_{bc}$ a function of $\varepsilon_{c}$. While it is not feasible to evaluate Eq.~\ref{eqn:powerseries}, one could still attempt to evaluate Eq.~\ref{eqn:presum.appendix} directly. Here we briefly suggest a couple methods to do so, for interested readers.

First, note that Eq.~\ref{eqn:presum.appendix} is a linear implicit equation for ${\rm d}\varepsilon_{b}/{\rm d}{\bf x}_{a}$, and so can be written as a linear matrix inversion problem. While the matrix involved would necessarily be large ($\mathcal{O}(N\times N)$), it is worth noting that it is also very sparse, since the ``cross terms'' (the pre-factors $\partial \boldsymbol{\mathcal{G}}_{bc}/\partial\varepsilon_{c}$ for $c\ne b$) are only non-vanishing for a subset of the interacting near-neighbors of each particle. If, for example, one adopts the tidal formulation with $\hat{\phi}=\tilde{\phi}$ and the ${\rm MAX}$-based symmetrization scheme: then this coefficient is non-vanishing only for $|{\bf x}_{c}-{\bf x}_{b}|<\varepsilon_{c}$ and $\varepsilon_{c}>\varepsilon_{b}$. This can therefore be solved in principle with schemes similar to those applied for standard projection schemes (used for preserving $\nabla \cdot {\bf B}=0$ in MHD; see \citealt{brackbill.barnes:projection.divb.mhd.control}) or implicit solvers for multi-band radiation-hydrodynamics or cosmic ray-hydrodynamics \citep{petkova.springel.2009:otvet.gadget,pakmor:2016.arepo.semi.implicit.solver.implementation,hopkins:gizmo.diffusion}. 

However for problems with very large dynamic range in timesteps, such schemes are known to be quite computationally inefficient. A second approach which may be more efficient is to evolve the variables ${\rm d}\varepsilon_{b}/{\rm d}{\bf x}_{a}$ directly. Specifically, if one adopts the time integration scheme from \S~\ref{sec:time.variant}, where values of quantities like ${\bf G}_{a}$ are used from a previous timestep $n-1$ for the computation of the forces in timestep $n$, one can replace $({\rm d}\varepsilon_{b}/{\rm d}{\bf x}_{a})^{n} \rightarrow ({\rm d}\varepsilon_{b}/{\rm d}{\bf x}_{a})^{n-1}$ in the right-hand side of Eq.~\ref{eqn:presum.appendix}, and compute the sum directly akin to the $\zeta$ terms. This is akin to taking a single step of an iterative approximate method for the solution to the inversion scheme described above \citep[see discussion in][for projection schemes in MHD]{hopkins:cg.mhd.gizmo}. However, this could become a large burden in memory, depending on the size of the neighbor lists. 

We only discuss these in hypothetical terms as we have not actually explored versions of these integration methods in the experiments here, given the weak effects we see comparing the approach below (\S~\ref{sec:deriv.details:ignore}) to above (\S~\ref{sec:deriv.details:eliminate}).

\subsection{Ignore It}
\label{sec:deriv.details:ignore}

That said, if one wishes to adopt something like $\hat{\phi} = \tilde{\phi}$ in a method like the tidal scheme with a symmetrization rule like the ${\rm MAX}$ or Eq.~\ref{eqn:hsymm.weight.exp} rule for $\tilde{\phi}$, so $\mathcal{S}_{2}$ does not vanish identically as in \S~\ref{sec:deriv.details:ignore}, we actually see relatively minor effects on energy conservation if we simply ignore the $\mathcal{S}_{2}$ terms (i.e.\ approximate $\| \mathcal{S}_{2} \| \rightarrow 0$, as opposed to trying to integrate it per \S~\ref{sec:deriv.details:integrate}). We test this explicitly in \S~\ref{sec:extra.tests}.

The series expansion of Eq.~\ref{eqn:powerseries} is helpful in understanding why this should be the case. After the ``leading order'' term (in $\sum \partial\boldsymbol{\mathcal{G}}_{k_{1} k_{2}} / \partial {\bf x}_{k_{0}}$) which we always retain, the terms which define $\mathcal{S}_{2}$ form a power series, with each successive term having the dimensions of the leading order term suppressed by an additional power of the term inside the product ($\boldsymbol{\Pi}$) in Eq.~\ref{eqn:powerseries}, which we will temporarily denote $\eta$ (the term in $1/\Omega_{k_{j}}^{\prime}...$). From a strictly dimensional-analysis point of view, we have $\mathcal{O}({\rm d}\varepsilon_{k_{1}}/{\rm d}{\bf x}_{k_{0}}) \sim \mathcal{O}({\rm d}\varepsilon_{k_{1}}/{\rm d}{\bf x}_{k_{0}})_{\mathcal{S}_{2} \rightarrow 0} + \sum_{n=1}^{\infty} \mathcal{O}( \eta^{n} )\,  \mathcal{O}({\rm d}\varepsilon_{k_{1}}/{\rm d}{\bf x}_{k_{0}})_{\mathcal{S}_{2} \rightarrow 0} $ where
\begin{align}
\eta \sim 
\left| 
\sum_{k_{j}=1}^{N} \frac{1}{\Omega_{k_{j}}^{\prime}}
\frac{\partial \mathcal{E}_{k_{j}}}{\partial \mathbf{G}_{k_{j}}}
\cdot
\frac{\partial \boldsymbol{\mathcal{G}}_{k_{j-1} k_{j}}}{\partial \varepsilon_{k_{j}}}
\right| 
\end{align}
If $\eta \ll 1$ typically, then $\| \mathcal{S}_{2} \| \ll 1$, and since these are corrections to the already leading-order correction term (i.e.\ we multiply the extra terms needed for energy conservation by powers of $1 + \mathcal{O}(\eta^{n})$), they will only make the energy conservation slightly more accurate (at a level which can easily be swamped by other sources of integration error).

But it is easy to see that $\eta \ll 1$ is the natural expectation in the valid $N$-body limit, for the same reasons that we noted the coefficients were sparse in \S~\ref{sec:deriv.details:integrate}. Recall $\boldsymbol{\mathcal{G}}_{k_{j-1} k_{j}}$ represents the contribution to ${\bf G}_{k_{j}}$ from particle $k_{j-1}$ and $\varepsilon_{k_{j}} = \mathcal{E}_{k_{j}}$ by definition; and as noted in \S~\ref{sec:deriv.details:integrate}, $\partial \boldsymbol{\mathcal{G}}_{k_{j-1} k_{j}} / \partial \varepsilon_{k_{j}}$ is non-vanishing for only a subset of interacting particles inside their mutual softening radius of compact support. So $\eta$ scales approximately as the fraction of ${\bf G}$ (which for the gravity-based methods, represents the {\em total} gravitational potential or acceleration or tidal tensor) coming from just a subset of the nearest particles inside the softening kernel, which should generically be $\ll1$ for $N$-body approaches to be accurate. 

To illustrate this, consider an example of practical relevance: adopting the tidal scheme (\S~\ref{sec:ex.T}), with $\hat{\phi}\rightarrow \tilde{\phi}$, ${\rm MAX}$-based symmetrization, and the Wendland $C^{2}$ kernel. This gives: 
\begin{align}
\eta \sim &  \left| 
\sum_{k_{j}}
\frac{\varepsilon_{k_{j}}}{6\,{\Omega^{\prime}_{k_{j}}} \| {\bf T}_{k_{j}}^{2} \|} {\bf T}_{k_{j}}^{\alpha\beta} \partial_{\alpha}\,\partial_{\beta} \, m_{k_{j}} \frac{\partial \hat{\phi}_{k_{j-1} k_{j}}}{\partial \varepsilon_{k_{j}}}
\right| \\
\nonumber &= 
\left| 
\sum_{k_{j} \in\, {\rm kernel},\, \varepsilon_{k_{j}} > \varepsilon_{k_{j-1}}}
\frac{
{\bf T}_{k_{j}}^{\alpha\beta} \varpi_{k_{j-1} k_{j}}^{\alpha\beta}\, {\rm d}{\bf T}^{k_{j-1}}_{k_{j},\,\alpha\beta}
}
{
\| {\bf T}_{k_{j}}^{\alpha\beta} 
{\bf T}_{k_{j},\,\alpha\beta} \|
}
\right| \\
\nonumber &\sim \frac{2}{3} 
\left| 
\sum_{k_{j} \in\, {\rm kernel},\, \varepsilon_{k_{j}} > \varepsilon_{k_{j-1}}}
\frac{
{\bf T}_{k_{j}}^{\alpha\beta} {\rm d}{\bf T}^{k_{j-1}}_{k_{j},\,\alpha\beta}
}
{
\| {\bf T}_{k_{j}}^{\alpha\beta} 
{\bf T}_{k_{j},\,\alpha\beta} \|
}
\right| \\ 
\nonumber & 
\lesssim \mathcal{O}\left( \frac{ \| \sum_{k_{j} \in\, {\rm kernel},\, \varepsilon_{k_{j}} > \varepsilon_{k_{j-1}}} {\rm d}{\bf T}_{k_{j-1} k_{j}} \| }{\| {\bf T}_{k_{j}} \|} \right)
\end{align}
In the second line, we have defined ${\rm d}{\bf T}^{k_{j-1}}_{k_{j},\,\alpha\beta} \equiv (m_{k_{j}}/2) \partial_{\alpha}\partial_{\beta} \hat{\phi}_{k_{j-1} k_{j}}$ such that ${\bf T}_{k_{j-1}}^{\alpha\beta}=\sum_{k_{j}} {\rm d}{\bf T}^{k_{j-1}}_{k_{j},\,\alpha\beta}$, i.e.\ this represents the fractional contribution to ${\bf T}$ from a single particle pair, and we collect all other dimensionless $\mathcal{O}(1)$ constants into $\varpi_{k_{j-1} k_{j}}^{\alpha\beta}$ (noting that $\partial \hat{\phi}/\partial \varepsilon$ can be written as some $C\,\hat{\phi}/\varepsilon$). It is easy to see from the properties of $\hat{\phi}$ (for our model choice here) that $\varpi_{k_{j-1} k_{j}}^{\alpha\beta}=0$ unless $|{\bf x}_{k_{j}}-{\bf x}_{k_{j-1}}| < \varepsilon_{k_{j}}$ and $\varepsilon_{k_{j}} > \varepsilon_{k_{j-1}}$, i.e.\ it will be non-zero for (on average) about half the neighbors within the mutual softening kernel ($k_{j} \in \, {\rm kernel}$). In the next line, we assume that the contributions to this sum are dominated by particles within the ``core'' of the kernel (as otherwise the derivative terms here become small rapidly), and use the properties of the kernel to evaluate, for which we simply obtain $\varpi_{k_{j-1} k_{j}}^{\alpha\beta} \sim -2/3$. The last inequality notes the positive-definite denominator terms and assumes ${\bf T}$ is reasonably smooth, to summarize the point that $\eta$ will generally be about equal to or smaller than the fractional contribution of the particles just within the kernel (or actually about half of these, on average) to the total tidal tensor, which is by definition $\ll 1$ in the test-particle limit.

\end{appendix}

\end{document}